\documentclass[journal]{IEEEtran}
\usepackage{amsmath,amssymb}
\usepackage[latin1]{inputenc}
\usepackage[numbers, square, comma, compress]{natbib}
\usepackage{epsfig}
\usepackage{epstopdf}
\usepackage{tikz}

\usepackage{enumitem}
\usepackage{needspace}
\usepackage{chngcntr}

\makeatletter
\renewcommand{\@IEEEsectpunct}{.\ \,}
\makeatother

\newcommand{\Q}{\mathbb{Q}}
\newcommand{\C}{\mathbb{C}}
\newcommand{\Y}{y}
\newcommand{\Z}{z}
\newcommand{\W}{w}

\newcommand{\Hh}{h}
\newcommand{\X}{x}
\newcommand{\M}{M}
\newcommand{\R}{\mathbb{R}}

\newcommand{\I}{\mathbb{I}}

\newcommand{\ZZ}{{\mathbb Z}}
\newcommand{\D}{{\mathcal D}}
\newcommand{\A}{{\mathcal A}}

\newcommand{\bbar}[1]{\setbox0=\hbox{$#1$}\dimen0=.2\ht0 \kern\dimen0 \overline{\kern-\dimen0 #1}}
\newcommand{\OO}{{\mathcal O}}

\providecommand{\mb}[1]{\ensuremath{\mathbf #1}}

\providecommand{\abs}[1]{\ensuremath{\left\lvert #1 \right\rvert}}

\providecommand{\norm}[1]{\ensuremath{\left\Vert #1 \right\Vert}}

\providecommand{\vv}[1]{\textquotedblleft #1\textquotedblright}

\DeclareMathOperator*{\diag}{diag}
\DeclareMathOperator*{\pdet}{pdet}
\DeclareMathOperator*{\Np}{Np}
\DeclareMathOperator{\Tr}{Tr}

\DeclareMathOperator{\tr}{tr}
\DeclareMathOperator{\Gal}{Gal}
\DeclareMathOperator{\dB}{dB}
\DeclareMathOperator{\rd}{rd}
\DeclareMathOperator*{\argmax}{argmax}
\DeclareMathOperator*{\argmin}{argmin}

\newtheorem{thm}{Theorem}[section]
\newtheorem{theorem}[thm]{Theorem}
\newtheorem{cor}[thm]{Corollary}

\newtheorem{lem}[thm]{Lemma}
\newtheorem{prop}[thm]{Proposition}

\newtheorem{definition}[thm]{Definition}
\newtheorem{rem}[thm]{Remark}

\renewcommand{\IEEEQED}{\IEEEQEDopen}

\begin{document}
\title{Almost universal codes for MIMO wiretap channels}
\author{Laura Luzzi,  Roope Vehkalahti and Cong Ling 

\thanks{%
This work was supported in part by FP7 project PHYLAWS (EU FP7-ICT 317562), the Academy of Finland grants \#283135 and \#299916 and the Finnish Cultural Foundation. This work was presented in part at the IEEE International Symposium on
Information Theory (ISIT 2016), Barcelona, Spain \cite{LLV_ISIT2016}.}
\thanks{%
L. Luzzi is with ETIS, UMR 8051, Université Paris-Seine (Université de Cergy-Pontoise, ENSEA,
CNRS), 95014 Cergy-Pontoise, France (e-mail: laura.luzzi@ensea.fr).\par
R. Vehkalahti is now with the Department of Communications and Networking, FI-02150, Aalto University, Espoo, Finland (e-mail: roope.vehkalahti@aalto.fi). While this work was in progress, he was with the Department of Mathematics and Statistics, FI-20014, University of Turku, Finland. \par
C. Ling is with the Department of Electrical and
Electronic Engineering, Imperial College London, London SW7 2AZ,
United Kingdom (e-mail: cling@ieee.org).} 
}

\maketitle

\begin{abstract}
Despite several works on secrecy coding for fading and MIMO wiretap channels from an error probability perspective, the construction of information-theoretically secure codes over such channels remains an open problem. In this paper, we consider a fading wiretap channel model where the transmitter has only partial statistical channel state information. Our channel model includes static channels, i.i.d. block fading channels, and ergodic stationary fading with fast decay of large deviations
for the eavesdropper's channel. 

We extend the flatness factor criterion from the Gaussian wiretap channel to fading and MIMO wiretap channels, and establish a simple design criterion where the normalized product distance / minimum determinant of the lattice and its dual should be maximized simultaneously.

Moreover, we propose concrete lattice codes satisfying this design criterion, which are built from algebraic number fields with constant root discriminant in the single-antenna case, and from division algebras centered at such number fields in the multiple-antenna case. The proposed lattice codes achieve strong secrecy and semantic security for all rates $R<C_b-C_e-\kappa$, where $C_b$ and $C_e$ are Bob and Eve's channel capacities respectively, and $\kappa$ is an explicit constant gap. Furthermore, these codes are almost universal in the sense that a fixed code is good for secrecy for a wide range of fading models.

Finally, we consider a compound wiretap model with a more restricted uncertainty set, and show that rates $R<\bar{C}_b-\bar{C}_e-\kappa$ are achievable, where $\bar{C}_b$ is a lower bound for Bob's capacity and $\bar{C}_e$ is an upper bound for Eve's capacity for all the channels in the set.
\end{abstract}


\section{Introduction}
The wiretap channel model was introduced by Wyner \cite{Wyner}, who showed that secure and reliable communication can be achieved simultaneously over noisy channels even without the use of secret keys. 
Wyner's secrecy condition, which is sometimes called the \emph{weak secrecy} condition, requires that the normalized mutual information $\frac{1}{k}\I(M;Z^k)$ between the confidential message $M$ and the channel output $Z^k$ should vanish when the code length $k$ tends to infinity. However, certain weak secrecy schemes exhibit security flaws \cite{BlochBarros2011}, and today the most widely accepted secrecy metric in the information theory community is Csiszár's \emph{strong secrecy} \cite{Csiszar}, i.e. 
$\I(M;Z^k)$ 
should tend to zero when 
$k$ tends to infinity.

While in the information theory community confidential messages are often assumed to be uniformly distributed, this assumption is not accepted in cryptography. A cryptographic treatment of the wiretap channel was proposed in \cite{Bellare_Tessaro_Vardy} to combine the requirements of the two communities, establishing that achieving \emph{semantic security} in the cryptographic sense is equivalent to achieving strong secrecy for all distributions of the message. This equivalence holds to some extent also for continuous channels \cite[Proposition 1]{LLBS}.

\subsection{Known results on the secrecy capacity of wiretap channels}

The original work of Wyner considered discrete channels; the Gaussian wiretap channel was first studied in
\cite{LeungYanCheong_Hellman} 
where it was proven that its (weak) secrecy capacity is $C_b-C_e$, 
where $C_b$ and $C_e$ are the capacities of Bob and Eve's channels respectively. 
Ergodic fading 
models were first considered in \cite{Liang_Poor_Shamai_2008, Gopala_Lai_ElGamal_2008} and their secrecy capacity was investigated under the assumption of perfect channel state information at the transmitter (CSIT); 
\cite{Gopala_Lai_ElGamal_2008} also considered the scenario where the CSI of the legitimate channel is perfectly known, but there is only statistical information about the wiretapper's channel, and gave some degrees of freedom results in this case. All these early works were under the assumption of weak secrecy. 

Clearly, the assumption of perfect CSIT about Eve's channel is unrealistic in most cases, and for fast fading channels, even the assumption of perfect CSIT about Bob's channel may be problematic. 
A general (non-explicit) formula for the secrecy capacity of a fading wiretap channel with imperfect CSIT was given in \cite{Bloch_Laneman_PartialCSI} for an intermediate secrecy metric based on variational distance. 
With statistical CSIT only, 
the weak secrecy capacity is $C_b-C_e$ for i.i.d. Rayleigh fading wiretap channels such that Bob and Eve's channels are independent \cite{Lin_Lin_MISOSE}.

Note that while it is often argued that it is possible to obtain strong secrecy from weak secrecy \vv{for free} using the privacy amplification technique in \cite{Maurer_Wolf}, applying this technique to fading channels without CSIT seems to be an open problem; see the discussion in \cite{He_Yener_2014}. 

The weak secrecy capacity of multiple-input multiple output (MIMO) Gaussian wiretap channels was investigated in \cite{Khisti_Wornell_2010, Oggier_Hassibi_2011, BustinLiuPoorShamai2009, LiuShamai2009} 
assuming perfect 
CSIT. 
In this setting, it was also shown in \cite{Khina_Kochman_Khisti} that the MIMO channel can be decomposed into parallel channels, allowing to use scalar Gaussian codebooks to achieve strong secrecy. 

The case of fading channels where only statistical CSIT is available is less well-understood.   In \cite{Lin_Lin} it was shown that the weak secrecy capacity for i.i.d. Rayleigh fading MIMO wiretap channels is $C_b-C_e$ 
 if Bob and Eve's channels are independent. 

Yet these channel models are rather restrictive, since knowing the channel statistics of the wiretapper is a strong assumption. 
 A more general model is the \emph{compound channel}\footnote{
We note that typically the compound model in the MIMO literature refers to a scenario where the uncertainty set consists of static channels.
However, one can consider more general uncertainty sets which contain both static and time-varying channels.}, where Bob and Eve's channels belong to a certain uncertainty set $(\mathcal{D}_b, \mathcal{D}_e)$. Following the  standard convention in
\cite{Lapidoth_Narayan}  we say that a sequence of wiretap codes achieve rate $R$   if it achieves the strong secrecy rate $R$ for each of the pairs $(D_b, D_e)\in (\mathcal{D}_b, \mathcal{D}_e) $ in the uncertainty set \emph{uniformly}, i.e. Eve's leakage and Bob's error probability tend to zero uniformly.  The compound capacity can then be defined  as the
maximal achievable rate. 
The ultimate goal of code design would then be to find a code that uniformly achieves rate $C_b-C_e$ for all the channel pairs 
such that Bob's capacity is lower bounded by $C_b$ and Eve's capacity is
 upper bounded by $C_e$.  Unfortunately it is not known whether this goal is achievable. 
The secrecy capacity of degraded compound wiretap channels was studied in
\cite{LiangKramerPoorShamai2009, Ekrem_Ulukus_2010, BjelakovicBocheSommerfeld2013}. 
 An arbitrarily varying MIMO channel with no CSI about the wiretapper was considered in \cite{He_Yener_2014}, assuming 
 perfect CSI of the legitimate channel and that the wiretapper has less antennas than the legitimate receiver. For arbitrarily varying wiretap channels, the deterministic compound capacity was shown to be discontinuous with respect to small variations in the uncertainty set \cite{Boche_Schaefer_Poor}.

If we relax the constraint of uniform convergence in the definition of compound capacity as in Han's definition \cite{Han_2003} (see also the discussion in \cite{Loyka_Charalambous_2016}), then we can deal with more general uncertainty sets. With Han's definition, a wiretap   code  achieves rate $R$ over the compound channel if it achieves this rate for any channel pair $(D_b, D_e)\in (\mathcal{D}_b, \mathcal{D}_e)$ individually. Obviously the compound capacity in Han's sense cannot be smaller than the standard compound capacity, but again it is not known for general uncertainty sets.

\subsection{Previous code constructions}

\subsubsection*{Coding for wiretap channels} 
In the case of discrete memoryless channels, the first wiretap code constructions were based on polar codes \cite{Mahdavifar_Vardy} and LDPC codes \cite{Subramanian} for degraded and symmetric wiretap channels. The polar code construction was extended to general wiretap channels in \cite{Gulcu_Barg,Wei_Ulukus}. 

Lattice codes for the Gaussian wiretap channel under an error probability criterion were first proposed in \cite{Belfiore_Oggier,OSB}. Subsequent works on algebraic lattice codes extended the error probability approach to fading and MIMO channels \cite{BO_ICC,BO_TComm,KHHV,Ong_Oggier}.

In the case of Gaussian wiretap channels, \cite{LLBS} considered the problem of designing lattice codes which achieve strong secrecy and semantic security. Following an approach by Csiszár \cite{Csiszar, Bloch_Laneman}, strong secrecy is guaranteed if the output distributions of the eavesdropper's channel corresponding to different messages are indistinguishable in the sense of variational distance. To this aim, the \emph{flatness factor} of a lattice was proposed in \cite{LLBS} as a fundamental criterion which implies that conditional outputs are indistinguishable. Using random coding arguments, it was shown that there exist families of lattice codes which are \emph{good for secrecy}, meaning that their flatness factor is vanishing, and achieve strong secrecy and semantic security for rates up to $1/2$ nat from the secrecy capacity.
The work \cite{Hamed} adopted the flatness factor as a design criterion in MIMO wiretap channels, yet it is unclear whether that approach achieves strong secrecy.\\
Other non-algebraic lattice code constructions with strong secrecy include polar lattices for Gaussian wiretap channels \cite{Liu_Yan_Ling_secrecy}. A different approach (not based on lattices) in \cite{Tyagi_Vardy_2014} achieves the strong secrecy capacity of the Gaussian wiretap channel using 2-universal hash functions.

\subsubsection*{Universal codes for fading channels} 
Several previous works considered the problem of designing universal codes for fading and MIMO channels without secrecy constraints. Division algebras were first used to obtain MIMO codes that are \vv{approximately universal} from the point of view of the diversity-multiplexing gain trade-off in \cite{Tavildar_Viswanath}. 
Lattice codes with precoded-integer forcing were shown to achieve constant gap to MIMO capacity in slow fading channels in \cite{Ordentlich_Erez}. Most closely related to the present work, \cite{LV} proposed a construction of algebraic lattices based on number field towers which 
are almost universal 
over static and ergodic fading MIMO channels. More recently, 
random lattice codes from Generalized Construction A were shown to achieve compound capacity for the uncertainty set of static MIMO channels \cite{Campello_Ling_Belfiore}. After this paper was first submitted, the Generalized Construction A was extended to a MIMO wiretap setting \cite{Campello_Ling_Belfiore_secrecy}.

\subsection{Main contributions}
\subsubsection*{Main results}
We consider a MIMO fading wiretap channel model where the transmitter has only access to partial statistical CSI, while the legitimate receiver has perfect knowledge of its own channel, and the eavesdropper has perfect knowledge of both channels. 
All static, i.i.d. fading and i.i.d. block fading, and all ergodic fading models are allowed for the main channel. For the eavesdropper's channel, our results 
hold for static channels, i.i.d. fading and block fading channels, and stationary ergodic channels with faster than linear convergence in the law of large numbers.

We propose an algebraic construction of lattices which achieve strong secrecy and semantic security for all secrecy rates $R < C_b-C_e -\kappa$, where $C_b$ and $C_e$ are Bob and Eve's channel capacities respectively, and $\kappa$ is an explicit constant gap which depends on the geometric invariants of the chosen lattices\footnote{For stationary ergodic eavesdropper's channel models with slow convergence in the law of large numbers,
we can guarantee weak secrecy for the same rates.}. 

Our codes are \emph{almost universal} in the sense that given $C_b$ and $C_e$, the \emph{same} code is good for secrecy for a wide range of fading models. Since for many of the channel models we consider we don't know the actual strong secrecy capacity, the achievable rate $C_b-C_e -\kappa$ provides a lower bound. 

Thanks to the universality property, our codes do achieve a constant gap to the compound capacity in Han's sense. The gap is at most $\kappa$ because for some wiretap channels in the compound set, the achievable rate is at most $C_b-C_e$. For individual fading channel pairs where the capacity is known to be $C_b-C_e$, our gap to capacity is exactly $\kappa$.

We also consider a compound channel model with the standard definition of compound capacity, and prove that if we consider a more restrictive uncertainty set, then we can guarantee uniform bounds for the error probability and the leaked information, and our codes achieve a constant gap $\kappa$ to the standard compound capacity. 

Unfortunately, for the best currently known families of lattices from algebraic number fields the value of $\kappa$ turns out to be very large: $9.75$ bits per complex channel use, which for an i.i.d. Rayleigh fading channel corresponds to an SNR advantage of approximately $30 \dB$ for the legitimate receiver. Some perspectives to improve this gap are discussed in the conclusion of the paper.

\subsubsection*{Design criteria} We extend the secrecy criterion based on the flatness factor in \cite{LLBS} to the case of fading and MIMO channels and propose a family of concrete lattice codes from algebraic number fields satisfying this criterion. Intuitively, a vanishing flatness factor, to be defined precisely in our paper, implies that the output distributions of the eavesdropper's channel corresponding to different messages converge to the same distribution (which depends on the eavesdropper's channel). Hence no information is leaked to the eavesdropper asymptotically, even if she knows her channel as well as the legitimate user's channel. 

The key feature to guarantee secrecy is that the \emph{dual} of the faded lattice at the eavesdropper should have a good minimum distance, so that the flatness factor of the faded lattice vanishes with high probability. At the same time, to guarantee reliability, the faded lattice at the legitimate receiver should have a good minimum distance when the channel is not in outage. 

More precisely, we establish a simple design criterion where the \emph{normalized product distance} / \emph{normalized minimum determinant} of the lattice and its dual should be maximized simultaneously; in the case of the Gaussian wiretap channel, the packing density of the lattice and its dual should be maximized\footnote{The dual code also plays a role in the design of wiretap codes for discrete memoryless channels, such as
LDPC codes for  binary erasure wiretap channels \cite{Subramanian}.}. The gap $\kappa$ to the secrecy capacity only depends on these geometric invariants.

\subsubsection*{Lattice construction} 
Our wiretap lattice codes are constructed from a particular sequence of algebraic number fields with constant root discriminant\footnote{Coincidentally, the sequences of number fields that we consider are also used in lattice-based cryptography \cite{Peikert_Rosen}.}. 
These lattices were already used in \cite{ISIT2015_SISO, LV} to design almost universal codes for fading and MIMO channels without secrecy constraints. 
In this paper, we show that the underlying multiplicative structure and constant root discriminant property guarantee that the 
lattices and their duals satisfy our joint design criteria for secrecy. Compared to \cite{LV}, we also improve the coding rate 
by replacing spherical shaping with a discrete Gaussian distribution over the infinite lattice as in \cite{LLBS}.

\subsection{Organization of the paper} To make the paper reader-friendly, we present our methodology firstly for single-antenna fading wiretap channels, then for MIMO wiretap channels, since the latter requires division algebras which are more technical. The rest of the paper is accordingly organized as follows. In Section II, we introduce some technical tools, such as the lattice Gaussian distribution, the flatness factor, and ideal lattices. Section III is devoted to code construction and security proofs for single-antenna fading wiretap channels. The proposed lattice codes can be generalized 
to the MIMO case using the multi-block matrix lattices from division algebras in \cite{LV}. This is accomplished in Section \ref{MIMO_lattices} 
and 
\ref{MIMO_results}, which 
may be skipped
in the first reading. 
In Section \ref{compound}, we extend our achievability results to the compound model. In Section \ref{design_criteria}, we discuss the implications of our results in terms of code design criteria. 
Finally, Section \ref{conclusion} concludes the paper and presents some open problems.

\section{Preliminaries} \label{preliminaries}

\subsection{Basic lattice definitions}
In this section we recall some basic notions about lattices and define the corresponding notations. 

Consider $\C^{k}$ as a $2k$-dimensional real vector space with the real inner product
\begin{equation} \label{inner_product_C}
\langle \mb{x},\mb{y}\rangle=\Re(\mb{x}^{\dagger}\mb{y}).
\end{equation}
This inner product naturally defines a metric on the space
 $\C^k$ by setting $\norm{\mb{x}}= \sqrt{\langle \mb{x},\mb{x}\rangle}$. With this inner product, we can identify $\C^k$ with $\R^{2k}$ with the canonical real inner product, through the isometry 
 \begin{equation} \label{isometry2}
\phi(z_1,\ldots,z_k)=(\Re(z_1),\ldots,\Re(z_k),\Im(z_k),\ldots,\Im(z_k)).
\end{equation}
An $n$-dimensional lattice $\Lambda$ is a discrete subgroup of $\R^n$ defined by 
$$\Lambda=\{M_G\mb{x}\;:\; \mb{x} \in \mathbb{Z}^n\},$$
where the columns of the generator matrix $M_G \in M_n(\R)$ are linearly independent. 

We consider lattices of even dimension $n=2k$ in the Euclidean space $\R^{2k}$, which is identified with the complex space $\C^k$ through (\ref{isometry2}). Given a lattice $\Lambda \subset \C^k$, we define the \emph{dual lattice} as
\begin{equation} \label{dual_lattice_definition}
\Lambda^*=\{\mb{x} \in \C^k \;|\; \forall \mb{y} \in \Lambda , \;\; \langle \mb{x},\mb{y} \rangle \in \mathbb{Z}\}.
\end{equation}
A \emph{fundamental region} of the lattice $\Lambda$ is a measurable set $\mathcal{R}(\Lambda) \subset \R^n$ such that $\R^n$ is the disjoint union of the translates of $\mathcal{R}(\Lambda)$, i.e. $\R^n= \dot\bigcup_{\boldsymbol\lambda \in \Lambda} (\mathcal{R}(\Lambda) + \boldsymbol\lambda)$. We denote by $V(\Lambda)$ the volume of any fundamental region of $\Lambda$, and by $\lambda_1(\Lambda)$ the minimum distance of the lattice, i.e. the smallest norm of a non-zero vector:
$$\lambda_1(\Lambda)=\min_{\boldsymbol\lambda \in \Lambda \setminus \{\mb{0}\}} \norm{\boldsymbol\lambda}.$$

\subsection{Flatness factor and discrete Gaussian distribution}
In this section, we define some fundamental lattice parameters that will be used in the rest of the paper. For more background about the smoothing parameter and the flatness factor in information theory and cryptography, we refer the reader to \cite{Micciancio_Regev,LLBS, Peikert}. 

Let $f_{\sqrt{\Sigma},\mathbf{c}}(\mathbf{z})$ denote the $k$-dimensional circularly symmetric complex normal distribution with mean $\mathbf{c}$ and covariance matrix $\Sigma$:
$$f_{\sqrt{\Sigma},\mathbf{c}}(\mathbf{z})=\frac{1}{\pi^k \det(\Sigma)} e^{-(\mb{z}-\mb{c})^{\dagger}\Sigma^{-1} (\mb{z}-\mb{c})} \quad \forall \mb{z} \in \C^k.$$
We 
use the notation $f_{\sigma,\mb{c}}(\mb{z})$ for $f_{\sigma I,\mb{c}}(\mb{z})$ and $f_{\sqrt{\Sigma}}$ for $f_{\sqrt{\Sigma},0}$.

Given a lattice $\Lambda \subset \C^k$, we consider the $\Lambda$-periodic function
$$f_{\sqrt{\Sigma},\Lambda}(\mb{z})=\sum_{\boldsymbol\lambda \in \Lambda} f_{\sqrt{\Sigma},\boldsymbol\lambda}(\mb{z}), \quad \forall \mb{z} \in \C^k.$$
Note that the restriction of $f_{\sqrt{\Sigma},\Lambda}(\mb{z})$ to any fundamental region $\mathcal{R}(\Lambda)$ is a probability distribution. 
\begin{definition}
Given a complex lattice $\Lambda \subset \C^k$ and a positive definite matrix $\Sigma \in M_{n}(\C)$, the \emph{flatness factor} $\epsilon_{\Lambda}(\sqrt{\Sigma})$ is defined as the maximum deviation of $f_{\sqrt{\Sigma},\Lambda}$ from the uniform distribution over a fundamental region $\mathcal{R}(\Lambda)$ of $\Lambda$, with volume $V(\Lambda)$:
$$\epsilon_{\Lambda}(\sqrt{\Sigma})=\max_{\mathbf{z} \in \mathcal{R}(\Lambda)} \abs{V(\Lambda)f_{\sqrt{\Sigma},\Lambda}(\mathbf{z})-1}.$$
\end{definition}
Compared to \cite{LLBS}, in this paper we use an extended version of the flatness factor for correlated Gaussians, related to the extended notion of the smoothing parameter in \cite{Peikert}. We also extend the definition to the case of complex lattices. In the case of scalar matrices we write $\epsilon_{\Lambda}(\sigma)=\epsilon_{\Lambda}(\sigma I)$. 

Note that correlations can be absorbed by the lattice in the sense that $\epsilon_{\Lambda}(\sqrt{\Sigma})=\epsilon_{\sqrt{\Sigma}^{-1}\Lambda}(I)$, and that $\epsilon_{\Lambda}(\sqrt{\Sigma_1}) \leq \epsilon_{\Lambda}(\sqrt{\Sigma_2})$ if $\Sigma_1$ and $\Sigma_2$ are two
positive definite matrices with $\Sigma_1 \succeq \Sigma_2$.
\begin{definition}
Given a lattice $\Lambda$ and $\varepsilon>0$, the \emph{smoothing parameter}\footnote{We define the smoothing parameter per complex dimension, which differs by a factor $\sqrt{2}$ from the definition in \cite{Micciancio_Regev}. We have adjusted the bounds on $\eta_{\varepsilon}(\Lambda)$ accordingly.} $\eta_{\varepsilon}(\Lambda)$ is the smallest
$s$
such that $\sum_{\boldsymbol\lambda^* \in \Lambda^*\setminus \{\mb{0}\}}
e^{-\frac{\pi}{2} s^2 \norm{\boldsymbol\lambda^*}^2} \leq \varepsilon$, where $\Lambda^*$ is the dual lattice.
\end{definition}
For scalar covariance matrices the smoothing parameter is related to the flatness factor as follows \cite{LLBS}:
$$\sqrt{2\pi}\sigma=\eta_{\varepsilon}(\Lambda) \quad \text{if and only if} \quad \epsilon_{\Lambda}(\sigma I)=\varepsilon.$$
More generally, for $\Sigma \succeq 0$ we can say that
\begin{equation} \label{smoothing_parameter_relation}
 \sqrt{2\pi\Sigma} \succeq \eta_{\varepsilon}(\Lambda) \quad \text{if} \quad \epsilon_{\Lambda}(\sqrt{\Sigma}) \leq \varepsilon.
\end{equation}
The smoothing parameter is upper bounded by the minimum distance of the dual lattice \cite{Micciancio_Regev}. More precisely, we have the following corollary of a result by Banaszczyk \cite{Banaszczyk}:
\begin{lem} \label{Banaszczyk_Corollary}
Suppose that $\Lambda$ is an $n$-dimensional lattice, and consider two constants $c>\frac{1}{\sqrt{2\pi}}$, $C=c\sqrt{2\pi e} e^{-\pi c^2}<1$. 

If $\tau >\frac{\sqrt{n}c}{\lambda_1(\Lambda)}$, then 
\begin{equation} \label{Banaszczyk_sum}
\sum_{\lambda \in \Lambda \setminus \{\mb{0}\}} e^{-\tau^2 \pi \norm{\lambda}^2} \leq \frac{C^n}{1-C^n}.
\end{equation}
Therefore the smoothing parameter of the dual lattice is bounded as follows:
\begin{equation} \label{Banaszczyk_eta}
\eta_{\varepsilon}(\Lambda^*) \leq \frac{\sqrt{2n}c}{\lambda_1(\Lambda)} \quad \text{for} \quad \varepsilon=\frac{C^n}{1-C^n}.
\end{equation}
Equivalently, in terms of the flatness factor,
\begin{equation} \label{Banaszczyk_epsilon}
\epsilon_{\Lambda^*}\left(\frac{\sqrt{n}c}{\sqrt{\pi}\lambda_1(\Lambda)}\right) \leq \frac{C^n}{1-C^n}.
\end{equation}
\end{lem}
\begin{IEEEproof}
Let $\mathcal{B}$ be the open unit ball, and $\rho(A)=\sum_{x \in A} e^{-\pi x^2}$. From Lemma 1.5 in \cite{Banaszczyk} we have that 
$$ \forall c \geq \frac{1}{\sqrt{2\pi}}, \quad \rho(\Lambda \setminus c \sqrt{n} \mathcal{B}) < C^n \rho (\Lambda),$$
where $C=c\sqrt{2\pi e}\, e^{-\pi c^2}$. Then we can write
\begin{align*}
&\rho(\Lambda \! \setminus c \sqrt{n} \mathcal{B}) < C^2 \rho(\Lambda) =C^n \rho(\Lambda \! \setminus c \sqrt{n} \mathcal{B}) + C^n \rho(\Lambda \cap c \sqrt{n} \mathcal{B})\\
& \Rightarrow \quad \rho(\Lambda \setminus c \sqrt{n} \mathcal{B}) < \frac{C^n}{1-C^n} \rho(\Lambda \cap c \sqrt{n} \mathcal{B}).
\end{align*}
Now suppose that $\tau >\frac{c\sqrt{n}}{\lambda_1(\Lambda)}$ and consequently $\tau \Lambda \setminus c \sqrt{n} \mathcal{B} = \tau \Lambda \setminus \{\mb{0}\}$. We have
\begin{align}
&\sum_{\boldsymbol\lambda \in \Lambda \setminus \{\mb{0}\}} e^{-\tau^2 \pi \norm{\boldsymbol\lambda}^2} = \sum_{\tau \boldsymbol\lambda \in \tau \Lambda \setminus \{\mb{0}\}} e^{-\pi \norm{\tau \boldsymbol\lambda}^2}= \rho (\tau \Lambda \setminus \{\mb{0}\}) \notag \\
&=\rho(\tau \Lambda \setminus c \sqrt{n} \mathcal{B}) < \frac{C^n}{1-C^n} \rho(\Lambda \cap c \sqrt{n} \mathcal{B})\notag =\frac{C^n}{1-C^n} \rho(\{\mb{0}\}) \notag \\ &=\frac{C^n}{1-C^n}. \tag*{\IEEEQED}
\end{align}
\let\IEEEQED\relax%
\end{IEEEproof}

The second tool that we need to define our lattice coding schemes is the notion of discrete Gaussian distribution. \\
Given $\mb{c} \in \C^k$ and $\Sigma \succeq 0$, the \emph{discrete Gaussian distribution} over the (shifted) lattice $\Lambda - \mb{c} \subset \C^{k}$ is the following
discrete distribution taking values in $\Lambda - \mb{c}$:
$$D_{\Lambda-\mb{c},\sqrt{\Sigma}}(\boldsymbol\lambda-\mb{c})=\frac{f_{\sqrt{\Sigma}}(\boldsymbol{\lambda}-\mb{c})}{\sum_{\boldsymbol\lambda' \in \Lambda}f_{{\sqrt{\Sigma}}}(\boldsymbol\lambda'-\mb{c})}.$$

The following result is a generalization of Regev's lemma~\cite[Claim~3.9]{Re09} (see also \cite[Lemma 8]{LLBS}) to correlated Gaussian distributions. The proof is given in Appendix \ref{app:GeneralizedRegevLemma}.

\begin{lem} \label{extended_Regev_Lemma}
Let $X_1$ be sampled according to the discrete Gaussian distribution $D_{\Lambda + \mb{c},\sqrt{\Sigma_1}}$ and $X_2$ be sampled according to the continuous Gaussian $f_{\sqrt{\Sigma_2}}$. Let $\Sigma_0=\Sigma_1+\Sigma_2$ and $\Sigma^{-1}=\Sigma_1^{-1}+\Sigma_2^{-1}$. Denote by $g(\mb{x})$ the density of the random variable $X=X_1+X_2$. If
\begin{equation} \label{epsilon_condition}
\epsilon_{\Lambda}(\sqrt{\Sigma}) \leq \varepsilon \leq \frac{1}{2},
\end{equation}
then the $L^1$ distance 
$\mathbb{V}(\,,\,)$ 
between the distributions 
$g$ 
and $f_{\sqrt{\Sigma_0}}$ is bounded as follows:
$$\mathbb{V}(g,f_{\sqrt{\Sigma_0}})\leq 4 \varepsilon.$$
\end{lem}

We will also need a basic result concerning linear transformations of discrete Gaussian distributions, which is proven in Appendix \ref{proof_linear}. 
\begin{lem} \label{linear_transformation}
Let $X$ be sampled according to the $k$-dimensional discrete Gaussian distribution $D_{\Lambda+\mb{c},\sqrt{\Sigma}}$, and let $A \in M_k(\C)$ an invertible matrix. Then the distribution of $Y=AX$ is $D_{A(\Lambda+\mb{c}),\sqrt{A\Sigma A^{\dagger}}}$. 
\end{lem}
Finally, we introduce subgaussian random variables, whose tails behave similarly to the Gaussian tail distributions:
\begin{definition}
A random vector $\mb{z}$ taking values in $\C^{k}$ is \emph{$\delta$-subgaussian} with parameter $\sigma$ if $\forall \mb{t} \in \C^{k}$, $\mathbb{E}[e^{\Re(\mb{t}^{\dagger}\mb{z})}] \leq e^{\delta}e^{\frac{\sigma^2}{4}\norm{\mb{t}}^2}$. 
\end{definition}
For a complex Gaussian vector $\mb{z} \sim \mathcal{N}_{\C}(0,\Sigma)$, $\mathbb{E}[e^{\Re(\mb{t}^{\dagger}\mb{z})}]=e^{\frac{1}{2} \mb{t}^{\dagger}\Sigma \mb{t}}$. 

The following result
holds (see also \cite[Lemma 2.8]{Micciancio_Peikert}):
\begin{lem} \label{lemma_subgaussian}
Let $\mb{x} \sim D_{\Lambda + \mb{c},\sigma}$ be a $k$-dimensional discrete complex Gaussian random variable, and let $A \in M_k(\C)$. Suppose that $\epsilon_{\Lambda}(\sigma)<1$. Then $\forall \mb{t} \in \C^k$,
$$\mathbb{E}[e^{\Re(\mb{t}^{\dagger} A \mb{x})}] \leq \left(\frac{1+\epsilon_{\Lambda}(\sigma)}{1-\epsilon_{\Lambda}(\sigma)}\right) e^{\frac{\sigma^2}{4}\norm{A^{\dagger} \mb{t}}^2}.$$
\end{lem}
The proof can be found in Appendix \ref{proof_subgaussian}.

\subsection{Ideal lattices from number fields with constant root discriminant} \label{ideal_lattices}
Let us first 
formalize some properties of algebraic number fields that are relevant for our construction of algebraic lattice codes in the single-antenna case.
We refer the reader to \cite{Cohn} for the relevant notions about number fields. \\
Let $F$ be a totally complex number field of degree $[F:\Q]=2k$, with ring of integers $\mathcal{O}_F$. We denote by $d_F$ the discriminant of the number field.
The \emph{relative canonical embedding} of $F$ into $\C^k$ is given by
$$\psi(x)=(\sigma_1(x),\ldots,\sigma_k(x)),$$
where $\{\sigma_1,\ldots,\sigma_k\}$ is a set of $\Q$-embeddings $F \to \C$ such that we have chosen one from each complex conjugate pair. \\
Assume that $\mathcal{I}$ is a \emph{fractional ideal} of $F$, that is, there exists some integer $a$ such that $a\mathcal{I}$ is a proper ideal of $\mathcal{O}_F$. Then $\Lambda=\psi(\mathcal{I})$ is a $2k$-dimensional lattice in $\C^k$. In particular, $\psi(\mathcal{O}_F)$ is a lattice.

We define the \emph{codifferent} of $F$ as
$$\mathcal{O}_F^{\vee}=\{ x \in F: {\Tr}_{F/\Q}(x\mathcal{O}_F) \subseteq \mathbb{Z}\}.$$
The codifferent is a fractional ideal,  and its algebraic norm is the inverse of the discriminant:
\begin{equation} \label{discriminant}
N(\mathcal{O}_F^{\vee})=1/d_F.
\end{equation}
 The codifferent embeds as the complex conjugate of the dual lattice:
\begin{equation} \label{dual_lattice}
\Lambda^*=2\overline{\psi(\mathcal{O}_F^{\vee})}.
\end{equation}
Using Lemma \ref{Banaszczyk_Corollary}, equation (\ref{Banaszczyk_eta}), we have that $\forall c > \frac{1}{\sqrt{2\pi}}$
\begin{equation} \label{dual}
\eta_{\varepsilon_k}(\Lambda) \leq \frac{\sqrt{4k}c}{\lambda_1(\Lambda^*)}
= \frac{\sqrt{k}c}{\lambda_1(\overline{\psi(\mathcal{O}_F^{\vee})})}. 
\end{equation}
where $\varepsilon_k=\frac{C^{2k}}{1-C^{2k}} \to 0$ as $k \to \infty$.\footnote{A similar result is shown in \cite[Lemma 6.2]{Peikert_Rosen} for $\varepsilon=2^{-2k}$. In this paper we prefer to consider general $\varepsilon$ in order to get the best possible secrecy rates.} 

Due to the arithmetic mean -- geometric mean inequality, for any fractional ideal $\mathcal{I}$ of $\mathcal{O}_F$, $\lambda_1(\psi(\mathcal{I})) \geq \sqrt{k} (N(\mathcal{I}))^{\frac{1}{2k}}.$
In particular, from (\ref{discriminant}) we get
\begin{equation} \label{lambda1}
\lambda_1(\overline{\psi(\mathcal{O}_F^{\vee})})=\lambda_1(\psi(\mathcal{O}_F^{\vee}))\geq \sqrt{k}/\abs{d_F}^{\frac{1}{2k}}.
\end{equation}
Combining equations (\ref{dual}) and (\ref{lambda1}), we find that the smoothing parameter of $\Lambda$ is upper bounded by the root discriminant: 
\begin{equation} \label{Peikert_Rosen_bound}
\eta_{\varepsilon_k}(\Lambda) \leq c \abs{d_F}^{\frac{1}{2k}} \quad \text{for} \quad \varepsilon_k=\frac{C^{2k}}{1-C^{2k}}.
\end{equation}
Note that as long as $c > \frac{1}{\sqrt{2 \pi}}$, we have $C <1$ and $\varepsilon_k \to 0$ exponentially fast, but the rate of convergence will get slower if $C$ is very close to $1$.\\
In order to have small smoothing parameter when the dimension $k$ is large, we need the discriminant $\abs{d_F}$ to grow as slowly as possible with $k$. \\
Given a family $\mathcal{F}=\{F_k\}$ of number fields  with $[F_k:\Q] \to \infty$ as $k \to \infty$, we define the \emph{asymptotic root discriminant} \cite{Xing} of $\mathcal{F}$ as 
\begin{equation} \label{rd}
\rd_{\mathcal{F}}=\limsup_{k \to \infty} \abs{d_K}^{\frac{1}{[F_k:\Q]}}.
\end{equation}

The following theorem by Martinet \cite{Martinet} proves the existence of infinite towers of totally complex number fields with constant root discriminant:
\begin{theorem}[Martinet] \label{Martinet_theorem}
There exists an infinite tower of totally complex number fields $\mathcal{F}_C=\{F_k\}$ of degree $2k=5\cdot2^t$, such that
\begin{equation} \label{G}
 \abs{d_{F_k}}^{\frac{1}{2k}}=G\,\approx 92.368.
\end{equation}
Consequently, $\rd_{\mathcal{F}_C} \approx 92.368$.
\end{theorem}

The value for $\rd_{\mathcal{F}}$
in Theorem \ref{Martinet_theorem} is not the best known possible; the existence of a family of totally complex number fields $\mathcal{F}_{HM}$ with $\rd_{\mathcal{F}_{HM}}< 82.2$
was proved in \cite{Hajir_Maire_2002}. However, for the number fields in this family, the root discriminant is not constant although it remains bounded.

\begin{rem}
Although in principle the number fields in the families $\mathcal{F}_C$ and $\mathcal{F}_{HM}$
can be computed explicitly for fixed degree $k$, at present an efficient algorithm to do so is not available; see the discussion in \cite{LV}.  
\end{rem}

Given a sequence $\mathcal{F}=\{F_k\}$ of number fields, we denote by 
$\{\Lambda_{\mathcal{F}}^{(k)}\}=\{\psi(\mathcal{O}_{F_k})\}$ 
the corresponding sequence of lattices in $\C^{k}$, with volume 
\begin{equation*} 
V(\Lambda_{\mathcal{F}}^{(k)})=2^{-k}\sqrt{\abs{d_F}}.
\end{equation*}

\subsection{Ideal lattices and normalized product distance}
Given an element $\mb{x}=(x_1,\dots, x_k)\in \C^k$ we will use the notation $\mathrm{p}(\mb{x})=\prod_{i=1}^k |x_i|$, and define 
$$
\mathrm{p}(\Lambda)=\inf_{\mathbf{x} \in \Lambda \setminus \{\mb{0}\}} \mathrm{p}(\mb{x}).
$$

A classically used parameter to design lattices for the Rayleigh fast fading channel \cite{GB96} is the \emph{normalized product distance} 
\begin{equation} \label{Np}
\Np(\Lambda)= \frac{\mathrm{p}(\Lambda)}{V(\Lambda)^{\frac{1}{2}}}.
\end{equation}

The proof of the following will be given in Appendix \ref{app:numberfield}.
\begin{lem}\label{complex}
Let  $F/\Q$ be a totally complex extension of degree $2k$ and  let $\psi$ be the relative canonical embedding and $\mathcal{I}$ a fractional ideal of $F$. Then
$$\Np(\psi(\mathcal{I}))\geq\frac{2^{\frac{k}{2}}}{|d_F|^{\frac{1}{4}}}, \quad \Np(\psi(\mathcal{I})^*)\geq\frac{2^{\frac{k}{2}}}{|d_F|^{\frac{1}{4}}}.$$
\end{lem}

In other words, given a fixed number field $F$, the product distances of all its ideal lattices and their duals are lower bounded by the same value
$2^{\frac{k}{2}}/|d_F|^{\frac{1}{4}}$, which only depends on the size of the discriminant of the field $F$.


This property of number fields immediately implies a result concerning the euclidean distance of lattice points in ideal lattices. 

\begin{definition} \label{Hermite_invariant}
Given a $2k$-dimensional lattice  $\Lambda$ in $\C^k$, its \emph{Hermite invariant} is defined as 
\begin{equation*} 
h(\Lambda)=\inf_{\mathbf{x} \in \Lambda \setminus \{\mb{0}\}} \frac{||\mb{x}||^2}{V(\Lambda)^{\frac{1}{k}}}=\frac{\lambda_1(\Lambda)^2}{V(\Lambda)^{\frac{1}{k}}}.
\end{equation*}
\end{definition}

Using the arithmetic -- geometric mean inequality,
we have for all  $2k$-dimensional lattices that
\begin{equation}\label{AMGM}
(\Np(\Lambda))^2\leq \frac{h(\phi(\Lambda))^k}{k^{k}}.
\end{equation}
Therefore, given a fixed number field $F$, for any ideal $\mathcal{I}$ we have that
\begin{equation}\label{secrecy}
h(\psi(\mathcal{I}))\geq \frac{2k}{|d_F|^{1/2k}}, \quad h(\psi(\mathcal{I})^*)\geq \frac{2k}{|d_F|^{1/2k}}.
\end{equation}

In other words, given a number field with small discriminant, then all the ideal lattices and  their duals have large Hermite invariants.

\section{Single-antenna fading wiretap channel} \label{SISO_section}

\subsection{Channel model}
\begin{figure}[tb]
\begin{center}
\begin{footnotesize}
\begin{tikzpicture}[
nodetype1/.style={
    rectangle,
    rounded corners,
    minimum width=6mm,
    minimum height=7mm,
    draw=black
},
nodetype2/.style={
    rectangle,
    rounded corners,
    minimum width=10mm,
    minimum height=7mm,
    draw=black
},
tip2/.style={-latex,shorten >=0.4mm}
]
\matrix[row sep=0.6cm, column sep=0.8cm, ampersand replacement=\&]{
\node (Alice) {\textsc{Alice}};  \& \node (encoder) [nodetype2]   {\textsc{enc}}; \&
\node (H_b) [nodetype1] {$H_b$}; \&
\node (W) {$\bigoplus$}; \&
\node (decoder) [nodetype2] {\textsc{dec}}; \&
\node (Bob) {\textsc{Bob}};\\
\& (invisible) \&
\node (H_e) [nodetype1] {$H_e$}; \&
\node (We)  {$\bigoplus$}; \&
\node (Eve) {\textsc{Eve}}; \&  \\};
\draw[->] (Alice) edge[tip2] node [above] {$M , M'\,$} (encoder) ;
\draw[->] (encoder) edge[tip2] node [above] (X) {$\mb{x}$} (H_b) ;
\draw[->] (H_b) edge[tip2] (W);
\draw[->] (W) edge[tip2] node [above] {$\mb{y}$}  (decoder) ;
\draw[->] (decoder) edge[tip2] node [above] {$\hat{M}$} (Bob) ;
\draw[->] (We) edge[tip2] node [above] {$\mb{z}$}  (Eve) ;
\draw[->,>=latex] (X) |- node [anchor=east] {} (H_e);
\draw[->] (H_e) edge[tip2] (We);
\node[above=0.5cm] (Nb) at (W.north) {$\mb{w}_b$};
\draw[->] (Nb) edge[tip2] (W);
\node[below=0.5cm] (Ne) at (We.south) {$\mb{w}_e$};
\draw[->] (Ne) edge[tip2] (We);
\end{tikzpicture}
\end{footnotesize}
\caption{The fading wiretap channel.}
\label{fig:wiretapchannel}
\end{center} 
\end{figure}
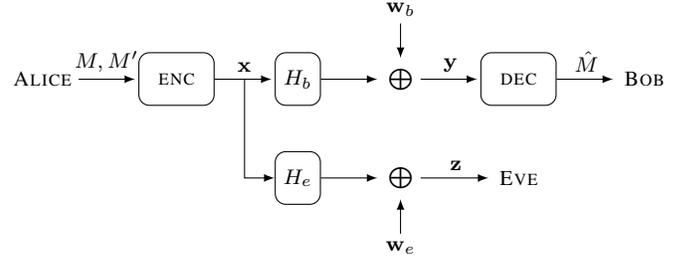

We consider the single-antenna ergodic fading channel model illustrated in Figure \ref{fig:wiretapchannel}, where the outputs $\mb{y}$ and $\mb{z}$ at Bob and Eve's end are given by
\begin{equation} \label{model}
\begin{cases}
{\Y}_i={\Hh}_{b,i}\X_i + \W_{b,i},\\
{\Z}_i={\Hh}_{e,i}\X_i + \W_{e,i},
\end{cases} \quad i=1,\ldots,k
\end{equation}
where $\W_{b,i}$, $\W_{e,i}$ are i.i.d. complex Gaussian vectors with zero mean and variance
$\sigma_b^2$, $\sigma_e^2$ per complex dimension. 
A confidential message $\M$ and an auxiliary message $\M'$ with rate $R$ and $R'$ respectively 
are encoded into $\mb{x}$. We denote by $\hat{\M}$ the estimate of the confidential message at Bob's end.
We define $H_e=\diag(\Hh_{e,1},\ldots,\Hh_{e,k})$, $H_b=\diag(\Hh_{b,1},\ldots,\Hh_{b,k})$.
The input $\mb{x}$ satisfies the average power constraint
\begin{equation} \label{power_constraint}
\frac{1}{k} \sum_{i=1}^{k} \abs{\X_i}^2 \leq P.
\end{equation}

We suppose that $\Hh_{b,i}$, $\Hh_{e,i}$ are isotropically invariant channels such that the channel capacities $C_b$ and $C_e$ are well-defined. All rates are expressed in nats per complex channel use. 

We assume that the weak law of large numbers (LLN) holds for Bob's channel: $\forall \delta>0$
\begin{equation}
\lim_{k \to \infty} \mathbb{P}\left\{ \abs{\frac{1}{k} \sum_{i=1}^k \ln \left(1+\frac{P\abs{h_{b,i}}^2}{\sigma_b^2}\right) - C_b}>\delta\right\}=0, \label{LLN_Bob}
\end{equation}
This general setting includes the Gaussian channel, i.i.d. block fading channels where the size of the blocks is fixed and the number of blocks tends to infinity as well as all ergodic fading channels. 

Moreover, we require a stricter condition for Eve's channel, i.e. the asymptotic rate of convergence in the LLN should be faster than $o\left(\frac{1}{k}\right)$: $\forall \delta'>0,$
\begin{equation}
\!\lim_{k \to \infty} k \; \mathbb{P}\left\{ \abs{\frac{1}{k} \sum_{i=1}^k \ln \left(1+\frac{P\abs{h_{e,i}}^2}{\sigma_e^2}\right)\!- C_e}\!>\delta'\!\right\}=0 \label{LLN_Eve}
\end{equation}
This condition is satisfied for static channels, i.i.d. fading channels and i.i.d. block fading channels, and ergodic channels whose decay of large deviations
is vanishing with rate $o\left(\frac{1}{k}\right)$.
\footnote{
This condition was missing in the conference version of this paper \cite{LLV_ISIT2016}, where it was stated that Corollary \ref{main} holds whenever Eve's channel is ergodic. Actually ergodicity is not sufficient with the current approach. Here we make that statement more precise.}
Recall that in the ergodic case, $C_b=\mathbb{E}_{h_b}\left[\ln\left(1+\frac{P\abs{h_b}^2}{\sigma_b^2}\right)\right]$ and $C_e=\mathbb{E}_{h_e}\left[\ln\left(1+\frac{P\abs{h_e}^2}{\sigma_e^2}\right)\right]$, where $h_b$ and $h_e$ are random variables with the same first order distribution as the processes $\{h_{b,i}\}$,$\{h_{e,i}\}$ \cite{Telatar}.

We suppose that Bob has perfect CSI of his own channel, and Eve has perfect CSI of both channels. Alice has no instantaneous CSI, apart from partial knowledge of channel statistics. More precisely, the knowledge of $C_b$ and $C_e$ and of the properties (\ref{LLN_Bob}) and (\ref{LLN_Eve}) is sufficient for Alice.

\begin{definition}
A coding scheme achieves \emph{strong secrecy} if
\begin{align*}
&\lim_{k \to \infty} \mathbb{P}\{\hat{\M} \neq \M\} = 0, \\
&\lim_{k \to \infty} \mathbb{I}(\M; \mb{z}, H_e)=0. 
\end{align*}
\end{definition}

\begin{definition}
A coding scheme achieves \emph{weak secrecy} if
\begin{align*}
&\lim_{k \to \infty} \mathbb{P}\{\hat{\M} \neq \M\} = 0,  \\
&\lim_{k \to \infty} \frac{1}{k} \mathbb{I}(\M; \mb{z}, H_e)=0. 
\end{align*}
\end{definition}

\begin{rem} \label{remark_CSI}
Even if Eve knows Bob's channel $H_b$, and even though $H_e$ and $H_b$ are possibly correlated, the leakage can still be expressed as $\mathbb{I}(\M; \mb{z}, H_e)$. In fact, the Markov chain $\mb{z} - H_e - H_b$ always holds, and using the chain rule for mutual information twice we get
\begin{align*}
&\mathbb{I}(M; \mathbf{z}|H_e,H_b)=\mathbb{I}(M, H_b; \mb{z}|H_e)-\mathbb{I}(H_b; \mb{z}|H_e)\\
&=\mathbb{I}(M, H_b ; \mb{z}|H_e)=\mathbb{I}(M; \mb{z}|H_e)-\mathbb{I}(H_b; \mb{z}|M, H_e)\\
&=\mathbb{I}(M; \mb{z}|H_e).
\end{align*}
\end{rem}

\begin{rem} \label{remark_secrecy_capacity}
To the best of our knowledge, in the case of statistical CSIT only, for general channels the strong and weak secrecy capacities $C_s$ and $C_s^{w}$ are not known. In \cite{Lin_Lin_MISOSE} the equality $C_s^{w} = C_b-C_e$ was shown in the case of i.i.d. Rayleigh fading channels where Bob and Eve's channels are independent\footnote{Note that the weak secrecy capacity is an upper bound for the strong secrecy capacity.}. 
In \cite[Lemma 2]{Bloch_Laneman_2008}, it was shown that $C_s^{w} \geq C_b-C_e$ for arbitrary wiretap channels. In \cite{Bloch_Laneman} (Corollary 2 and remarks about Theorem 3) it was noted that this result extends to the strong secrecy metrics for i.i.d. channels provided that exponential convergence holds in the Chernoff bound; \cite{Bloch_Laneman_PartialCSI} deals with general ergodic channels but considers an intermediate secrecy metrics (stronger than weak secrecy but weaker than strong secrecy).
\end{rem}

\subsection{Lattice wiretap coding} \label{lattice_wiretap_coding}
Let $\Lambda_e^{(k)} \subset \Lambda_b^{(k)}$ be a pair of nested lattices in $\C^k$ with nesting ratio $\abs{\Lambda_b/\Lambda_e}=e^{kR}$, and  volumes 
\begin{equation} \label{alpha_e}
V(\Lambda_e) = \frac{(\pi e P)^k}{e^{kR'}},\quad V(\Lambda_b) = \frac{(\pi e P)^k}{e^{k(R+R')}},
\end{equation}
where $R'>0$. To simplify the notation, we will omit the dependence on $k$ of the lattices unless necessary.\\
Let $\mathcal{R}(\Lambda_e)$ be a fundamental region of $\Lambda_e$. We consider the secrecy scheme in \cite{LLBS}, where each confidential message $m \in \mathcal{M}=\{1,\ldots,e^{kR}\}$ is associated to a 
coset leader $\boldsymbol{\lambda}_m \in \Lambda_b \cap \mathcal{R}(\Lambda_e)$. To transmit the message $m$, Alice samples $\mb{x} \in \Lambda_b$ from the discrete Gaussian $D_{\Lambda_e + \boldsymbol{\lambda}_m,\sigma_s}$ with 
$\sigma_s^2=P$. 
We denote this lattice coding scheme by $\mathcal{C}(\Lambda_b,\Lambda_e)$. 
\begin{rem}[Power constraint and rate of auxiliary message] \label{theta_t_remark}
Let $\theta_t=\frac{\pi-t}{\pi} 
\to 1$ as $t \to 0$. It follows from \cite[Lemma 6 and Remark 6]{LLBS} that $\forall\, 0<t<\pi$, if 
$\varepsilon_k=\epsilon_{\Lambda_e^{(k)}}(  \sqrt{ \theta_t P})<1 $, 
$$\abs{\mathbb{E}[\norm{\mathbf{x}}^2]-k P} \leq \frac{2 \pi \varepsilon_k}{1-\varepsilon_k} P.$$
Thus as $k \to \infty$, the variance per complex dimension of $\mb{x}$ tends to $P$
provided that
\begin{equation} \label{Lemma6_condition}
\lim_{k \to \infty} \epsilon_{\Lambda_e^{(k)}}( \sqrt{\theta_t P}) = 0,
\end{equation}
and the power constraint (\ref{power_constraint}) is verified asymptotically\footnote{More precisely, one can choose any $\sigma_s^2<P$, so that the power constraint is verified for $k$ large enough. We omit this step to simplify the notation.}.
From \cite[Lemma 7 and Remark 7]{LLBS}, 
the information rate 
$\mathbb{H}(M')$
of the auxiliary message $M'$ (corresponding to the choice of a point in $\Lambda_e$) is bounded by
\begin{align*}
&\abs{ \mathbb{H}(M') - \left(\ln(\pi e P)-\frac{1}{k} \ln V(\Lambda_e)\right)} \leq \nu_t(\varepsilon_k)\\
&=-\log(1-\varepsilon_k)+\frac{\pi}{1-\varepsilon_k} \varepsilon_k(1+1/t^4),
\end{align*}
where $\nu_t(\epsilon_k) \to 0$ as $\epsilon_k \to 0$. 
Therefore we have
\begin{equation*} 
\abs{\mathbb{H}(M')-R'}\leq \nu_t(\epsilon_k).
\end{equation*}
If $\epsilon_k \to 0$, the entropy rate of the auxiliary message tends to $R'$ as $k \to \infty$.
\end{rem}

\subsubsection*{Coding scheme based on number fields with constant root discriminant}
Given a sequence $\mathcal{F}=\{F_k\}$ of number fields, let  $\{\Lambda_{\mathcal{F}}^{(k)}\}$ be the family of lattices defined in Section \ref{ideal_lattices}. We consider scaled versions $\Lambda_b=\alpha_b \Lambda_{\mathcal{F}}^{(k)}$, $\Lambda_e=\alpha_e \Lambda_{\mathcal{F}}^{(k)}$ such that (\ref{alpha_e}) holds.

Since the choice of $R$ and $R'$ determines the scaling factors $\alpha_b$ and $\alpha_e$, we will denote the corresponding lattice coding scheme by $\mathcal{C}(\Lambda_{\mathcal{F}},R,R')$. 

\subsection{Achievable secrecy rates}
We now state our main result, which will be proven in sections \ref{SISO_secrecy} and \ref{SISO_reliability}:
\begin{thm} \label{main_general}
Consider the wiretap scheme $\mathcal{C}(\Lambda_b,\Lambda_e)$ in Section \ref{lattice_wiretap_coding},
and suppose that there exist positive constants $\mathrm{t}_b, \mathrm{t}_e$ such that 
\begin{equation} \label{dual_criterion}
\liminf_{k \to \infty} \Np(\Lambda_b)^{2/k}\geq \mathrm{t}_b, \quad \liminf_{k \to \infty} \Np(\Lambda_e^*)^{2/k} \geq \mathrm{t}_e.
\end{equation} 
where $\Np$ is the normalized product distance defined in (\ref{Np}).\\
If the main channel and the eavesdropper's channel verify the conditions (\ref{LLN_Bob}) and (\ref{LLN_Eve}), then the codes $\mathcal{C}(\Lambda_b,\Lambda_e)$
achieve strong secrecy for any message distribution $p_{M}$, and thus they achieve semantic security, if 
\begin{align} 
\label{rate_conditions}
\begin{split}
 R' > C_e + \ln\left(\frac{e}{\pi}\right)-\ln \mathrm{t}_e, \\ 
 R + R' <C_b -\ln\left(\frac{4}{\pi e}\right)+\ln \mathrm{t}_b. 
\end{split}
\end{align}
Thus, any strong secrecy rate
$$R < C_b -C_e -2\ln \left(\frac{2}{\pi}\right)+\ln \mathrm{t}_b\mathrm{t}_e$$
is achievable with the proposed lattice codes.
\end{thm}

Then, we can state the following Corollary. 

\begin{cor} \label{main}
Let $\mathcal{F}=\{F_k\}$ be a sequence of number fields with 
$\rd_{\mathcal{F}}< \infty$, 
where $\rd_{\mathcal{F}}$ is the asymptotic root discriminant defined in (\ref{rd}).
If the main channel and the eavesdropper's channel verify the conditions (\ref{LLN_Bob}) and (\ref{LLN_Eve}) respectively, then the wiretap coding scheme 
$\mathcal{C}(\Lambda_{\mathcal{F}},R,R')$ 
achieves strong secrecy and semantic security if 
\begin{equation} \label{rate_conditions_main}
R' > C_e + \ln\left(\frac{e \rd_{\mathcal{F}}}{2\pi}\right), \quad R + R' <C_b -\ln\left(\frac{2\rd_{\mathcal{F}}}{\pi e}\right).
\end{equation}
Thus, any strong secrecy rate
$$R < C_b -C_e -2\ln\left(\rd_{\mathcal{F}}/\pi\right)$$
is achievable with the proposed lattice codes.
\end{cor}

\begin{IEEEproof}[Proof of the Corollary]
By using the definition of normalized product distance  and  Lemma \ref{complex} we find that for the number field lattices 
$\mathcal{C}(\Lambda_{\mathcal{F}},R,R')$ 
we have
$\liminf_{k \to \infty}\Np(\Lambda_e)^{\frac{2}{k}}\geq 2/\rd_{\mathcal{F}}$
and 
$\liminf_{k \to \infty}\Np(\Lambda_e^*)^{\frac{2}{k}}\geq 2/\rd_{\mathcal{F}}.$ 
\end{IEEEproof}

\begin{rem}
Let $\mathcal{S}(C_b,C_e)$ denote the set of all ergodic stationary isotropically invariant fading processes $\{(H_{b},H_{e})\}$ such that (\ref{LLN_Bob}) and (\ref{LLN_Eve}) hold. The proposed codes are \emph{almost universal} in the sense that a \emph{fixed} coding scheme $\mathcal{C}(\Lambda^{(k)},R,R')$ with rates satisfying (\ref{rate_conditions}) achieves strong secrecy and semantic security over \emph{all} channels in the set $\mathcal{S}(C_b,C_e)$. Moreover, it is clear from the statement of Corollary \ref{main} that this fixed code will also achieve secrecy over all fading processes in $\mathcal{S}(C_b',C_e')$ for all $C_b' \geq C_b$ and for all $C_e' \leq C_e$. 
\end{rem}

Although a rate of convergence of the order $o\left(\frac{1}{k}\right)$ in the law of large numbers for Eve's channel seems to be necessary for strong secrecy, any rate of convergence is enough to guarantee weak secrecy:
\begin{prop} \label{prop_weak_secrecy}
Suppose that (\ref{dual_criterion}) holds for the wiretap scheme $\mathcal{C}(\Lambda_b,\Lambda_e)$. If the condition (\ref{LLN_Bob}) holds for the main channel and $\forall \delta'>0$ we have
\begin{equation}
\lim_{k \to \infty} \mathbb{P}\left\{ \abs{\frac{1}{k} \sum_{i=1}^k \ln \left(1+\frac{P\abs{h_{e,i}}^2}{\sigma_e^2}\right)\!- C_e}\!>\delta'\!\right\}=0 \label{LLN_Eve_weak}
\end{equation}
for the eavesdropper's channel, then  $\mathcal{C}(\Lambda_b,\Lambda_e)$ achieves weak secrecy for all rates (\ref{rate_conditions}).\\ 
In particular, if $\rd_{\mathcal{F}}< \infty$, any weak secrecy rate 
$R < C_b -C_e -2\ln\left(\rd_{\mathcal{F}}/\pi\right)$
is achievable with the codes $\mathcal{C}(\Lambda_{\mathcal{F}},R,R')$.
\end{prop}

A sketch of the proof of Proposition \ref{prop_weak_secrecy} can be found in Appendix \ref{weak_secrecy}. 

\begin{figure}[tbp]
\centering\includegraphics[width=0.45\textwidth]{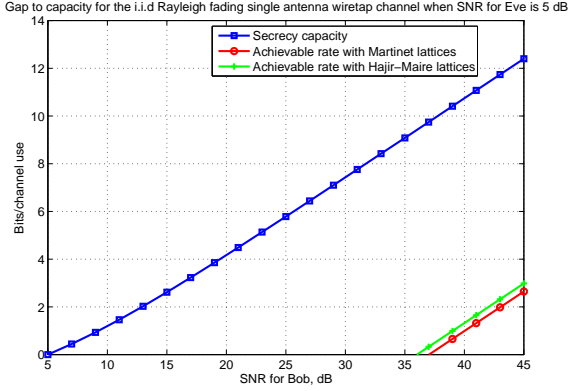}
\caption{Achievable rate for a single-antenna i.i.d. Rayleigh wiretap channel, where the SNR for Eve is fixed at $5 \dB$. \label{Rayleigh_figure}}
\end{figure}

\begin{rem}
At least in the settings in which the secrecy capacity is known and is equal to $C_s=C_b-C_e$, when using the Martinet family of number fields $\mathcal{F}_{C}$ the proposed lattice schemes incur a gap to secrecy capacity of $2 \ln (G/\pi)$ nats per channel use with $G=\rd_{\mathcal{F}_C} \approx 92.368$, i.e. approximately $6.76$ nats (or $9.76$ bits) per channel use. When the main channel and eavesdropper's channel are i.i.d. Rayleigh channels, this corresponds to an SNR gap of approximately $30 \dB$ (see Figure \ref{Rayleigh_figure}).
\end{rem}

\subsection{Proof of Theorem \ref{main_general}: Secrecy} \label{SISO_secrecy}

Let $\mb{x} \in \Lambda_b$ be the lattice point sampled by Alice from the discrete Gaussian $D_{\Lambda_e + \boldsymbol{\lambda}_M,\sigma_s}$.
Then, the received signal $\mb{z}$ at Eve's end is $\mb{z}=H_e\mb{x} + \mb{w}_e$.  Since the message
$\M$ and the channel
$H_e$ are independent, the leakage can be expressed as follows:
{\allowdisplaybreaks
\begin{align}
&\I(\M;\mb{z},H_e)=\I(\M;H_e)+\I(\M;\mb{z}|H_e)=\I(\M;\mb{z}|H_e) \notag\\
&=\mathbb{E}_{H_e}\left[\I(p_{\M|H_e};p_{\mb{z}|H_e})\right]=\mathbb{E}_{H_e}\left[\I(p_{\M};p_{\mb{z}|H_e})\right]. \label{leakage}
\end{align}
}

We want to show that the \emph{average} leakage with respect to the fading is small. In order to do so, we will show that for any confidential message $m$, the output distributions $p_{\mb{z}|H_e,M=m}$ are close to a Gaussian distribution with high probability. 
\subsubsection{Fixed channel sequence}
First, we prove a bound for the leakage for a fixed channel sequence $H_e=\diag(h_{e,1},\ldots,h_{e,k})$. 

\begin{prop}[Bound for the leakage] \label{prop_leakage_SISO}
Suppose that $\Np(\Lambda_e^*)^{\frac{2}{k}} \geq \mathrm{t}_e$ for the $2k$-dimensional lattice $\Lambda_e$, and that $H_e=\diag(h_{e,1},\ldots,h_{e,k})$ is fixed and such that $\frac{1}{k} \sum_{i=1}^k \ln\left(1+\frac{P}{\sigma_e^2} \abs{h_{e,i}}^2\right) \leq \bar{C}_e$. \\
Then $\forall c>\frac{1}{\sqrt{2\pi}}$, if $R'\geq \bar{C}_e +\ln(2c^2e)-\ln \mathrm{t}_e$, for sufficiently large $k \geq \bar{k}(c)$, the leakage is bounded by $\I(p_{\M};p_{\mb{z}|H_e}) \leq 8 k\varepsilon_k R  - 8 \varepsilon_k \ln 8 \varepsilon_k$, where $\varepsilon_k=\frac{C^{2k}}{1-C^{2k}}$ and $C=c\sqrt{2\pi e} e^{-\pi c^2}$. 
\end{prop}

\begin{IEEEproof}
For a fixed channel realization $H_e$ and a fixed message $M=m$, from Lemma \ref{linear_transformation} we have that $H_e \mb{x} \sim D_{H_e \Lambda_e + H_e \boldsymbol{\lambda}_m, \sqrt{H_e H_e^{\dagger}}\sqrt{P}}$. Using Lemma \ref{extended_Regev_Lemma} with $\Sigma_1=H_e H_e^{\dagger} P$, $\Sigma_2=\sigma_e^2 I$,  we have
\begin{equation*} 
\mathbb{V}(p_{\mb{z}|H_e,M=m},f_{\sqrt{\Sigma_0}})\leq 4\varepsilon_k
\end{equation*}
provided that
\begin{equation} \label{fading_secrecy_condition}
\epsilon_{H_e \Lambda_e}(\sqrt{\Sigma})=\epsilon_{\sqrt{\Sigma}^{-1} H_e \Lambda_e}(1) \leq \varepsilon_k \leq \frac{1}{2},
\end{equation}
where we define
$$\Sigma_0=H_e H_e^{\dagger} P + \sigma_e^2 I, \quad \Sigma^{-1}=\frac{(H_e H_e^{\dagger})^{-1}}{P} + \frac{I}{\sigma_e^2}.$$
Recalling the upper bound (\ref{Banaszczyk_eta}) in Lemma \ref{Banaszczyk_Corollary}, we have that for any $c > \frac{1}{\sqrt{2\pi}}$, $C=c\sqrt{2\pi e} e^{-\pi c^2}$, $\varepsilon_k=\frac{C^{2k}}{1-C^{2k}}$,
\begin{align}
 & \eta_{\varepsilon_k}(\sqrt{\Sigma^{-1}}H_e \Lambda_e) \leq \frac{2c\sqrt{k}}{\lambda_1((\sqrt{\Sigma^{-1}}H_e\Lambda_e)^*)} \notag \\ 
 & = \frac{2c\sqrt{k}}{\lambda_1(\sqrt{\Sigma}(H_e^{\dagger})^{-1}\Lambda_e^*)}. \label{faded_smoothing_parameter} 
 \end{align}
Using (\ref{dual_lattice}) and the arithmetic mean -- geometric mean inequality,
{\allowdisplaybreaks
\begin{align*}
&\lambda_1(\sqrt{\Sigma}(H_e^{\dagger})^{-1}\Lambda_e^*)\\
& \geq \sqrt{k} \prod_{i=1}^k \Bigg(\frac{P\sigma_e^2}{\sigma_e^2 +P\abs{h_{e,i}}^2}\Bigg)^{\frac{1}{2k}} \min_{\mb{x} \in \Lambda_e^* \setminus \{\mb{0}\}} \prod_{i=1}^k \abs{x_i}^{\frac{1}{k}} \\
&=\sqrt{k} \prod_{i=1}^k \Bigg(\frac{P\sigma_e^2}{\sigma_e^2 +P\abs{h_{e,i}}^2}\Bigg)^{\frac{1}{2k}} \mathrm{p}(\Lambda_e^*)^{\frac{1}{k}}
\end{align*}
}%
Replacing in (\ref{faded_smoothing_parameter}), we find that 
\begin{equation} \label{eta_bound}
\eta_{\varepsilon_k}(\sqrt{\Sigma^{-1}}H_e \Lambda_e) \leq \frac{2c \prod\nolimits_{i=1}^k (1 +\frac{P}{\sigma_e^2}\abs{h_{e,i}}^2)^{\frac{1}{2k}}}{\mathrm{p}(\Lambda_e^*)^{\frac{1}{k}}\sqrt{P}}.
\end{equation}  
Equivalently, in terms of flatness factor we have
$$\epsilon_{\sqrt{\Sigma^{-1}}H_e \Lambda_e} \left(\frac{2c\prod\nolimits_{i=1}^k (1+\frac{P}{\sigma_e^2}\abs{h_{e,i}}^2)^{\frac{1}{2k}}}{\mathrm{p}(\Lambda_e^*)^{\frac{1}{k}}\sqrt{2\pi P}}\right)\leq \varepsilon_k$$
for fixed fading $H_e$.
Now suppose that 
\begin{equation} \label{sigma_fading}
\frac{2c e^{\frac{\bar{C}_e}{2}}}{\mathrm{p}(\Lambda_e^*)^{\frac{1}{k}}\sqrt{2\pi P}} \leq 1. 
\end{equation}
Then (\ref{fading_secrecy_condition}) holds for sufficiently large $k$ (depending only on $c$) and it follows from \cite[Lemma 2]{LLBS} that
\begin{equation} \label{theorem4}
\I(p_{\M};p_{\mb{z}|H_e}) \leq 8 k\varepsilon_k R  - 8 \varepsilon_k \ln 8 \varepsilon_k.
\end{equation}
Recalling the definition of normalized product distance and the scaling condition (\ref{alpha_e}), 
we have
$$\Np(\Lambda_e^*)=\frac{\mathrm{p}(\Lambda_e^*)}{\sqrt{V(\Lambda_e^*)}}=\mathrm{p}(\Lambda_e^*)\sqrt{V(\Lambda_e)}=\mathrm{p}(\Lambda_e^*)\frac{\sqrt{\pi e P}^k}{e^{kR'/2}}.$$
Thus we can rewrite the condition (\ref{sigma_fading}) as 
$$\frac{2ec^2e^{\bar{C}_e}}{\Np(\Lambda_e^*)^{\frac{2}{k}}e^{R'}} \leq 1$$
In particular if the bound (\ref{dual_criterion}) holds for $\Np(\Lambda_e^*)^{2/k}$, this condition will be guaranteed if
$$\frac{2ec^2e^{\bar{C}_e}}{\mathrm{t}_e e^{R'}} \leq 1.$$
or equivalently if 
$R' \geq \bar{C}_e + \ln\left(2ec^2\right) - \ln\mathrm{t}_e.$
\end{IEEEproof}

\subsubsection{Random channel sequence}
For a random channel sequence $H_e=\diag(h_{e,1},\ldots,h_{e,k})$, we can bound the leakage as follows:
\begin{flalign}
&\mathbb{E}_{H_e}\left[\I(p_{\M};p_{\mb{z}|H_e})\right] \notag \\
& \leq \mathbb{P}\Big\{\prod\limits_{i=1}^k \Big(1+\frac{P\abs{h_{e,i}}^2}{\sigma_e^2} \Big)^{\frac{1}{k}}>e^{C_e+\delta}\Big\} (kR) \notag \\
&
+\mathbb{E}_{H_e}\!\!\left[\I(p_{\M};p_{\mb{z}|H_e}) \,\Big|\, \prod\limits_{i=1}^k \!\Big(1+\frac{P\abs{h_{e,i}}^2}{\sigma_e^2} \Big)^{\frac{1}{k}} \!\!\leq e^{C_e+\delta}\right]\!. \!\! \! &\label{leakage_sum}
\end{flalign}
Given $\delta>0$, the law of large numbers (\ref{LLN_Eve}) implies that 
\begin{equation} \label{upper_bound_capacity}
\mathbb{P}\left\{\prod_{i=1}^k \left(1+\frac{P}{\sigma_e^2}\abs{h_{e,i}}^2\right)^{\frac{1}{k}}>e^{C_e+\delta}\right\} \to 0.
\end{equation}
Therefore the first term vanishes when $k \to \infty$. If the bound (\ref{dual_criterion}) holds, then $\forall \gamma>0$, for sufficiently large $k$, $\Np(\Lambda_e^*)^{2/k}>\mathrm{t}_e-\gamma$.
Using Proposition \ref{prop_leakage_SISO}, $\forall \gamma>0$, the second term in (\ref{leakage_sum}) tends to zero and the scheme achieves strong secrecy provided that 
$$R'\geq C_e + \delta +\ln(2c^2e)-\ln(\textrm{t}_e-\gamma).$$
Since this is true for any $\delta, \gamma >0$ and any $c > \frac{1}{\sqrt{2\pi}}$, we find that a rate
\begin{equation} \label{R_prime}
R' > C_e+ \ln\left(\frac{e}{\pi}\right)-\ln\mathrm{t}_e
\end{equation}
is required for strong secrecy.

\begin{rem}
In equation (\ref{R_prime}), we improve the gap compared to the conference version of this paper, due to considering general $c> \frac{1}{\sqrt{2\pi}}$ rather than $c=1$. 
\end{rem}

\begin{rem} \label{one_sided_Eve}
In this proof 
we are only using the fact that the probability to have a good channel for Eve is vanishing faster than $\frac{1}{k}$. Consequently, in the case when Alice does not know Eve's channel capacity $C_e$ but only knows an upper bound $\bar{C}_e\geq C_e$ such that 
\begin{equation} \label{one_sided_LLN_bound}
\!\lim_{k \to \infty} k\; \mathbb{P}\left\{ \frac{1}{k} \sum_{i=1}^k \ln \left(1 + \frac{P\abs{h_{e,i}}^2}{\sigma_e^2} \right) \!> \bar{C}_e +\delta'\right\}=0
\end{equation}
holds, strong secrecy is still guaranteed provided that 
$R' > \bar{C}_e+ \ln\left(\frac{e}{\pi}\right)-\ln\mathrm{t}_e.$
\end{rem}


\subsubsection{Power constraint and rate of auxiliary message} 
We still need to check that the condition (\ref{Lemma6_condition}) holds. This is required for the power constraint (\ref{power_constraint}), and implies that the information rate of the auxiliary message tends to $R'$ asymptotically (see Remark \ref{theta_t_remark}).
\begin{prop}[Bound for the flatness factor] \label{prop_flatness_SISO}
Suppose that $\Np(\Lambda_e^*)^{\frac{2}{k}} \geq \mathrm{t}_e$ for the $2k$-dimensional lattice $\Lambda_e$. Let $0 < t < \pi$, $\theta_t=\frac{\pi-t}{\pi}$ and $c>\frac{1}{\sqrt{2\pi}}$. If $R'\geq \ln (2ec^2)-\ln \mathrm{t}_e-\ln \theta_t$, then $\epsilon_{\Lambda_e}(\sqrt{\theta_t P}) \leq \varepsilon_k=\frac{C^{2k}}{1-C^{2k}}$, where $C=c\sqrt{2\pi e}e^{-\pi c^2}$.
\end{prop}
\begin{IEEEproof}
By the arithmetic--geometric mean inequality, 
\begin{align*}
&\lambda_1(\Lambda_e^*) \geq \sqrt{k}p(\Lambda_e^*)^{\frac{1}{k}}=\sqrt{k} \frac{\Np(\Lambda_e^*)^{\frac{1}{k}}}{V(\Lambda_e^*)^{\frac{1}{k}}}\\
&\geq 
\sqrt{k} \Np(\Lambda_e^*)^{\frac{1}{k}} \frac{e^{R'/2}}{\sqrt{\pi e P}} \geq \sqrt{k} \sqrt{\mathrm{t}_e}  \frac{e^{R'/2}}{\sqrt{\pi e P}}
\end{align*}
Then for $\varepsilon_k=\frac{C^{2k}}{1-C^{2k}}$, we have 
\begin{align*}
\eta_{\varepsilon_k}(\Lambda_e) \leq \frac{2c\sqrt{k}}{\lambda_1(\Lambda_e^*)} \leq \frac{2c\sqrt{\pi e P}}{\sqrt{\mathrm{t}_e} e^{\frac{R'}{2}}}
\end{align*} 
Therefore $\varepsilon_{\Lambda_e}(\sqrt{\theta_t P})=\varepsilon_k \to 0$ provided that 
$$\sqrt{\theta_t P} \geq \frac{2c\sqrt{\pi e P}}{\sqrt{\mathrm{t}_e} e^{R'/2} \sqrt{2\pi}},$$
or equivalently $R' \geq \ln (2c^2 e) -\ln \mathrm{t}_e -\ln \theta_t$. 
\end{IEEEproof}

For $c \to \frac{1}{\sqrt{2\pi}}$ and $t \to 0$, $\theta_t \to 1$, we find the condition
\begin{equation} \label{variance_condition}
R' > \ln \left(\frac{e}{\pi}\right) -\ln \mathrm{t}_e,
\end{equation}
which is weaker than (\ref{R_prime}). 

\subsection{Proof of Theorem \ref{main_general}: Reliability} \label{SISO_reliability}
Recall that to transmit the message $m$, Alice samples $\mb{x}$ from the discrete Gaussian $D_{\Lambda_e + \boldsymbol{\lambda}_m,\sigma_s}$. \\
Let $\mathbf{y}=H_b\mb{x}+\mb{w}_b$ be the received signal at Bob. Note that if Bob correctly decodes $\mb{x}$, he can also identify the right coset of $\Lambda_e$ in $\Lambda_b$, and consequently the confidential message $m$. 

\subsubsection{Fixed channel sequence}
First of all, we prove an upper bound for Bob's finite-length error probability for a given sequence of channels $H_b=\diag(h_{b,1},\ldots,h_{b,k})$.
\begin{prop}[Bound for the error probability] \label{prop_Pe_SISO}
Suppose that $\Np(\Lambda_b)^{\frac{2}{k}} \geq \mathrm{t}_b$, $\Np(\Lambda_e^*)^{\frac{2}{k}}\geq \mathrm{t}_e$ for the $2k$-dimensional lattices $\Lambda_b$ and $\Lambda_e$, and that $H_b=\diag(h_{b,1},\ldots,h_{b,k})$ is given with $\frac{1}{k}\sum_{i=1}^k \ln\left(1+\frac{P}{\sigma_b^2}\abs{h_{b,i}}^2\right)\geq \bar{C}_b$.\\
Then $\forall c>\frac{1}{\sqrt{2\pi}}$, for code rates $R+R'<\bar{C}_b -\ln \left(\frac{8c^2}{e}\right)+\ln \mathrm{t}_b$, $R'> \ln(2ec^2)-\ln \mathrm{t}_e$, the ML error probability for Bob is bounded by 
$$P_e \leq \frac{1+\varepsilon_k}{1-\varepsilon_k} \varepsilon_k,$$
where $\varepsilon_k=\frac{C^{2k}}{1-C^{2k}}$, and $C=c\sqrt{2\pi e}e^{-\pi c^2}$.
\end{prop}

\begin{IEEEproof}
We suppose that Bob performs MMSE-GDFE preprocessing as in \cite{ElGamal_Caire_Damen}: let $\rho_b=\frac{P}{\sigma_b^2}$, and consider the QR decomposition
$$\widetilde{H}_b=\left(\begin{array}{c} H_b \\ \frac{1}{\sqrt{\rho_b}}I\end{array}\right)=\left(\begin{array}{c} Q_1 \\ Q_2 \end{array}\right)R,$$
where $R, Q_1 \in M_k(\C)$.
Observe that $\tilde{H}_b^{\dagger}\tilde{H}_b=H_b^{\dagger}H_b+\frac{I}{\rho_b}=R^{\dagger}R$, and
\begin{align*}
&\norm{\mathbf{y}-H_b \mathbf{x}}^2 + \frac{1}{\rho_b} \norm{\mathbf{x}}^2 \\
&=\mb{x}^{\dagger} H_b H_b \mb{x} +\mb{y}^{\dagger}H_b\mb{x}-\mb{x}^{\dagger}H_b^{\dagger} \mb{y} +\mb{y}^{\dagger}\mb{y} +\frac{\mb{x}^{\dagger}\mb{x}}{\rho_b}\\
&=\mb{x}^{\dagger} R^{\dagger} R \mb{x} -\mb{y}^{\dagger} Q_1R\mb{x}-\mb{x}^{\dagger} R^{\dagger} Q_1^{\dagger} \mb{y} +\mb{y}^{\dagger} \mb{y}  \\
&=\norm{Q_1^{\dagger} \mathbf{y} - R \mathbf{x}}^2+C,
\end{align*}
where $C$ is a constant which does not depend on $\mathbf{x}$. 

Since the distribution of $\mathbf{x}$ is not uniform, MAP decoding is not equivalent to ML. However, similarly to \cite[Theorem 5]{LLBS}, for fixed $H_b$ which is known at the receiver, the result of MAP decoding can be written as
{\allowdisplaybreaks
\begin{align*}
&\hat{\mathbf{x}}_{\text{MAP}}=\argmax_{\mathbf{x} \in \Lambda_b} p(\mathbf{x} | \mathbf{y})= \argmax_{\mathbf{x} \in \Lambda_b} \left(p(\mathbf{x}) p(\mathbf{y}|\mathbf{x})\right)\\
&=\argmax_{\mathbf{x} \in \Lambda_b} \left(e^{-\frac{\norm{\mathbf{x}}^2}{2P}}e^{-\frac{\norm{\mathbf{y}-H_b\mathbf{x}}^2}{2\sigma_b^2}}\right)\\
&=\argmin_{\mb{x} \in \Lambda_b} \left(\frac{1}{\rho_b} \norm{\mb{x}}^2 + \norm{\mb{y} - H_b \mb{x}}^2\right)\\
&=\argmin_{\mb{x} \in \Lambda_b} \norm{Q_1^{\dagger}\mb{y}-R\mb{x}}^2.
\end{align*}
}
Thus, Bob can compute
\begin{equation} \label{MMSE_model}
\mb{y}'=Q_1^{\dagger}\mb{y}=R\mb{x}+\mb{v},
\end{equation}
where $\mb{v}=Q_1^{\dagger} \mb{w}_b -\frac{1}{\rho_b}(R^{-1})^{\dagger} \mb{x}$ \cite{ElGamal_Caire_Damen}. 

Clearly, the error probability for the original system model with optimal (MAP) decoding is upper bounded by the ML error probability for the system model (\ref{MMSE_model}).

The noise $\mathbf{v}$ is the sum of a discrete Gaussian 
and of a continuous Gaussian. 
We will show that its tails behave similarly to a Gaussian random variable. 

Suppose that a fixed message $m$ has been transmitted, so that $\mb{x} \sim D_{\Lambda_e + \boldsymbol{\lambda}_m,\sqrt{P}}$. 
It follows from Lemma \ref{lemma_subgaussian} that $\mb{x}$ is $\delta$-subgaussian with parameter $\sqrt{P}$ for $\delta=\ln\left(\frac{1+\varepsilon}{1-\varepsilon}\right)$ provided that
\begin{equation} \label{flatness_factor_Bob}
\varepsilon=\epsilon_{\Lambda_e}(\sqrt{P})<1,
\end{equation}
which is guaranteed by (\ref{variance_condition}). This is weaker than the condition (\ref{R_prime}) we have already imposed for secrecy, so it doesn't affect the achievable secrecy
rate. Consequently, for the equivalent noise $\mathbf{v}$,
\begin{align*}
&\mathbb{E}[e^{\Re(\mb{t}^{\dagger}\mb{v})}]=\mathbb{E}\left[e^{\Re(\mb{t}^{\dagger} Q_1^{\dagger} \mb{w}_{b})}\right]\mathbb{E}\left[e^{-\Re\left(\frac{1}{\rho_b} \mb{t}^{\dagger} (R^{-1})^{\dagger} \mb{x}\right)}\right] \\
& \leq \left(\frac{1+\varepsilon}{1-\varepsilon}\right) e^{\frac{\sigma_b^2}{4}\mb{t}^{\dagger}\left(Q_1^{\dagger}Q_1+\frac{1}{\rho_b} (R^{-1})^{\dagger}R^{-1}\right)\mb{t}}= \left(\frac{1+\varepsilon}{1-\varepsilon}\right) e^{\frac{\sigma_b^2}{4}\norm{t}^2}
\end{align*}
since
$$Q_1^{\dagger}Q_1+\frac{1}{\rho_b} (R^{-1})^{\dagger}R^{-1}=(R^{-1})^{\dagger}\left(H_{b}^{\dagger} H_{b}+\frac{1}{\rho_b} I\right)R^{-1}=I.$$
Therefore, $\mb{v}$ is $\delta$-subgaussian with parameter $\sigma_b$.

For fixed $R$, from the union bound for the error probability we get
\begin{align*}
P_e(R) \leq \sum_{\mb{x}' \in \Lambda_b, \mb{x}' \neq \mb{x}} \mathbb{P}\left\{ \mb{x} \to \mb{x}' | R\right\}
\end{align*}
Note that we have
\begin{align*}
&\mathbb{P}\left\{ \mb{x} \to \mb{x}' | R\right\} = \mathbb{P} \left\{ \norm{\mb{v} - R(\mb{x} -\mb{x}')}^2 \leq \norm{\mb{v}}^2\right\} \\
&= \mathbb{P}\left\{ 2 \langle R(\mb{x}-\mb{x}'),\mb{v}\rangle \geq \norm{R(\mb{x}-\mb{x}')}^2\right\}\\
&=\mathbb{P} \left\{ a \geq \frac{1}{2} \norm{R(\mb{x}-\mb{x}')}\right\}
\end{align*}
where $a= \Re \left(\frac{(R(\mb{x}-\mb{x}'))^{\dagger} \mb{v}}{\norm{R(\mb{x}-\mb{x}')}}\right)$ is a real scalar random variable with zero mean. By subgaussianity of $\mb{v}$, $\forall t>0$
$$\mathbb{E}[e^{ta}] \leq e^{\delta} e^{\frac{\sigma_b^2}{4}t^2}.$$
Using the Chernoff bound, we find that $\forall t>0$
\begin{align*}
&\mathbb{P}\left\{ a \geq \frac{1}{2} \norm{R(\mb{x}-\mb{x}')}\right\} \leq \mathbb{E}[e^{ta}] e^{-\frac{t}{2}\norm{R(\mb{x}-\mb{x}')}} \\
& \leq  e^{\delta} e^{\frac{\sigma_b^2}{4}t^2} e^{-\frac{t}{2}\norm{R(\mb{x}-\mb{x}')}}
\end{align*}
The tightest bound is obtained for $t=\norm{R(\mb{x}-\mb{x}')}/\sigma_b^2$ and yields
$$\mathbb{P}\left\{ \mb{x} \to \mb{x}' | R\right\} \leq e^{\delta} e^{-\frac{\norm{R(\mb{x}-\mb{x}')}^2}{4\sigma_b^2}}.$$
Therefore we find
\begin{align*}
P_e(R) \leq e^{\delta} \sum_{\boldsymbol{\lambda} \in R\Lambda_b \setminus \{\mb{0}\}} e^{-\frac{\norm{\boldsymbol{\lambda}}^2}{4\sigma_b^2}}
\end{align*}
Due to Lemma \ref{Banaszczyk_Corollary}, equation (\ref{Banaszczyk_sum}), $P_e(R) \to 0$ as long as
\begin{equation} \label{condition_tau}
\tau^2=\frac{1}{4 \pi \sigma_b^2} > \frac{2 c^2 k}{\lambda_1(R\Lambda_b)^2}.
\end{equation}
The minimum distance in the received lattice is lower bounded as follows using the arithmetic -- geometric mean inequality:
{\allowdisplaybreaks
\begin{align*}
&\lambda_1(R\Lambda_b)^2=\min_{\mb{x} \in \Lambda_b \setminus \{\mb{0}\}} \sum_{i=1}^k \abs{R_i x_i}^2 \notag\\
&\geq   k \prod_{i=1}^k \left(\frac{1}{\rho_b}+\abs{h_{b,i}}^2\right)^{\frac{1}{k}} \mathrm{p}(\Lambda_b)^{\frac{2}{k}}. 
\end{align*}
}%
From the scaling condition (\ref{alpha_e}), we have
$$\mathrm{p}(\Lambda_b)=\Np(\Lambda_b) \sqrt{V(\Lambda_b)}=\Np(\Lambda_b) \frac{\sqrt{\pi e P}^k}{e^{k(R+R')/2}}.$$
Replacing in (\ref{condition_tau}), we find that $P_e(R) \to 0$ when $k \to \infty$ as long as
$$e^{R+R'} < \frac{\Np(\Lambda_b)^{\frac{2}{k}}e}{8c^2}\prod_{i=1}^k \left(1+\frac{P}{\sigma_b^2} \abs{h_{b,i}}^2\right)^{\frac{1}{k}}.$$
Using the assumption (\ref{dual_criterion}), a sufficient condition is that 
$$e^{R+R'} < \frac{\mathrm{t}_be}{8c^2}\prod_{i=1}^k \left(1+\frac{P}{\sigma_b^2} \abs{h_{b,i}}^2\right)^{\frac{1}{k}}.$$
Recalling the hypothesis $\frac{1}{k}\sum_{i=1}^k \ln\left(1+\frac{P}{\sigma_b^2}\abs{h_{b,i}}^2\right)\geq \bar{C}_b$, this concludes the proof.
\end{IEEEproof}

\subsubsection{Random channel sequence} We now consider the average error probability for a random sequence of channels $H_b=\diag(h_{b,1},\ldots,h_{b,k})$. By the law of total probability, $\forall \eta>0$,
\begin{multline*}
P_e \leq \mathbb{P}\Big\{  \prod_{i=1}^k \Big(1+\frac{P}{\sigma_b^2} \abs{h_{b,i}}^2\Big)^{1/k} < e^{C_b-\eta}\Big\}+  \\
+\mathbb{P}\Big\{ \hat{\mb{x}} \neq \mb{x} \; \big| \; \prod_{i=1}^k \Big(1+\frac{P}{\sigma_b^2} \abs{h_{b,i}}^2\Big)^{1/k} \geq e^{C_b-\eta}\Big\}.
\end{multline*}
The first term vanishes when $k \to \infty$ due to
the law of large numbers (\ref{LLN_Bob}). 
If the bound (\ref{dual_criterion}) holds, then $\forall \gamma>0$, for sufficiently large $k$, $\Np(\Lambda_b)^{\frac{2}{k}}>\mathrm{t}_b-\gamma$ and $\Np(\Lambda_e)^{\frac{2}{k}}>\mathrm{t}_e-\gamma$. 
Using Proposition \ref{prop_Pe_SISO}, the second term tends to $0$ if
\begin{align*}
&R+R' < C_b- \eta - \ln \left(\frac{8c^2}{e}\right)+\ln(\textrm{t}_b-\gamma), \\
&R' > \ln(2ec^2)-\ln(\mathrm{t}_e-\gamma).
\end{align*}
Since $\eta, \gamma>0$ and $c>\frac{1}{\sqrt{2\pi}}$ are arbitrary, any rate
\begin{equation} \label{R_b}
R+ R' < C_b- \ln \left(\frac{4}{\pi e}\right)+\ln \textrm{t}_b
\end{equation}
is achievable for Bob, with $R' > \ln(2ec^2)-\ln\mathrm{t}_e$. 

From equations (\ref{R_prime}) and (\ref{R_b}), the proposed coding scheme achieves strong secrecy for any message distribution (and thus semantic security) for any secrecy rate
$$R < C_b -C_e -2\ln\left(\frac{2}{\pi}\right)+\ln \mathrm{t}_b \mathrm{t}_e.$$
This concludes the proof of Theorem \ref{main_general}.

\begin{rem}
In the conference version of this paper \cite{LLV_ISIT2016}, the error probability estimate was based on the sphere bound, while in this paper it is based on the union bound. Both approaches give the same gap to Bob's capacity. 
\end{rem}

\begin{rem} \label{one_sided_Bob}
Note that in this proof we only need the one-sided law of large numbers 
$$ \lim_{k \to \infty} \mathbb{P}\left\{ \frac{1}{k} \sum_{i=1}^k \ln \left(1 + \frac{P}{\sigma_b^2} \abs{h_{e,i}}^2\right) < C_b -\delta\right\}=0.$$
Therefore if Alice does not know Bob's capacity $C_e$ but only knows an upper bound $\bar{C}_b\leq C_b$, reliability holds provided that 
$R+ R' < \bar{C}_b- \ln \left(\frac{4}{\pi e}\right)+\ln \mathrm{t}_b.$
\end{rem}

\begin{rem} \label{one_sided}
From Remarks \ref{one_sided_Eve} and \ref{one_sided_Bob}, we can conclude that if Alice does not know the exact capacities $C_b$ and $C_e$ but is provided with a lower bound $\bar{C}_b\leq C_b$ and an upper bound $\bar{C}_e\geq C_e$ such that (\ref{one_sided_LLN_bound}) holds, the scheme can still achieve strong secrecy rates 
$R < \bar{C}_b -\bar{C}_e -2\ln\left(\frac{2}{\pi}\right)+\ln \mathrm{t}_b \mathrm{t}_e.$
\end{rem}

\subsection{Gaussian wiretap channel} \label{Gaussian_section}
Although in our proofs we used the product distance properties of the lattices $\Lambda_b$ and  $\Lambda_e^*$, if we assume that the channels under consideration are Gaussian, we only need to know that the Hermite invariants of $\Lambda_b$ and $\Lambda_e^*$ are large.

Consider the special case of the channel model (\ref{model}) where $h_{b,i}$, $h_{e,i}$ are constant and equal to $1$ for all $i=1,\ldots,k$: 
\begin{equation}
\begin{cases}
{\Y}_i=\X_i + \W_{b,i},\\
{\Z}_i=\X_i + \W_{e,i},
\end{cases} \quad i=1,\ldots,k
\end{equation}

\begin{prop} \label{rate_Gaussian}
Consider the wiretap scheme $\mathcal{C}(\Lambda_b,\Lambda_e)$ in Section \ref{lattice_wiretap_coding}, and suppose that the Hermite invariants of $\Lambda_b$ and $\Lambda_e^*$ (see Definition \ref{Hermite_invariant}) are bounded by 
\begin{equation} \label{dual_criterion_Gaussian}
\liminf_{k \to \infty} \frac{h(\Lambda_b)}{k} \geq \mathrm{h}_b^2, \quad \liminf_{k \to \infty} \frac{h(\Lambda_e^*)}{k} \geq \mathrm{h}_e^2,
\end{equation} 
for some positive constants $\mathrm{h}_b, \mathrm{h}_e$. 
Then the codes $\mathcal{C}(\Lambda_b,\Lambda_e)$
achieve strong secrecy and semantic security if 
\begin{align*} 
& R' > \ln\left(1+\frac{P}{\sigma_e^2}\right) + \ln\left(\frac{e}{\pi}\right)-\ln \mathrm{h}_e,  \\
& R + R' <\ln\left(1+\frac{P}{\sigma_b^2}\right) -\ln\left(\frac{4}{\pi e}\right)+\ln \mathrm{h}_b. 
\end{align*}
Thus, any strong secrecy rate
$$R < \ln\left(1+\frac{P}{\sigma_b^2}\right)-\ln\left(1+\frac{P}{\sigma_e^2}\right) -2\ln \left(\frac{2}{\pi}\right)+\ln \mathrm{h}_b\mathrm{h}_e$$
is achievable with the proposed lattice codes.
\end{prop}
The proof of Proposition \ref{rate_Gaussian} is very similar to the proof of Theorem \ref{main_general}. A sketch is provided in Appendix \ref{proof_Gaussian}. 

\begin{figure}[tbp]
\centering\includegraphics[width=0.45\textwidth]{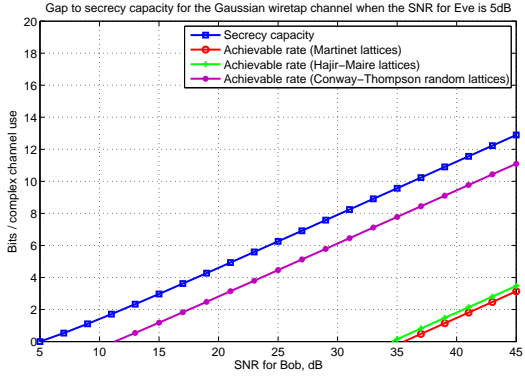}
\caption{Achievable rate on a single-antenna Gaussian wiretap channel, where the SNR for Eve is fixed at $5 \dB$. \label{AWGN_figure}}
\end{figure}

From the bound (\ref{secrecy}), 
we have 
$\mathrm{h}_b=\mathrm{h}_e=2/\rd_{\mathcal{F}}$ 
for the lattices $\Lambda_b=\alpha_b \Lambda_{\mathcal{F}}^{(k)}$, $\Lambda_e=\alpha_e \Lambda_{\mathcal{F}}^{(k)}$ and Proposition \ref{rate_Gaussian} gives achievable rates 
$R < \ln\left(1+P/\sigma_b^2\right)-\ln\left(1+P/\sigma_e^2\right)-2 \ln (\rd_{\mathcal{F}}/\pi)$
for the wiretap coding scheme $\mathcal{C}(\Lambda_{\mathcal{F}},R,R')$. This is the same result that we obtain if we apply directly Corollary \ref{main}. 
For the Martinet sequence $\mathcal{F}_C$ of number fields, recalling that $\abs{d_F}^{1/2k}=\rd_{\mathcal{F}_C}=G \approx 92.368$, 
we get a rather large gap to capacity of $9.75$ bits per complex channel use, or $4.875$ bits per real channel use, corresponding to around $30 \dB$ (see Figure \ref{AWGN_figure}). Thus, a legitimate receiver with an SNR of $35 \dB$ could only be protected against eavesdroppers with an SNR of $5 \dB$ or less.

However, for general lattices the condition (\ref{dual_criterion_Gaussian}) is easier to satisfy than the condition (\ref{dual_criterion}) in Theorem \ref{main_general}.
Using an analogue of the Minkowski-Hlawka theorem for inner product spaces, Conway and Thompson showed the existence of self-dual lattices with large Hermite invariants \cite[Theorem 9.5]{Milnor_Husemoller}:
\begin{theorem}[Conway-Thompson]
For all $n$, there exists a rank $n$ self-dual lattice $\tilde{\Lambda}_n$ with Hermite invariant $h(\tilde{\Lambda}_n) \geq K(n)$, where $K(n) \sim \frac{n}{2 \pi e}$ as $n \to \infty$.
\end{theorem}

Observe that identifying $2k$-dimensional real lattices with $k$-dimensional complex lattices as in (\ref{isometry2}) does not affect the Hermite invariant and dual Hermite invariant, since duality is defined with respect to the real inner product as in (\ref{dual_lattice_definition}). With this identification, for a wiretap scheme $\mathcal{C}(\Lambda_b,\Lambda_e)$ built from the Conway-Thompson sequence of lattices $\Lambda_b=\alpha_b \tilde{\Lambda}_{2k}$, $\Lambda_e=\alpha_e \tilde{\Lambda}_{2k}$  we have $\mathrm{h_b}\mathrm{h_e}=\frac{\sqrt{h(\Lambda_b)h(\Lambda_e)}}{k} \sim \frac{1}{\pi e}$ and applying Proposition \ref{rate_Gaussian} we obtain achievable rates
$$R< \ln\left(1+\frac{P}{\sigma_b^2}\right)-\ln\left(1+\frac{P}{\sigma_e^2}\right) -\ln \frac{4e}{\pi},$$
i.e. a gap of $1.24$ nats or $1.79$ bits per complex channel use from the secrecy capacity, or a loss of approximately $6 \dB$ (see Figure \ref{AWGN_figure}). This is slightly worse than the gap of $1/2$ nat per real channel use (or $1$ nat per complex channel use) obtained in \cite{LLBS} for random lattices using the Minkowski-Hlawka theorem. On the other hand, the design criterion (\ref{dual_criterion_Gaussian}) based on the Hermite invariant, though suboptimal, is more practical to analyze the performance of non-random lattices.

\section{Algebraic lattice constructions for multi-antenna channels} \label{MIMO_lattices}
In this section, we will recall the algebraic constructions of lattice codes for multiple antenna wireless channels, which will be needed for the wiretap coding scheme in the MIMO case.

\subsection{Matrix lattices} \label{section_matrix_lattices}
The space $M_{nk\times n}(\C)$
is a $2n^2k$-dimensional real vector space endowed with a real inner product
\begin{equation} \label{inner_product_MnC}
\langle X,Y\rangle=\Re(\Tr (X^{\dagger}Y)),
\end{equation}
where $\Tr$ is the matrix trace. This inner product defines a metric on the space $M_{nk\times n}(\C)$ by setting $||X||= \sqrt{\langle X,X\rangle}$.
\begin{rem} \label{isometry}
Consider the function $\xi: M_{nk\times n}(\C) \to \C^{n^2k}$ which vectorizes each matrix by stacking its columns. Note that $\xi$ is an isometry between $M_{nk\times n}(\C)$ with the previously defined inner product and $\C^{n^2 k}$ with the inner product (\ref{inner_product_C}).

Given $H \in M_{nk \times nk}(\C)$ and $X \in M_{nk \times n}(\C)$, we have
\begin{equation} \label{tensor}
\xi(HX)=\mathcal{H}\xi(X), \quad \mathcal{H}=H \otimes I_n.
\end{equation}
\end{rem}
Given a matrix $X \in M_{nk\times n}(\C)$ of the form
\begin{equation} \label{X}
X=\begin{pmatrix}X_1 \\ \vdots \\ X_k\end{pmatrix},
\end{equation}
we introduce the notation
$$X^{h} \doteqdot \begin{pmatrix} X_1^{\dagger} \\ \vdots \\ X_k^{\dagger} \end{pmatrix}.$$
We also define the \emph{product determinant} as follows:
\begin{equation} \label{pdet}
\pdet(X)=\prod_{i=1}^k \det(X_i).
\end{equation}
\begin{rem} \label{inequality}
For $X$ of the form (\ref{X}), we have
\begin{align} \label{pdet_bound}
& \norm{X}^2\! =\sum_{i=1}^k \norm{X_i}^2 \stackrel{(a)}{\geq} n \sum_{i=1}^k \abs{\det(X_i)}^{\frac{2}{n}} \notag \\
& \stackrel{(b)}{\geq} nk \prod_{i=1}^k \abs{\det(X_i)}^{\frac{2}{nk}} = nk \abs{\pdet(X)}^{\frac{2}{nk}}.
\end{align}
Here $(a)$ follows from the inequality $\norm{A}^n \geq \abs{\det(A)} n^{n/2}$ for any $A \in M_n(\C)$, and (b) follows from the arithmetic -- geometric mean inequality.
\end{rem}

\begin{definition}
A {\em matrix lattice} $L \subseteq M_{nk\times n}(\C)$ has the form
$$
L=\ZZ B_1\oplus \ZZ B_2\oplus \cdots \oplus \ZZ B_r,
$$
where the matrices $B_1,\dots, B_r$ are linearly independent over $\R$, i.e., form a lattice basis, and $r$ is
called the \emph{rank}  or the \emph{dimension} of the lattice.
\end{definition}
The {\it Gram matrix} of an  $r$-dimensional lattice $L\subset M_{nk\times n}(\C)$ is defined as
$$\textrm{Gr}(L)=\left( \langle X_i,X_j\rangle\right)_{1\le i,j\le r},$$
where $\{ X_i \}_{1 \leq i \leq r}$ is a basis of $L$.  The volume of the fundamental parallelotope of $L$ is then given by
$$V(L)=\sqrt{|\det(\textrm{Gr}(L))|}.$$
\begin{definition}
Given a lattice $L$ in $M_{nk\times n}(\C)$, the \emph{dual lattice} is defined as
$$
L^*=\{ X\in M_{nk\times n}(\C) \;|\;  \forall Y \in L, \;\langle X,Y \rangle \in \ZZ \}.
$$
\end{definition}
We also define the \emph{product determinant} and \emph{normalized minimum determinant} of the matrix lattice $L \subset M_{kn \times n}(\C)$ as follows:
\begin{align*}
\begin{split} \label{normalized_minimum_determinant}
&\pdet(L)=\min_{X \in L \setminus \{0\}} \pdet(X),\\
&\delta(L)= \frac{\pdet(L)}{V(L)^{\frac{1}{2n}}}.
\end{split}
\end{align*}

\subsection{MIMO lattices from division algebras}
We will first recall the construction of single-block space-time codes from cyclic division algebras (see for example \cite{OBV}). Due to space constraints, we refer the reader to \cite{Reiner} for algebraic definitions.
\begin{definition}\label{cyclic}
Let $F$ be an algebraic number field of degree $2k$ and assume   that $E/F$ is a cyclic Galois
 extension of degree $n$ with Galois group
$\Gal(E/F)=\left\langle \sigma\right\rangle$. We can define an associative $F$-algebra
$$
\mathcal{A}=(E/F,\sigma,\gamma)=E\oplus uE\oplus u^2E\oplus\cdots\oplus u^{n-1}E,
$$
where   $u\in\mathcal{A}$ is an auxiliary
generating element subject to the relations
$xu=u\sigma(x)$ for all $x\in E$ and $u^n=\gamma\in F \setminus \{\mb{0}\}$.\\
We call the resulting algebra a \emph{cyclic algebra}. Here $F$ is the center of the algebra $\A$.
\end{definition}

 \begin{definition}
 We  call $\sqrt{[\A:F]}$ the \emph{degree} of the algebra $\A$. It is easily verified that the degree of $\A$ is equal to $n$.
\end{definition}

We consider $\A$ as a right vector space over $E$. Every element $a=x_0+ux_1+\cdots+u^{n-1}x_{n-1}\in\mathcal{A}$, with $x_i \in E$ for all $i=0,\ldots,n-1$,
has the following representation as a matrix:
\[
\phi(a)=\begin{pmatrix}
x_0& \gamma\sigma(x_{n-1})& \gamma\sigma^2(x_{n-2})&\cdots &
\gamma\sigma^{n-1}(x_1)\\
x_1&\sigma(x_0)&\gamma\sigma^2(x_{n-1})& &\gamma\sigma^{n-1}(x_2)\\
x_2& \sigma(x_1)&\sigma^2(x_0)& &\gamma\sigma^{n-1}(x_3)\\
\vdots& & & \ddots & \vdots\\
x_{n-1}& \sigma(x_{n-2})&\sigma^2(x_{n-3})&\cdots&\sigma^{n-1}(x_0)\\
\end{pmatrix}
\]

The mapping $\phi$ is called the \emph{left regular representation} of $\A$ and allows us to embed any cyclic algebra into $M_n(\C)$. Under such an embedding $\phi(\A)$ forms an $2kn^2$-dimensional $\Q$-vector space.

We are particularly interested in algebras $\A$ for which $\phi(a)$ is invertible for all non-zero $a\in\A$.

\begin{definition}\label{divisionalgebra}
A cyclic $F$-algebra $\D$ is a \emph{division algebra} if every
non-zero element of $\D$ is invertible.
\end{definition}
In order to code over several fading blocks, we will next define a multi-block lattice construction based on a cyclic division algebra. A multi-block embedding was constructed in \cite{YB07,Lu} for division algebras whose center $F$ contains an imaginary quadratic field. In this paper we consider a more general multi-block embedding proposed in \cite{LSV}, which applies to any totally complex center $F$. \\
Let $F$ be totally complex of degree $[F:\Q]=2k$. $F$ admits $2k$ $\Q$-embeddings $\alpha_i: F\hookrightarrow\C$ in complex conjugate pairs: $\alpha_i=\bbar{\alpha_{i+k}}$, for $1\leq i\leq k$. Each $\alpha_i$ can be extended to an embedding $E\hookrightarrow \C$.
Given $a \in \D$, consider the mapping $\psi: \D\mapsto M_{nk\times n}(\C)$ given by
\begin{equation}\label{main_map}
\psi(a)=\begin{pmatrix} \alpha_1(\phi(a)) \\ \vdots \\ \alpha_{k}(\phi(a))\end{pmatrix},
\end{equation}
where each $\alpha_i$  is extended to an embedding $\alpha_i: M_n(E)\hookrightarrow M_n(\C)$.
\begin{rem}
For all $x \in \mathcal{D}$,
\begin{align} \label{pdet_norm}
&\pdet(\psi(a))=\prod_{i=1}^k \det(\alpha_i(\phi(a))) \stackrel{(a)}{=} \prod_{i=1}^k \alpha_i(\det(\phi(a))) \notag\\
& \stackrel{(b)}{=}(N_{F/\Q}(N_{\mathcal{D}/F}(a)))^{\frac{1}{2}}= (N_{\mathcal{D}/\Q}(a))^{\frac{1}{2}},
\end{align}
where (a) follows from the fact that the $\alpha_i$ are ring homomorphisms, and (b) follows from the definition of the reduced norm.
\end{rem}

In order to obtain a matrix lattice, we will consider a suitable discrete subset of the algebra called an order.
\begin{definition}
 A \emph{$\ZZ$-order} $\Gamma$ in $\D$ is a subring of $\D$ having the same identity element as
$\D$, and such that $\Gamma$ is a finitely generated
module over $\ZZ$ which generates $\mathcal{D}$ as a linear space over $\Q$.
\end{definition}

The following result was proven in \cite[Proposition 5]{LSV}:

\begin{prop}\label{reg2}
Let $\Gamma$ be a $\ZZ$-order in $\D$ and $\psi$ the previously defined embedding.
Then $\psi(\Gamma)$ is a $2kn^2$-dimensional lattice in $M_{nk \times n}(\C)$ which satisfies
$$
\min_{a \in \Gamma \setminus \{\mb{0}\}} \abs{\pdet(\psi(a))}= 1, \quad V(\psi(\Gamma))= 2^{-kn^2}\sqrt{|d(\Gamma/\ZZ)|}.
$$
\end{prop}
Here  $d(\Gamma/\ZZ)$ is a non-zero integer called the $\ZZ$-\emph{discriminant} of the order $\Gamma$. We refer the reader to \cite{Reiner} for the relevant definitions.

\subsection{Dual lattice and codifferent}

Let $\Gamma$ be a $\ZZ$-order in $\mathcal{D}$. We define the \emph{codifferent} of $\Gamma$ as
$$\Gamma^{\vee}=\{ x \in \mathcal{D} \;:\; \tr_{\mathcal{D}/\Q}(x\Gamma) \subseteq \ZZ\},$$
where $\tr_{\mathcal{D}/\Q}$ is the reduced trace.


The codifferent is an ideal of $\mathcal{D}$, and its reduced norm is related to the discriminant as follows \cite{Reiner}:
\begin{equation} \label{discriminant_MIMO}
N_{\mathcal{D}/\Q}(\Gamma^{\vee})=\frac{1}{d(\Gamma/\ZZ)^{\frac{1}{n}}}.
\end{equation}

Similarly to the commutative case, the codifferent of $\Gamma$ embeds as the complex conjugate of the dual lattice. 

\begin{lem} \label{lemma_codifferent}
$\psi(\Gamma)^*=2 \psi(\Gamma^{\vee})^h.$
\end{lem}

This Lemma is proven in Appendix \ref{proof_codifferent}.

\subsection{Orders with small discriminants and dense matrix lattices}

A family of division algebras with orders having particularly small discriminants was constructed in \cite{LV}. These orders yield dense lattices as shown in Proposition \ref{reg2}. \\
First, we need the following Theorem \cite[Theorem 6.14]{VHLR}:
\begin{theorem} \label{theorem_VHLR}
Let $F$  be a number field of degree $2k$ and $P_1$ and $P_2$ be two prime ideals of $F$.
Then there exists a degree $n$ division algebra $\D$ having an order $\Gamma$ with discriminant
\begin{equation}\label{discriminant2}
d(\Gamma/\ZZ)=(N_{F/\mathbb Q}(P_1) N_{F/\mathbb Q}(P_2))^{n(n-1)} (d_F)^{n^2}.
\end{equation}
\end{theorem}

Thanks to this property, a suitable family of division algebras can be chosen in two steps.

First, we should choose an infinite sequence of centers $\{F_k\}$ with small discriminants, such as Martinet's sequence $\mathcal{F}_C$ (Theorem \ref{Martinet_theorem}). Furthermore, one can choose suitable ideals in these number fields \cite[Lemma 7.9]{LV}:

\begin{lem}\label{norm}
Every number field $F_k$ in the Martinet family has ideals $P_1$ and $P_2$ such that
$$
N_{F/\mathbb Q}(P_1)\leq 23^{k/10} \,\,\mathrm{and}\,\, N_{F/\mathbb Q}(P_2) \leq 23^{k/10}.
$$
\end{lem}

This leads us to the main result in \cite{LV}:
\begin{theorem}\label{existence_of_lattices}
Given $n$, there exists a sequence of totally complex number fields $\{F_k\}$ of degree $2k$ and a sequence of division algebras $\mathcal{D}_k$ of index $n$ over $F_k$ having an order $\Gamma_k$ with discriminant
$$d(\Gamma_k/\ZZ) \leq \beta^{2kn(n-1)} G^{2kn^2},$$
where $G=\rd_{\mathcal{F}_C} \approx 92.368$ and $\beta=23^{\frac{1}{10}}$.
Consequently, $\{\Lambda^{(n,k)}\}=\{\psi(\Gamma_k)\}$ is a sequence of $2n^2k$-dimensional lattices with 
\begin{equation*} 
\pdet(\Lambda^{(n,k)}) =1, \quad V(\Lambda^{(n,k)}) \leq \beta^{kn(n-1)} \left(\frac{G}{2}\right)^{n^2k}.
\end{equation*}
\end{theorem}

\subsection{Flatness factor of multi-block matrix lattices from division algebras}

\begin{rem} \label{matrix_gaussian}
Due to the isometry $\xi$ between $M_{nk \times n}(\C)$ and $\C^{n^2k}$ (Remark \ref{isometry}), the definitions of flatness factor, smoothing parameter and discrete Gaussian distribution extend in a natural way for matrix lattices in $M_{nk \times n}(\C)$. \\
Given a lattice $\Lambda \subset M_{nk \times n}(\C)$, a multi-block matrix $\bar{X} \in M_{nk \times n}(\C)$ and a positive definite matrix $\Sigma \in M_{nk \times nk}(\C)$, we define
\begin{align*}
&\epsilon_{\Lambda}(\sqrt{\Sigma})\doteqdot \epsilon_{\xi(\Lambda)}(\sqrt{\Sigma} \otimes I_n),\\
&\eta_{\varepsilon}(\Lambda) \doteqdot \eta_{\varepsilon}(\xi(\Lambda)),\\
&D_{\Lambda -\bar{X},\Sigma}(X-\bar{X})\doteqdot D_{\xi(\Lambda - \bar{X}), \Sigma \otimes I_n} (\xi(X - \bar{X})) \quad \forall X  \in \Lambda.
\end{align*}
Note that these definitions are consistent with the previous ones: for example,
\begin{align*}
&\epsilon_{\sqrt{\Sigma^{-1}}\Lambda}(I)=\epsilon_{\xi(\sqrt{\Sigma^{-1}\Lambda})}(I)=\epsilon_{(\sqrt{\Sigma} \otimes I_n)^{-1}\xi(\Lambda)}(I)\\
&=\epsilon_{\xi(\Lambda)}(\sqrt{\Sigma \otimes I})=\epsilon_{\Lambda}(\sqrt{\Sigma}).
\end{align*}
\end{rem}

We now focus on the sequence of $n^2k$-dimensional multi-block matrix lattices $\Lambda^{(n,k)}=\psi(\Gamma_k) \subset M_{nk \times n}(\C)$ in Theorem \ref{existence_of_lattices}. \\
Let $c> \frac{1}{\sqrt{2\pi}}$, $C=c\sqrt{2\pi e} e^{-\pi c^2}$, $\varepsilon=\frac{C^{2n^2k}}{1-C^{2n^2k}}$. 

From
(\ref{Banaszczyk_eta})
and Lemma \ref{lemma_codifferent}, we obtain
\begin{equation} \label{dual_MIMO}
\eta_{\varepsilon}(\Lambda^{(n,k)}) \leq \frac{n\sqrt{k}c}{\lambda_1(\psi(\Gamma_k^{\vee})^h)}.
\end{equation}

\section{MIMO wiretap channel} \label{MIMO_results}

\subsection{Channel model} \label{MIMO_model}
We consider a MIMO fading channel model where Alice is equipped with $n$ antennas, while Bob and Eve have $n_b$ and $n_e$ antennas respectively. In this paper, we always assume that $n_b \geq n$ and $n_e \geq n$.\\
Transmission takes place over $k$ quasi-static fading blocks of delay $T=n$, and the transmitted codeword is
of the form (\ref{X}),
where the matrix $X_i \in M_n(\C)$ is sent during the $i$-th block. 

The outputs $Y$ and $Z$ at Bob and Eve's end respectively are given by
\begin{equation} \label{MIMO_equation}
\begin{cases}
Y=H_{b}X + W_{b},\\
Z=H_{e}X + W_{e},
\end{cases} 
\end{equation}
where the channel matrices  $H_b=\diag(H_{b,1},\ldots,H_{b,k}) \in M_{n_b k \times nk}(\C)$, $H_e=\diag(H_{e,1},\ldots,H_{e,k}) \in M_{n_e k \times nk}(\C)$ are (possibly rectangular) block diagonal matrices. The coefficients of the noise matrices $W_{b}$ and $W_{e}$ are i.i.d. circularly symmetric complex Gaussian with zero mean and variance
$\sigma_b^2$, $\sigma_e^2$ per complex dimension. The input $X$ satisfies the average power constraint (per channel use)
\begin{equation} \label{power_constraint_MIMO}
\frac{1}{nk} \sum_{i=1}^{k} \norm{X_i}^2 \leq P.
\end{equation}
The average power per symbol is $\sigma_s^2=\frac{P}{n}$. We denote by $\rho_b=\frac{\sigma_s^2}{\sigma_b^2}$ and $\rho_e=\frac{\sigma_s^2}{\sigma_e^2}$ the signal-to-noise ratios for Bob and Eve respectively. \\
We suppose that
$\{H_{b,i}\}$, $\{H_{e,i}\}$ are isotropically invariant channels such that the channel capacities $C_b$ and $C_e$ are well-defined and $\forall \gamma, \gamma'>0$,
{\allowdisplaybreaks
\begin{align}
\!\!\!\!\!\lim_{k \to \infty}& \mathbb{P}\!\left\{ \abs{\frac{1}{k} \sum_{i=1}^k \ln \det\! \left(I_{n_b}\!+\!\rho_bH_{b,i}^{\dagger}H_{b,i}\!\right) \!-\! C_b}\!>\gamma\!\right\}\!=0 \label{LLN_Bob_MIMO} \\
\!\!\!\!\!\lim_{k \to \infty}& k\, \mathbb{P}\!\left\{ \!\abs{\frac{1}{k}\! \sum_{i=1}^k \ln \det \! \left(I_{n_e}\!\!+\!\rho_eH_{e,i}^{\dagger}H_{e,i}\!\right) \!-\! C_e}\!>\!\gamma'\!\right\}\!=\!0 \label{LLN_Eve_MIMO}
\end{align}}
We suppose that Alice has no instantaneous CSI, Bob has perfect CSI of his own channel, and Eve has perfect CSI of her channel and of Bob's. \\
Similarly to the single-antenna case, condition (\ref{LLN_Bob_MIMO}) is satisfied for static channels, i.i.d. fading channels and i.i.d. block fading channels, and all ergodic
channels; condition (\ref{LLN_Eve_MIMO}) is more restrictive and holds for static, i.i.d. fading and block fading models, and ergodic channels for which the convergence in the law of large numbers is faster than $\frac{1}{k}$.

Recall that in the ergodic case with no instantaneous CSIT, where the transmitter uses uniform power allocation, the white-input capacities of Bob and Eve's channels are given by 
\begin{align*}
&C_b=\mathbb{E}_{\bar{H}_b}\left[\ln\det\left(I_{n_b}+\frac{\rho_b}{n} \bar{H}_b \bar{H}_b^{\dagger}\right)\right], \\ 
&C_e=\mathbb{E}_{\bar{H}_e}\left[\ln\det\left(I_{n_e}+\frac{\rho_e}{n} \bar{H}_e \bar{H}_e^{\dagger}\right)\right],
\end{align*}
where $\bar{H}_b \in M_{n_b \times n}(\mathbb{C})$ and $\bar{H}_e \in M_{n_e \times n}(\mathbb{C})$ are random matrices with the same first order distribution as the processes $\{H_{b,i}\}$,$\{H_{e,i}\}$.

A confidential message $M$ and an auxiliary message $M'$ with rate $R$ and $R'$ respectively are encoded into the multi-block codeword $X$.

As in the single-antenna case (Remark \ref{remark_CSI}), we have that $\mathbb{I}(M; Z|H_b,H_e)=\mathbb{I}(M; Z|H_e)$, i.e. the leakage is given by $\mathbb{I}(M; Z|H_e)$.
\begin{rem}
For general channels the strong secrecy capacity is not known in this setting (see Remark \ref{remark_secrecy_capacity} for the SISO case). In \cite{Lin_Lin} it was shown that the weak secrecy capacity
$$C_s^{w}=C_b-C_e$$
for i.i.d. fading wiretap channels such that Bob and Eve's fadings are independent. 
\end{rem}

\subsection{Multi-block lattice wiretap coding} \label{lattices_MIMO}
Let $\Lambda_e \subset \Lambda_b$ be a pair of nested multiblock matrix lattices in $M_{nk \times n}(\C)$ such that $\Lambda_e \subset \Lambda_b$ and $\abs{\Lambda_b/\Lambda_e}=e^{nkR}$, with volumes scaling as follows:
\begin{equation} \label{alpha_e_MIMO}
V(\Lambda_e) = \frac{(\pi e \sigma_s^2)^{n^2k}}{e^{nkR'}}, \quad V(\Lambda_b) = \frac{(\pi e \sigma_s^2)^{n^2k}}{e^{nk(R+R')}},
\end{equation}
where $R'>0$.
Each message $m \in \mathcal{M}=\{1,\ldots,e^{nkR}\}$ is mapped to a coset leader
$X^{(m)} \in \Lambda_b \cap \mathcal{R}(\Lambda_e)$, where $\mathcal{R}(\Lambda_e)$ is a fundamental region of $\Lambda_e$. In order to transmit the message $m$, Alice samples $X$ from the discrete Gaussian
$D_{\Lambda_e + X^{(m)},\sigma_s}$
where $\sigma_s^2=\frac{P}{n}$. We denote this coding scheme by $\mathcal{C}(\Lambda_b,\Lambda_e)$. \\
Similarly to Remark \ref{theta_t_remark}, it follows from \cite[Lemma 6 and Remark 6]{LLBS} that $\forall\, 0<t<\pi$, for $\theta_t=\frac{\pi-t}{\pi}$,
if $\varepsilon_k=\epsilon_{\Lambda_e^{(k)}}(\sqrt{\theta_t}\sigma_s)<1$, 
$$\abs{\mathbb{E}[\norm{X}^2]-n^2 k \sigma_s^2} \leq \frac{2 \pi \varepsilon_k}{1-\varepsilon_k} \sigma_s^2.$$
As $k \to \infty$, the variance per complex dimension of $X$ tends to $\sigma_s^2$
provided that
\begin{equation} \label{Lemma6_condition_MIMO}
\lim_{k \to \infty} \epsilon_{\Lambda_e^{(k)}}( \sqrt{\theta_t} \sigma_s) = 0,
\end{equation}
and the power constraint (\ref{power_constraint_MIMO}) is verified asymptotically.
From \cite[Lemma 7 and Remark 7]{LLBS}, 
the information rate per complex symbol of the auxiliary message  is bounded by
$$\abs{ \frac{\mathbb{H}(M')}{n} - \left(\ln(\pi e \sigma_s^2)-\frac{1}{n^2 k} \ln V(\Lambda_e)\right)} \leq \nu_t(\varepsilon_k),$$
where $\nu_t(\epsilon_k) \to 0$ as $\epsilon_k \to 0$. 
If $\epsilon_k \to 0$, the entropy rate of the auxiliary message tends to $R'$.

\subsubsection*{Coding scheme based on division algebras with constant root discriminant}
Let  $\{\Lambda^{(n,k)}\}=\{\psi(\Gamma_k)\}$ be the sequence of $n^2k$-dimensional multi-block matrix lattices in $M_{nk \times n}(\C)$ from Theorem \ref{existence_of_lattices}. We consider scaled versions $\Lambda_b=\alpha_b \Lambda^{(n,k)}$, $\Lambda_e=\alpha_e \Lambda^{(n,k)}$ such that $\Lambda_e \subset \Lambda_b$ and $\abs{\Lambda_b/\Lambda_e}=e^{nkR}$. Given rates $R,R'$, we denote the corresponding multi-block lattice coding scheme by $\mathcal{C}(\Lambda^{(n,k)},R,R')$.

\subsection{Achievable secrecy rates}
We now state the main result for MIMO wiretap channels, which will be proven in Sections \ref{MIMO_secrecy} and \ref{MIMO_reliability}.

\begin{thm} \label{main_MIMO}
Consider the  multi-block wiretap coding scheme $\mathcal{C}(\Lambda_b,\Lambda_e)$ in Section \ref{lattices_MIMO}, and suppose that
\begin{equation} \label{dual_condition_MIMO}
\liminf_{k \to \infty}\;\delta(\Lambda_e^*)^{\frac{2}{k}} \geq \mathrm{d}_e, \quad \liminf_{k \to \infty}\;\delta(\Lambda_b)^{\frac{2}{k}} \geq \mathrm{d}_b
\end{equation}
for some positive constants $\mathrm{d}_e$, $\mathrm{d}_b$. \\
If the main channel and the eavesdropper's channel verify the conditions (\ref{LLN_Bob_MIMO}) and (\ref{LLN_Eve_MIMO}) respectively, then $\mathcal{C}(\Lambda_b,\Lambda_e)$ achieves strong secrecy for any message distribution $p_{M}$ (and thus semantic security) if 
\begin{align}
&R' >C_e + n \ln \left(\frac{ne}{\pi}\right) -\ln \mathrm{d}_e, \label{rate_conditions_MIMO_1} \\
& R+ R' < C_b- n \ln \left(\frac{4n}{\pi e}\right) + \ln \mathrm{d}_b.  \label{rate_conditions_MIMO_2}
\end{align}
Thus, any strong secrecy rate
\begin{equation} \label{secrecy_rate_MIMO}
R < C_b -C_e - 2n \ln \left(\frac{2n}{\pi}\right) + \ln \mathrm{d}_b \mathrm{d}_e
\end{equation}
is achievable with the proposed lattice codes.
\end{thm}

\begin{cor} \label{corollary_MIMO}
If the main channel and the eavesdropper's channel verify the conditions (\ref{LLN_Bob_MIMO}) and (\ref{LLN_Eve_MIMO}) respectively, then the multi-block wiretap coding scheme $\mathcal{C}(\Lambda^{(n,k)},R,R')$ achieves strong secrecy and semantic security if 
\begin{align*}
&R' >C_e+ n\ln\left(\frac{ne\beta^{\frac{n-1}{n}}G}{2\pi}\right),  \\
& R+ R' < C_b- n\ln \left(\frac{2n\beta^{\frac{n-1}{n}}G}{\pi e}\right), 
\end{align*}
where $G=\rd_{\mathcal{F}_C} \approx 92.368$. Thus, any strong secrecy rate
\begin{equation*} 
R < C_b -C_e -2n\ln\left(\frac{nG\beta^{\frac{n-1}{n}}}{\pi}\right)
\end{equation*}
is achievable with the proposed lattice codes.
\end{cor}
\begin{IEEEproof}[Proof of the Corollary]
From Theorem \ref{existence_of_lattices} we get
$$\delta(\Lambda_b)^{\frac{2}{k}}=\delta(\Lambda^{(n,k)})^{\frac{2}{k}}=\frac{2^n}{\beta^{(n-1)}G^n}$$
On the other side, for the dual lattice we have
\begin{align*}
&\pdet((\Lambda^{(n,k)})^*)\stackrel{(a)}{=}\sqrt{N_{\mathcal{D}/\Q}(2\psi(\Gamma_k^{\vee})}\stackrel{(b)}{=}\frac{1}{d(\Gamma_k/\mathbb{Z})^{\frac{1}{2n}}}\\
&\stackrel{(c)}{=}\frac{2^{nk}}{\beta^{k(n-1)}G^{kn}},
\end{align*}
where (a) follows from (\ref{pdet_norm}), (b) follows from (\ref{discriminant_MIMO}) and (c) from Theorem \ref{existence_of_lattices}. The normalized minimum determinant of $\Lambda_e^*$ is 
\begin{align*}
&\delta(\Lambda_e^*)=\delta((\Lambda^{(n,k)})^*)=\frac{\pdet((\Lambda^{(n,k)})^*)}{V((\Lambda^{(n,k)})^*)^{\frac{1}{2n}}}\\
&=\pdet((\Lambda^{(n,k)})^*)V(\Lambda^{(n,k)})^{\frac{1}{n}}=\frac{2^{\frac{kn}{2}}}{\beta^{\frac{k(n-1)}{2}}G^{\frac{kn^2}{2}}},
\end{align*}
and so we find that
\begin{align}
\delta(\Lambda_e^*)^{\frac{2}{k}}=\frac{2^n}{\beta^{(n-1)}G^n}. \tag*{\IEEEQED}
\end{align}
\let\IEEEQED\relax
\end{IEEEproof}

\begin{rem}
Let $\mathcal{S}(C_b,C_e)$ denote the set of all ergodic stationary isotropically invariant fading processes $\{(H_{b},H_{e})\}$ such that (\ref{LLN_Bob_MIMO}) and (\ref{LLN_Eve_MIMO}) hold. Similarly to the single antenna case, a fixed lattice code sequence $\mathcal{C}(\Lambda^{(n,k)},R,R')$ with rates satisfying (\ref{rate_conditions_MIMO_1}) and (\ref{rate_conditions_MIMO_2}) universally achieves strong secrecy and semantic security over \emph{all} channels in the set  $\mathcal{S}(C_b',C_e')$ for all $C_b' \geq C_b$ and for all $C_e' \leq C_e$. 
\end{rem}

Finally, the condition (\ref{LLN_Eve_MIMO}) can be relaxed if only weak secrecy is required:
\begin{prop} \label{prop_weak_secrecy_MIMO}
If the condition (\ref{LLN_Bob_MIMO}) holds for the main channel and $\forall \gamma'>0$ we have
\begin{equation*}
\lim_{k \to \infty} \mathbb{P}\!\left\{ \abs{\frac{1}{k} \sum_{i=1}^k \ln \det \left(I+\rho_eH_{e,i}^{\dagger}H_{e,i}\right) - C_e}>\gamma'\right\}=0 
\end{equation*}
for the eavesdropper's channel, then the wiretap coding scheme $\mathcal{C}(\Lambda_b,\Lambda_e)$ achieves weak secrecy if conditions (\ref{dual_condition_MIMO}), (\ref{rate_conditions_MIMO_1}) and (\ref{rate_conditions_MIMO_2}) hold. In particular, any weak secrecy rate $R < C_b -C_e -2n\ln\left(\frac{nG\beta^{(n-1)}}{\pi}\right)$ is achievable with the lattice codes $\mathcal{C}(\Lambda^{(n,k)},R,R')$.
\end{prop}

The proof of Proposition \ref{prop_weak_secrecy_MIMO} is very similar to the proof of Theorem \ref{main_MIMO} and is omitted. 

\subsection{Proof of Theorem \ref{main_MIMO}: Secrecy} \label{MIMO_secrecy}
The proof follows the same steps as in the single antenna case (Section \ref{SISO_secrecy}). 

\subsubsection{Fixed channel} First, we prove an upper bound for the finite-length leakage when the eavesdropper's channel $H_e$ is fixed.
\begin{prop}[Bound for the leakage] \label{prop_MIMO_leakage}
Suppose that $\delta(\Lambda_e^*)^{\frac{2}{k}} \geq \textrm{d}_e$ for the $2n^2k$-dimensional lattice $\Lambda_e$, and that $H_e$ is fixed and such that $\frac{1}{k} \sum_{i=1}^k \ln \det (I+\rho_e H_e^{\dagger} H_e) \leq \bar{C}_e$. Then if 
$R'>\bar{C}_e  - \ln \mathrm{d}_e + 2n \ln(c \sqrt{2ne}),$
for sufficiently large $k \geq\bar{k}(c)$, the leakage is bounded by
\begin{equation} \label{theorem4_MIMO}
\I(p_{M};p_{Z|H_e}) \leq 8 n^2 k\varepsilon_k R  - 8 \varepsilon_k \ln 8 \varepsilon_k,
\end{equation}
where $\varepsilon_k=\frac{C^{2n^2k}}{1-C^{2n^2k}}$, and $C=c\sqrt{2\pi e} e^{-\pi c^2}$.
\end{prop}

\begin{IEEEproof}
We distinguish two cases: the symmetric case ($n_e=n$) and the asymmetric case ($n_e>n$).
\paragraph{Case $n_e=n$}
The received signal at Eve's end is $Z=H_eX+W_e$. As in equation (\ref{leakage}), the leakage can be written as
{\allowdisplaybreaks
\begin{align*}
&\I(M;Z,H_e)=
\mathbb{E}_{H_e}\left[\I(p_{M};p_{Z|H_e})\right]
\end{align*}
}

For a fixed realization $H_e=\diag(H_{e,1},\ldots,H_{e,k})$, we have
$$H_e X \sim D_{H_e \Lambda_e + H_e X^{(m)}, \sqrt{H_e H_e^{\dagger}}\sigma_s},$$
recalling the notation in Remark \ref{matrix_gaussian}.  Using Lemma \ref{extended_Regev_Lemma} with $\Sigma_1=H_e H_e^{\dagger} \sigma_s^2$, $\Sigma_2=\sigma_e^2 I_{nk}$,  we have
\begin{equation} \label{V_MIMO}
\mathbb{V}(p_{Z|H_e,M=m},f_{\sqrt{\Sigma_0}})\leq 4\varepsilon_k
\end{equation}
provided that
\begin{equation} \label{flatness_fading}
\epsilon_{H_e \Lambda_e}(\sqrt{\Sigma})=\epsilon_{\sqrt{\Sigma}^{-1}H_e\Lambda_e}(1) \leq \varepsilon_k \leq \frac{1}{2},
\end{equation}
where we define
$\Sigma_0=H_e H_e^{\dagger} \sigma_s^2 + \sigma_e^2 I_{nk},\quad \Sigma^{-1}=\frac{(H_e H_e^{\dagger})^{-1}}{\sigma_s^2} + \frac{I_{nk}}{\sigma_e^2}.$ Note that $\Sigma=\sigma_s^2\sigma_e^2 (\sigma_e^2I_{nk} + \sigma_s^2 H_e H_e^{\dagger})^{-1}H_e H_e^{\dagger}.$

Using (\ref{Banaszczyk_eta}), for $\varepsilon_k=\frac{C^{2n^2k}}{1-C^{2n^2k}}$, the smoothing parameter of the faded lattice is upper bounded by
\begin{align}
 & \eta_{\varepsilon_k}(\sqrt{\Sigma^{-1}}H_e \Lambda_e) \leq 
 \frac{2cn\sqrt{k}}{\lambda_1(\sqrt{\Sigma}(H_e^{\dagger})^{-1}(\Lambda_e)^{*})}. \label{faded_smoothing_parameter_MIMO} 
 \end{align}

Using Remark \ref{inequality}, we find
{\allowdisplaybreaks
\begin{align*}
\lambda_1(\sqrt{\Sigma}(H_e^{\dagger})^{-1}\Lambda_e^*)  \geq  nk \prod_{i=1}^k \frac{(\sigma_s^2)^{\frac{1}{k}}\mathrm{pdet}(\Lambda_e^*)^{\frac{2}{nk}}}{\det(I\! +\rho_e H_{e,i}H_{e,i}^{\dagger})^{\frac{1}{nk}}} 
\end{align*}
}
Replacing in the bound (\ref{faded_smoothing_parameter_MIMO}), we have
\begin{align*}
&\eta_{\varepsilon_k}(\!\sqrt{\Sigma^{-1}}\!H_e \Lambda_e) \!\leq\! \frac{2c\sqrt{n}}{\mathrm{pdet}(\Lambda_e^*)^{\frac{1}{nk}} \sigma_s }\prod_{i=1}^k\! \det\!\left(\!I\! + \rho_e H_{e,i}H_{e,i}^{\dagger}\!\right)\!^{\frac{1}{2nk}} \\
&\leq  \frac{2c\sqrt{n}}{\mathrm{pdet}(\Lambda_e^*)^{\frac{1}{nk}} \sigma_s } e^{\frac{\bar{C}_e}{2n}}.
\end{align*}
Suppose that 
\begin{equation} \label{sigma_fading_MIMO}
\frac{1}{\sqrt{2\pi}}\frac{2c\sqrt{n}e^{\frac{\bar{C}_e}{2n}}}{\mathrm{pdet}(\Lambda_e^*)^{\frac{1}{nk}}\sigma_s} \leq 1.
\end{equation}
Then (\ref{flatness_fading}) holds for sufficiently large $k$ (depending only on $c$), and  
it follows from \cite[Lemma 2]{LLBS} that
\begin{equation*} 
\I(p_{M};p_{Z|H_e}) \leq 8 n^2 k\varepsilon_k R  - 8 \varepsilon_k \ln 8 \varepsilon_k.
\end{equation*}

Recalling the definition of normalized minimum determinant and the scaling condition (\ref{alpha_e_MIMO}), 
\begin{align*}
&\pdet(\Lambda_e^*)^{\frac{1}{nk}}=\left(\delta(\Lambda_e^*) V(\Lambda_e^*)^{\frac{1}{2n}}\right)^{\frac{1}{nk}}=
\frac{\delta(\Lambda_e^*)^{\frac{1}{nk}}}{V(\Lambda_e)^{\frac{1}{2n^2k}}} \\
& =\frac{\delta(\Lambda_e^*)^{\frac{1}{nk}} e^{\frac{R'}{2n}}}{\sqrt{\pi e} \sigma_s}.
\end{align*}

In particular if $\delta(\Lambda_e^*)^{\frac{2}{k}}\geq \textrm{d}_e$, the sufficient condition (\ref{sigma_fading_MIMO}) for secrecy is satisfied if 
\begin{equation*}
R'>\bar{C}_e + 2n \ln(c \sqrt{2ne}) - \ln \mathrm{d}_e. 
\end{equation*}
\paragraph{Case $n_e>n$} As before, the received signal is $Z=H_eX+W_e \in H_e \Lambda_e + H_e X^{(m)} + W_e$. If $H_e$ is full rank, the lattice $H_e \Lambda_e$ is a $2n^2k$-dimensional lattice contained in a $2nn_ek$-dimensional space. Consider the QR decomposition
$$H_e=Q_e R_e$$
where $Q_e \in M_{n_e k \times nk}(\C)$ is unitary and $R_e \in M_{nk \times nk}(\C)$ is upper triangular. We have $Q_e=[Q_e' Q_e'']$, where $Q_e' \in M_{n_ek \times nk}(\C)$ is such that $(Q_e')^{\dagger} Q_e'=I_{nk}$, and $R_e=\left[\begin{array}{c} R_e' \\ 0 \end{array} \right]$, $R_e'=\diag(R_{e,1}', \ldots, R_{e,k}') \in M_{nk \times nk}(\C)$.
Multiplying Eve's channel equation in (\ref{MIMO_equation}) by $Q_e^{\dagger}$, we obtain
$$Q_e^{\dagger} Z= R_e X + Q_e^{\dagger} W_e=\left[\begin{array}{c} R_e' X + (Q_e')^{\dagger} W_e \\ (Q_e'')^{\dagger} W_e \end{array}\right]=\left[ \begin{array}{c} Z' \\ Z'' \end{array}\right]$$
Therefore, the second component is pure noise and contains no information about the message. Since $Q_e'$ is unitary, $W_e'=(Q_e')^{\dagger} W_e$ has the same distribution as $W_e$ and is independent of $X$ and $H_e$.
Consequently, we can rewrite the leakage as
$$\mathbb{I}(M;Z|H_e)=\mathbb{I}(M; Z'|H_e)=\mathbb{I}(M; Z'|R_e').$$
The rest of the proof then proceeds exactly as in the case $n_e=n$, by replacing $Z$ with $Z'$ and $H_e$ with $R_e'$. Observe that $H_e^{\dagger}H_e=(R_e')^{\dagger}R_e'$ and so
\begin{align}
&\prod_{i=1}^k \det \left( I +\rho_e (R_{e,i}')^{\dagger} R_{e,i}'\right)^{\frac{1}{k}}=\det\left(I + \rho_e (R_e')^{\dagger} R_e'\right)^{\frac{1}{k}} \notag\\
&=\det\!\left(I \!+\! \rho_e (H_e)^{\dagger} H_e\right)^{\frac{1}{k}} \!=\!\prod_{i=1}^k \det\! \left( I \!+\!\rho_e (H_{e,i})^{\dagger} H_{e,i}\right)^{\frac{1}{k}} \tag*{\IEEEQED}.
\end{align}
\let \IEEEQED \relax
\end{IEEEproof}

\subsubsection{Random channel} Thanks to Proposition \ref{prop_MIMO_leakage}, we can now bound the average leakage for random fading $H_e$ when $k \to \infty$. Due to the law of large numbers (\ref{LLN_Eve_MIMO}), $\forall \eta>0$ 
$$\mathbb{P}\left\{\prod_{i=1}^k \det\left(I+\rho_eH_{e,i}H_{e,i}^{\dagger}\right)^{\frac{1}{k}}>e^{C_e+\eta}\right\} \to 0.$$
The average leakage is bounded as follows:
\begin{align}
&\mathbb{E}_{H_e}\left[\I(p_{M};p_{Z|H_e})\right]\leq \notag &\\
&\!\!\!\! \leq \mathbb{P}\Big\{\prod_{i=1}^k \det\left(I+\rho_eH_{e,i}H_{e,i}^{\dagger}\right)^{\frac{1}{k}}>e^{C_e+\eta}\Big\} (n^2kR)+ \notag \\
&\!\!\!\!+
\mathbb{E}_{H_e}\!\!\left[\I(p_{M};p_{Z|H_e}) \Big| \prod_{i=1}^k \det\!\left(\!I\!+\!\rho_eH_{e,i}H_{e,i}^{\dagger}\right)^{\frac{1}{k}} \!\!\!\leq e^{C_e+\eta}\!\right]\! \!\! \label{leakage_sum_MIMO}
\end{align}
The first term vanishes when $k \to \infty$ due to the condition (\ref{LLN_Eve_MIMO}).\\
If the bound (\ref{dual_condition_MIMO}) holds, then $\forall \gamma >0$, for sufficiently large $k$, $\delta(\Lambda_e^*)^{\frac{2}{k}}>\mathrm{d_e}-\gamma$.
Using Proposition \ref{prop_MIMO_leakage}, the second term in (\ref{leakage_sum_MIMO}) tends to zero when $k \to \infty$ and the scheme achieves strong secrecy provided that
$$R'>C_e + \eta + 2n \ln(c \sqrt{2ne}) - \ln (\mathrm{d}_e-\gamma).$$
Since $\eta, \gamma>0$ and $c >\frac{1}{\sqrt{2\pi}}$ are arbitrary, any rate 
\begin{equation} \label{R_prime_MIMO}
R' >C_e + n \ln\left(\frac{ne}{\pi}\right) - \ln \mathrm{d}_e 
\end{equation}
is sufficient for strong secrecy.

\subsubsection{Power constraint and entropy of auxiliary message} 
We still need to check that the flatness factor condition (\ref{Lemma6_condition_MIMO}) holds, so that the power constraint is verified asymptotically and $R'$ is the rate of the auxiliary message. 

\begin{prop}[Bound for the flatness factor] \label{flatness_prop_MIMO}
Suppose that the $\delta(\Lambda_e^*)^{\frac{2}{k}} \geq \mathrm{d}_e$ for the $2n^2 k$ dimensional lattice $\Lambda_e$. Let $0 < t < \pi$ and $\theta_t=\frac{\pi-t}{\pi}.$ If $R' \geq n \ln (2nec^2) -\ln \mathrm{d}_e-n \ln \theta_t$, then $\epsilon_{\Lambda_e}(\sqrt{\theta_t}\sigma_s) \leq \varepsilon_k=\frac{C^{2n^2k}}{1-C^{2n^2k}}$, where $C=c\sqrt{2\pi e} e^{-\pi c^2}$.
\end{prop}

\begin{IEEEproof}
Using Remark \ref{inequality}, we have 
\begin{align*}
&\lambda_1(\Lambda_e^*) \geq \sqrt{nk}\pdet(\Lambda_e^*)^{\frac{1}{nk}} 
=\frac{\sqrt{nk} \delta(\Lambda_e^*)^{\frac{1}{nk}}e^{\frac{R'}{2n}}}{\sqrt{\pi e} \sigma_s} \\
& \geq \frac{\sqrt{nk} \mathrm{d}_e^{\frac{1}{2n}}e^{\frac{R'}{2n}}}{\sqrt{\pi e} \sigma_s}. 
\end{align*}
Then for $\varepsilon_k=\frac{C^{2n^2k}}{1-C^{2n^2k}}$ we have
\begin{align*}
\eta_{\varepsilon_k}(\Lambda_e) \leq \frac{2cn\sqrt{k}}{\lambda_1(\Lambda_e^*)} \leq \frac{2c\sqrt{n \pi e}\sigma_s}{\mathrm{d}_e^{\frac{1}{2n}}e^{\frac{R'}{2n}}}
\end{align*}
Therefore $\epsilon_{\Lambda_e}(\sigma_s) \leq \varepsilon_k$ 

provided that 
$$\sqrt{\theta_t}\sigma_s \geq \frac{2c\sqrt{n} \sqrt{\pi e} \sigma_s}{\mathrm{d}_e^{\frac{1}{2n}}e^{\frac{R'}{2n}} \sqrt{2\pi}}$$
or equivalently $R' \geq n \ln (2nec^2) -\ln \mathrm{d}_e -n \ln \theta_t$, as desired.
\end{IEEEproof}

In particular when $c \to \frac{1}{\sqrt{2\pi}}$ and $t \to 0$, $\theta_t \to 1$, we obtain the condition 
\begin{equation} \label{entropy_condition_MIMO}
R' > n \ln \left(\frac{ne}{\pi}\right)-\ln \mathrm{d}_e,
\end{equation}
which is weaker than (\ref{R_prime_MIMO}). 

\subsection{Proof of Theorem \ref{main_MIMO}: Reliability} \label{MIMO_reliability}
Recall that the received signal at Bob is $Y=H_b X + W_b$. 
\subsubsection{Fixed channel}
First of all, we prove the following uniform upper bound for the finite-length error probability of the code in the case of a fixed channel realization $H_b$:
\begin{prop}[Bound for the error probability] \label{prop_Pe_MIMO}
Suppose that $\delta(\Lambda_b)^{\frac{2}{k}} \geq \mathrm{d}_b$, $\delta(\Lambda_e^*)^{\frac{2}{k}} \geq \mathrm{d}_e$ for the $2n^2k$-dimensional lattices $\Lambda_b$ and $\Lambda_e$, and that $H_b$ is fixed with $\frac{1}{k} \sum_{i=1}^k \ln \det (I + \rho_b H_{b,i}^{\dagger} H_{b,i})\geq \bar{C}_b$. Then for code rates $R+R'<\bar{C}_b-n\ln\left(\frac{8c^2n}{e}\right)+\ln \mathrm{d_b}$, $R'\geq n \ln \left(2nec^2\right)-\ln \mathrm{d}_e$, the ML error probability for Bob is bounded by
$$P_e \leq \frac{1+\varepsilon_k}{1-\varepsilon_k} \varepsilon_k,$$
where $\varepsilon_k=\frac{C^{2n^2k}}{1-C^{2n^2k}}$ and $C=c\sqrt{2\pi e} e^{-\pi c^2}$. 
\end{prop}

\begin{IEEEproof}
Let $\rho_b=\frac{\sigma_s^2}{\sigma_b^2}$, and consider the \vv{thin} QR decomposition
$$\widetilde{H}_b=\left(\begin{array}{c} H_b \\ \frac{1}{\sqrt{\rho_b}}I_{nk}\end{array}\right)=QR_b=\left(\begin{array}{c} Q_1 \\ Q_2 \end{array}\right)R_b,$$
where $\tilde{H}_b, Q \in M_{k(n_b + n) \times kn}(\C)$, $Q_1 \in M_{kn_b \times kn}(\C)$. Note that $Q$ has orthonormal columns, $R_b \in M_{kn}(\C)$ is upper triangular and square block-diagonal, and
$$R_b^{\dagger}R_b=\widetilde{H}_b^{\dagger}\widetilde{H}_b=H_b^{\dagger}H_b+\frac{1}{\rho_b}I.$$
For the sake of simplicity, we consider the vectorized version of the received message: let $\mb{x}=\xi(X)$, $\mb{y}=\xi(Y)$, $\mb{w}_b=\xi(W_b)$. Then
$$\mathbf{y}=\mathcal{H}_b\mb{x}+\mb{w}_b,$$
where $\mathcal{H}_b=H_b \otimes I_n$. Note that if we set $\mathcal{Q}_1=Q_1 \otimes I_n$, $\mathcal{R}=R_b \otimes I_n$, we also have $\mathcal{H}_b=\mathcal{Q}_1 \mathcal{R}$.

Similarly to the single antenna case (Section \ref{SISO_reliability}), Bob can compute $$\mb{y}'=\mathcal{Q}_1^{\dagger}\mb{y}=\mathcal{R}\mb{x}+\mb{v},$$
where $\mb{v}=\mathcal{Q}_1^{\dagger} \mb{w}_b -\frac{1}{\rho_b}(\mathcal{R}^{-1})^{\dagger} \mb{x}$ \cite{ElGamal_Caire_Damen}. \\
Recall that $\mb{x}$ is sampled from $D_{\xi(\Lambda_e)+\xi(X^{(m)}),\sigma_s}$. %
Using Lemma \ref{lemma_subgaussian}, $\mb{x}$ is $\delta_k$-subgaussian with parameter $\sigma_s$ for $\delta_k=\ln\left(\frac{1+\varepsilon_k}{1-\varepsilon_k}\right)$ provided that
$\epsilon_{\Lambda_e}(\sigma_s) \leq \varepsilon_k<1$,
which is guaranteed by Proposition \ref{flatness_prop_MIMO}. 
With the same argument as in Section \ref{SISO_reliability}, we can show that
the equivalent noise $\mb{v}$ is $\delta_k$-subgaussian with parameter $\sigma_b$.  

Following the same steps as in Section \ref{SISO_reliability}, we have the union bound on the error probability for fixed $\mathcal{R}$:
\begin{align*}
P_e(\mathcal{R}) \leq e^{\delta_k} \sum_{\boldsymbol{\lambda} \in \mathcal{R}\Lambda_b \setminus \{\mb{0}\}} e^{-\frac{\norm{\boldsymbol{\lambda}}^2}{4\sigma_b^2}}=\frac{1+\varepsilon_k}{1-\varepsilon_k}\sum_{\boldsymbol{\lambda} \in \mathcal{R}\Lambda_b \setminus \{\mb{0}\}} e^{-\frac{\norm{\boldsymbol{\lambda}}^2}{4\sigma_b^2}}.
\end{align*}
Using Lemma \ref{Banaszczyk_Corollary}, $P_e(\mathcal{R}) \leq \frac{1+\varepsilon_k}{1-\varepsilon_k} \varepsilon_k $ if
\begin{equation} \label{condition_tau_MIMO}
\tau^2=\frac{1}{4 \pi \sigma_b^2} > \frac{2 c^2 n^2 k}{\lambda_1(\mathcal{R}\Lambda_b)^2}.
\end{equation}
The minimum distance in the received lattice is lower bounded as follows:
{\allowdisplaybreaks
\begin{align*}
&\lambda_1(\mathcal{R}\Lambda_b)^2=\min_{\bar{X} \in \Lambda_b \setminus \{\mb{0}\}} \norm{\mathcal{R} \xi(\bar{X})}^2=\min_{\bar{X} \in \Lambda_b \setminus \{\mb{0}\}} \norm{R_b\bar{X}}^2  \notag \\
& \stackrel{(a)}{\geq} \min_{\bar{X} \in \Lambda_b \setminus \{\mb{0}\}} nk \prod_{i=1}^k \abs{\det(R_{b,i} \bar{X}_i)}^{\frac{2}{nk}}\notag \\
& = \min_{\bar{X} \in \Lambda_b \setminus \{\mb{0}\}} nk \prod_{i=1}^k \abs{\det(\widetilde{H}_{b,i}^{\dagger} \widetilde{H}_{b,i})}^{\frac{1}{nk}} \prod_{i=1}^k \abs{\det\bar{X}_i}^{\frac{2}{nk}}\notag \\
&=  nk \prod_{i=1}^k \abs{\det(\widetilde{H}_{b,i}^{\dagger} \widetilde{H}_{b,i})}^{\frac{1}{nk}} \pdet(\Lambda_b)^{\frac{2}{nk}},  
\end{align*}
}%
where (a) follows from Remark \ref{inequality}. 
From the scaling condition (\ref{alpha_e_MIMO}), we get
$$\pdet(\Lambda_b)^{\frac{2}{nk}} 
=\frac{\delta(\Lambda_b)^{\frac{2}{nk}}\pi e \sigma_s^2}{e^{\frac{R+R'}{n}}}.$$
Replacing in the condition (\ref{condition_tau_MIMO}), we have that $P_e(\mathcal{R}) \leq \frac{1+\varepsilon_k}{1-\varepsilon_k} \varepsilon_k$
if
$$e^{\frac{R+R'}{n}} < \prod_{i=1}^k \det\left(\frac{I_{n}}{\rho_b}+ H_{b,i}^{\dagger}H_{b,i}\right)^{\frac{1}{nk}} \frac{\delta(\Lambda_b)^{\frac{2}{nk}} e \sigma_s^2}{8c^2 n \sigma_b^2}.$$
In particular, recalling the assumption $\delta(\Lambda_b)^{\frac{2}{k}}\geq \mathrm{d}_b$, a sufficient condition is 
\begin{equation*}
e^{\frac{R+R'}{n}} < \prod_{i=1}^k \det\left(I_{n}+\rho_b H_{b,i}^{\dagger}H_{b,i}\right)^{\frac{1}{nk}} \frac{\mathrm{d}_b^{\frac{1}{n}} e}{8c^2 n}. 
\end{equation*}
or equivalently $R+R'<\bar{C}_b-n\ln\left(\frac{8c^2n}{e}\right)+\ln \mathrm{d_b}$.
\end{IEEEproof}

\subsubsection{Random channel} Using the previous proposition, we now consider the behavior of the error probability for random channels $H_b$ when $k \to \infty$. 
By the law of total probability, $\forall \eta>0$,
\begin{multline*}
P_e \leq \mathbb{P}\Big\{  \prod_{i=1}^k \det \Big(I+\rho_b H_{b,i}^{\dagger}H_{b,i}\Big)^{1/nk} < e^{\frac{C_b-\eta}{n}}\Big\}  \\
+\mathbb{P}\Big\{ \hat{\mb{x}} \neq \mb{x} \; \big| \; \prod_{i=1}^k \det \Big(I+\rho_b H_{b,i}^{\dagger}H_{b,i}\Big)^{1/nk} \geq e^{\frac{C_b-\eta}{n}}\Big\}.
\end{multline*}
Due to the law of large numbers (\ref{LLN_Bob_MIMO}), the first term vanishes when $k \to \infty$. \\
If (\ref{dual_condition_MIMO}) holds, then $\forall \gamma>0$, for sufficiently large $k$, we have $\delta(\Lambda_b)^{\frac{2}{k}}\geq \mathrm{d}_b-\gamma$,  $\delta(\Lambda_e)^{\frac{2}{k}}\geq \mathrm{d}_e-\gamma$.
Then using Proposition \ref{prop_Pe_MIMO}, the error probability in the second term tends to $0$ if
\begin{equation} \label{R_b_MIMO}
R+R'<C_b-\eta-n\ln\left(\frac{8c^2n}{e}\right)+\ln(\mathrm{d_b}-\gamma)
\end{equation}
where $R'>n \ln(2nec^2)-\ln(\mathrm{d}_e-\gamma)$.

Since $\eta, \gamma>0$ and $c>\frac{1}{\sqrt{2\pi}}$ are arbitrary, from equations (\ref{R_prime_MIMO}) and (\ref{R_b_MIMO}), the proposed coding scheme achieves strong secrecy 
and
semantic security rates
$$ R < C_b -C_e -2n\ln\left(\frac{2n}{\pi}\right)+\ln \mathrm{d}_b\mathrm{d}_e.$$
This concludes the proof of Theorem \ref{main_MIMO}.

\section{Compound MIMO channel}\label{compound}

In this section, instead of assuming that fading is distributed according to a certain probability density function, we consider a 
setting where the main channel and eavesdropper's channel are unknown at the transmitter and are only known to belong to a certain \emph{uncertainty set} $\mathcal{S}$. \\
As in Section \ref{MIMO_model}, we consider a MIMO wiretap channel where Alice has $n$ antennas, and Bob and Eve have $n_b$ and $n_e$ antennas respectively. The received signals at Bob and Eve's end are given by
\begin{equation*} 
\begin{cases}
Y=H_{b}X + W_{b},\\
Z=H_{e}X + W_{e},
\end{cases} 
\end{equation*}
where $H_b=\diag(H_{b,1},\ldots,H_{b,k}) \in M_{n_b k \times nk}(\C)$, $H_e=\diag(H_{e,1},\ldots,H_{e,k}) \in M_{n_e k \times nk}(\C)$,  $W_{b}$ and $W_{e}$ have i.i.d. Gaussian entries with zero mean and variance $\sigma_b^2$, $\sigma_e^2$, and $X$ satisfies the average power constraint (\ref{power_constraint_MIMO}). As before, the average power per symbol is $\sigma_s^2=\frac{P}{n}$, and $\rho_b=\frac{\sigma_s^2}{\sigma_b^2}$ and $\rho_e=\frac{\sigma_s^2}{\sigma_e^2}$ are the signal-to-noise ratios for Bob and Eve. \\
We suppose that Bob has perfect CSI of his own channel, Eve has perfect CSI of both channels, and Alice only knows that $(H_b,H_e) \in \mathcal{S}$, where $\mathcal{S}$ is the uncertainty set. \\
We say that a coding scheme achieves \emph{strong secrecy} if $\forall (H_b,H_e) \in \mathcal{S}$,
\begin{align*}
&P_{e,k}=\sup_{(H_b,H_e) \in \mathcal{S}}\, \max_{m \in \mathcal{M}} \mathbb{P}\left\{ \hat{M} \neq m | H_b, H_e, M=m\right\} \to 0,\\
&L_{k}=\sup_{(H_b,H_e) \in \mathcal{S}} \mathbb{I}(M;Z,H_e) \to 0
\end{align*}
as $k \to \infty$.

\subsubsection*{Compound channel model}
In this model the channels are assumed to be held constant during transmission, i.e. $H_{b,i}=\bar{H}_b \in M_{n_b \times n}(\C)$, $H_{e,i}=\bar{H}_e \in M_{n_e \times n}(\C) \; \forall i=1,\ldots,k$, and $(\bar{H}_b,\bar{H}_e) \in \bar{\mathcal{S}} \subseteq \bar{\mathcal{S}}_b \times \bar{\mathcal{S}}_e$, where
\begin{align*} 
\begin{split}
&\bar{\mathcal{S}}_b=\left\{ \bar{H}_b \in M_{n_b \times n}:\; \ln \det (I+ \rho_b \bar{H}_{b} \bar{H}_{b}^{\dagger}) \geq C_b \right\},\\
&\bar{\mathcal{S}}_e=\left\{ \bar{H}_e \in M_{n_e \times n}:\; \ln \det (I+ \rho_e \bar{H}_{e} \bar{H}_{e}^{\dagger}) \leq C_e\right\} .
\end{split}
\end{align*}
for some $0 \leq C_e \leq C_b$. \\
Note that in this model, Eve's channel is not necessarily degraded with respect to Bob's channel.

\begin{rem}
The compound secrecy capacity for an uncertainty set $\bar{\mathcal{S}} \subseteq \bar{\mathcal{S}}_b \times \bar{\mathcal{S}}_e$ is not known in general, but has been computed in some special cases in \cite{Schaefer_Loyka_2015}. In particular if $\bar{\mathcal{S}}$ is compact, it follows from \cite[Corollary 2]{Schaefer_Loyka_2015} 
that the (strong) compound secrecy capacity is lower bounded as $C_c\geq C_b -C_e$. 
\end{rem}

\subsubsection*{Arbitrarily varying channel model}
In this model, the realizations $\{H_{b,i}\}$ and $\{H_{e,i}\}$ may change at each channel use in an arbitrary and unknown way \cite{Bjelakovic_Boche_Sommerfeld}, and $(H_b,H_e) \in \mathcal{S}^{(k)}=\mathcal{S}_b^{(k)} \times \mathcal{S}_e^{(k)}$, where   
\begin{align*}
\begin{split} 
&\mathcal{S}_b^{(k)}\!=\!\left\{ \!H_b\! \in M_{n_bk \times nk}\!:  \frac{1}{k} \sum_{i=1}^k \ln \det (I\!+\! \rho_b H_{b,i} H_{b,i}^{\dagger}) \geq C_b\! \right\},\\
&\mathcal{S}_e^{(k)}\!=\!\left\{\! H_e\! \in M_{n_ek \times nk}\!:  \frac{1}{k} \sum_{i=1}^k \ln \det (I\!+\! \rho_e H_{e,i} H_{e,i}^{\dagger}) \leq C_e\!\right\}.
\end{split}
\end{align*}

\begin{theorem} \label{compound_theorem}
Consider the multi-block wiretap coding scheme $\mathcal{C}(\Lambda_b,\Lambda_e)$ in Section \ref{lattices_MIMO}, and suppose that $\forall k$, $\liminf_{k \to \infty}\delta(\Lambda_b)^{\frac{2}{k}} \geq \mathrm{d}_b$, $\liminf_{k \to \infty}\delta(\Lambda_e^*)^{\frac{2}{k}} \geq \mathrm{d}_e$.  Then any strong secrecy rate
$$R < C_b -C_e - 2n \ln \left(\frac{2n}{\pi}\right) + \ln \mathrm{d}_b \mathrm{d}_e$$
is achievable both over the compound MIMO channel with uncertainty set $\bar{\mathcal{S}} \subseteq \bar{\mathcal{S}}_b \times \bar{\mathcal{S}}_e$ 
and over the arbitrarily varying MIMO channel with uncertainty set $\mathcal{S}^{(k)} \subseteq \mathcal{S}_b^{(k)} \times \mathcal{S}_e^{(k)}$. 
\end{theorem}

\begin{IEEEproof}
Let $c > \frac{1}{\sqrt{2\pi}}$ be a fixed parameter. Note that $\forall \gamma>0$, for sufficiently large $k$, we have $\delta(\Lambda_b)^{\frac{2}{k}}> \mathrm{d}_b-\gamma$,  $\delta(\Lambda_e)^{\frac{2}{k}}> \mathrm{d}_e-\gamma$.
\subsubsection*{Secrecy} It follows from Proposition \ref{prop_MIMO_leakage} that as long as $R'>C_e  - \ln (\mathrm{d}_e-\gamma) + 2n \ln(c \sqrt{2ne}),$
for sufficiently large $k \geq\bar{k}(c)$, for all channels $H_e \in \mathcal{S}_e^{(k)}$, the leakage is \emph{uniformly bounded} by  $\I(p_{M};p_{Z|H_e}) \leq 8 n^2 k\varepsilon_k R  - 8 \varepsilon_k \ln 8 \varepsilon_k$, where $\varepsilon_k=\frac{C^{2n^2k}}{1-C^{2n^2k}}$, and $C=c\sqrt{2\pi e} e^{-\pi c^2}$.
 
\subsubsection*{Reliability} It follows from Proposition \ref{prop_Pe_MIMO} that as long as $R+R'<C_b-n\ln\left(\frac{8c^2n}{e}\right)+\ln(\mathrm{d_b}-\gamma)$, for all channels $H_b \in \mathcal{S}_b^{(k)}$, the ML error probability for Bob is \emph{uniformly bounded} by $P_e \leq \frac{1+\varepsilon_k}{1-\varepsilon_k} \varepsilon_k$. 

Since the previous rates are achievable for all $c > \frac{1}{\sqrt{2\pi}}, \forall \gamma>0$, this concludes the proof.
\end{IEEEproof}

\section{Code design criteria for fading and MIMO wiretap channels} \label{design_criteria}
We will now discuss the implications of our results in terms of design of wiretap lattice codes. 

\subsection{Single antenna fading and Gaussian wiretap channels} 
Although in Corollary \ref{main} we focused on a particular sequence of nested lattices $\Lambda_e \subset \Lambda_b$ that were scaled versions of the same lattice $\Lambda^{(k)}$, Theorem \ref{main_general} suggests a more general design criterion for building promising lattice codes for fading channels. Namely, we should consider pairs of nested lattices $\Lambda_e \subset \Lambda_b$ for which the product 
$$
\Np(\Lambda_b)\Np(\Lambda_e^*)
$$
is maximized. As shown earlier, ideals from number fields with small discriminants give us promising candidates.

Here the term $\mathrm{t}_b=\Np(\Lambda_b)^{\frac{2}{k}}$ can be seen as  providing reliability for the communication between Alice and Bob while $\mathrm{t}_e=\Np(\Lambda_e^*)^{\frac{2}{k}}$ provides security against the wiretapper. 

While we mainly targeted general fading channels in this work, we also gained some intuition on code design in Gaussian wiretap channels. 
Proposition \ref{rate_Gaussian} suggests that in the Gaussian case 
one should maximize the product of the Hermite invariants
\begin{equation}\label{AWGNcriterion}
h(\Lambda_b)h(\Lambda_e^*).
\end{equation}
Rather than using number field lattices, in this particular case one might 
optimize \eqref{AWGNcriterion} for example by considering the densest self dual lattices.

\subsection{Code design for MIMO wiretap channels}
An analogous code design criterion can be given also in the MIMO case using the concept of normalized minimum determinant $\delta(\Lambda)$ of a matrix lattice, which was defined in Section \ref{section_matrix_lattices}.  

Using this concept, 
Theorem \ref{main_MIMO} suggests that for MIMO channels we should maximize $\delta(\Lambda_e^*)\delta(\Lambda_b)$.

\subsection{Comparison with earlier code design }
 The earliest work on lattice code design for the AWGN channel is based on an error probability approach  \cite{OSB}.  
 The main criterion for maximizing the confusion of the eavesdropper is that the theta function of $\Lambda_e$ should be minimized. As this function is hard to analyze, the authors discussed a  simplified criterion where one  should  maximize the Hermite invariant of $\Lambda_e$ \cite[eq. (48)]{OSB}.

In comparison, 
our criterion differs in two ways. First, we prove that following  our design principles the information leakage will be minimized. 
Second, our study emphasizes that the code design criterion for secrecy should be stated in terms of $\Lambda_e^*$   and not of $\Lambda_e$.

 The work \cite{LLBS} concentrates on 
 achieving strong secrecy 
 over the Gaussian wiretap channel. 
 Its results suggest that 
 the theta function of   $\Lambda_e^*$ should be minimized for secrecy. 
 Maximizing the Hermite invariant of $\Lambda_e^*$ can be seen as a first-order approximation of this criterion, which we now make rigorous in Proposition \ref{rate_Gaussian}.
When considering random lattices, this first order approximation yields slightly worse achievable strong secrecy rates ($1.24$ nats per complex channel use from secrecy capacity, versus $1$ nat per complex channel use in \cite{LLBS}, see Section \ref{Gaussian_section}).



Lattice code design for the fading wiretap channel was pioneered in \cite{BO_ICC} and \cite{BO_TComm} where the error probability approach led the authors to consider certain inverse determinant sums over the lattice $\Lambda_e$; both of these 
works suggest the use of number fields for wiretap coding. 
Similar conditions were derived also in \cite{Hamed, KKH}
to minimize the information leakage.
 

Compared to earlier works on fading wiretap channels, our criterion is the first which guarantees positive strong secrecy rates, 
in the sense that we prove that by maximizing $\Np(\Lambda_e^*)$ or $\delta(\Lambda_e^*)$ one can indeed push the leaked information to zero. Also similarly to the Gaussian case it 
seems to be better
to state the design criterion for $\Lambda_e^*$ instead of $\Lambda_e$.
\begin{rem}
We point out that in   the derivation of the code design criterion for $\Lambda_e$ in \cite[p. 5706]{OSB} the authors first obtain a condition for the theta function of $\Lambda_e^*$ and only after using Poisson summation they end up with a condition for $\Lambda_e$. So the authors could also have stated their criterion for   $\Lambda_e^*$. 

Obviously for lattices that are \emph{isodual} or even \emph{self-dual}  it is irrelevant whether the condition is given for  $\Lambda_e$ or for $\Lambda_e^*$. It is interesting to note that the authors in \cite{OSB} were  
concentrating on the analysis of iso-dual or self dual lattices with large Hermite invariants. For such lattices our criterion  agrees with theirs.
 In the fading case, \cite{BO_ICC} and \cite{BO_TComm} focused on number field and division algebra lattices. Therefore  
 their code design principles automatically lead to lattices for which $\delta(\Lambda_e^*)$ is non zero.
\end{rem}

\section{Conclusions and perspectives} \label{conclusion}

In this work, we have shown that algebraic lattice constructions based on number fields and division algebras can achieve strong secrecy and semantic security universally over a wide range of fading and MIMO wiretap channels. Universality is a very desirable property for practical applications, since the eavesdropper's channel is not known at the transmitter. 
\paragraph*{Relevance and limitations of the channel model} 
Our model assumes perfect CSI of the legitimate channel at the receiver.
This assumption is not realistic for a fast fading channel, since in practice 
most of the available time slots would have to be used to transmit training symbols for channel estimation.

However, our channel model 
is not limited to fast fading, but only assumes the weak law of large numbers for the channel statistics. 
This includes for example a block fading model, where some fraction of each block can be used for channel estimation and the rest is left for data transmission.
We have also provided some results for the arbitrarily varying fading model in Section \ref{compound}, where Bob's channel oscillates most of the time above a certain threshold and Eve's channel oscillates mostly below another threshold, without necessarily converging in mean.

In such slow fading models a long code spanning many fading blocks is required to approach capacity. Our codes readily work in such a scenario due to their universality; decoding will succeed as long as the sum capacity of the fading blocks exceeds the target rate (up to a constant gap). 

We also note that even in the stationary ergodic case we require fast convergence in the law of large numbers
only for the eavesdropper, while the rate of convergence
can be slower for the legitimate channel. Here perfect CSI at the eavesdropper is assumed as a worst-case scenario.

A more realistic wiretap channel model with imperfect CSI at the receiver under a secrecy outage metric is left for future work.

\paragraph*{Technical improvements} 
Several technical improvements are needed before our lattice code construction can be implemented in practice. In particular, although the proposed families of lattices are deterministic, their construction is not explicit since it requires the computation of Hilbert class fields of high degree, for which efficient algorithms are currently not available. 

Moreover, our construction incurs a large gap to the secrecy capacity.
This gap might be reduced by improving the nested lattice construction, for example by taking suitable ideals of the ring of integers in the number field case\footnote{See the discussion in the extended version of \cite{ISIT2015_SISO}, available at http://arxiv.org/abs/1411.4591v2.}, or ideals of orders in the division algebra case, in order to optimize the code design according to the criteria proposed in Section \ref{design_criteria}.

\section*{Acknowledgements}

We are grateful to Matthieu Bloch for useful discussions about the secrecy capacity of arbitrary wiretap channels.
We would also like to thank the two anonymous reviewers for their detailed comments and suggestions which helped to improve the paper. 

\appendix



\subsection{Proof of Lemma~\ref{extended_Regev_Lemma}} \label{app:GeneralizedRegevLemma}
We need the following elementary fact characterizing the product of two Gaussian functions (see, e.g., \cite[Fact 1]{Peikert}):

Let $\Sigma_1, \Sigma_2 \succ \mathbf{0}$ be positive definite matrices, let $\Sigma_0 = \Sigma_1 + \Sigma_2 \succ \mathbf{0}$ and 
$\Sigma^{-1} = \Sigma_1^{-1} +\Sigma_2^{-1} \succ \mathbf{0}$, let $X, \mathbf{c}_1, \mathbf{c}_2 \in \C^k$ 
be arbitrary, and let $\mathbf{c}_3 \in \C^k$
such that $\Sigma^{-1}\mathbf{c}_3 = \Sigma_1^{-1}\mathbf{c}_1 + \Sigma_2^{-1}\mathbf{c}_2$. Then $\forall \mb{x} \in \C^k$,
\begin{equation}\label{eq:product}
f_{\sqrt{\Sigma_1}}(\mb{x}- \mathbf{c}_1) f_{\sqrt{\Sigma_2}}(\mb{x}-\mathbf{c}_2) = f_{\sqrt{\Sigma_0}}(\mathbf{c}_1-\mathbf{c}_2) f_{\sqrt{\Sigma}}(\mb{x}-\mathbf{c}_3)
\end{equation}

Now, we are ready to generalize Regev's lemma to correlated Gaussian distributions. 
Let $\mathbf{c}_3 = \Sigma\Sigma_2^{-1}\mb{x}$. We have
{\allowdisplaybreaks
\begin{eqnarray*}
g(\mb{x}) &=& \sum_{\mb{x}_1 \in \Lambda+\mathbf{c}} {\frac{f_{\sqrt{\Sigma_1}}(\mb{x}_1)}{f_{\sqrt{\Sigma_1}}(\Lambda+\mathbf{c})} f_{\sqrt{\Sigma_2}}(\mb{x}-\mb{x}_1)}\\
   &\stackrel{(a)}{=}& \sum_{\mb{x}_1 \in \Lambda+\mathbf{c}} {\frac{f_{\sqrt{\Sigma_0}}(\mb{x})}{f_{\sqrt{\Sigma_1}}(\Lambda+\mathbf{c})} f_{\sqrt{\Sigma}}(\mb{x}_1-\mathbf{c}_3)} \\
   &=& f_{\sqrt{\Sigma_0}}(\mb{x}) \frac{ \Sigma_{\mb{x}_1 \in \Lambda+\mathbf{c}} {f_{\sqrt{\Sigma}}(\mb{x}_1-\mathbf{c}_3)}}{f_{\sqrt{\Sigma_1}}(\Lambda+\mathbf{c})} \\
   &=& f_{\sqrt{\Sigma_0}}(\mb{x}) \frac{f_{\sqrt{\Sigma}}(\Lambda+\mathbf{c}-\mathbf{c}_3)}{f_{\sqrt{\Sigma_1}}(\Lambda+\mathbf{c})} \\
   &\stackrel{(b)}{\in}& f_{\sqrt{\Sigma_0}}(\mb{x}) \left[ \frac{1-\varepsilon}{1+\varepsilon}, \frac{1+\varepsilon}{1-\varepsilon} \right] \\
   &\stackrel{(c)}{=}& f_{\sqrt{\Sigma_0}}(\mb{x}) \left[ {1-4\varepsilon},{1+4\varepsilon} \right] \\
\end{eqnarray*}}
where $(a)$ is due to (\ref{eq:product}), $(b)$ follows from the definition of the flatness factor for correlated Gaussian distributions, and $(c)$ is because  $\varepsilon \leq \frac{1}{2}$. More precisely, since $\sqrt{\Sigma_3} \succeq \eta_{\varepsilon}(\Lambda)$, $f_{\sqrt{\Sigma_3}}(\Lambda+\mathbf{c}-\mathbf{c}_3) \in [\frac{1-\varepsilon}{V(\Lambda)},\frac{1+\varepsilon}{V(\Lambda)}]$; moreover, since $\Sigma_1 \succ \Sigma_3$, we also have $f_{\sqrt{\Sigma_1}}(\Lambda+\mathbf{c}) \in [\frac{1-\varepsilon}{V(\Lambda)},\frac{1+\varepsilon}{V(\Lambda)}]$. \hfill \IEEEQED

\subsection{Proof of Lemma \ref{linear_transformation}} \label{proof_linear}

Let $\boldsymbol\mu \in A(\Lambda +\mb{c})$. Then 
\begin{align*}
&\mathbb{P}\left\{ Y=\boldsymbol\mu\right\}=\mathbb{P}\left\{X=A^{-1}\boldsymbol\mu\right\}=\frac{f_{\sqrt{\Sigma}}(A^{-1}\boldsymbol\mu)}{f_{\sqrt{\Sigma}}(\Lambda+\mb{c})}\\
&=\frac{e^{-\boldsymbol\mu^{\dagger} (A^{-1})^{\dagger}\Sigma^{-1}A^{-1}\boldsymbol\mu}}{\sum_{\mb{z} \in \Lambda + \mb{c}} e^{-\mb{z}^{\dagger} \Sigma^{-1}\mb{z}}}.
\end{align*}
The thesis follows since by definition
\begin{align}
&D_{A(\Lambda +\mb{c}), \sqrt{A\Sigma A^{\dagger}}}(\boldsymbol\mu)=\frac{f_{\sqrt{A\Sigma A^{\dagger}}}(\boldsymbol\mu)}{\sum_{\boldsymbol\mu'\in A(\Lambda+\mb{c})}f_{\sqrt{A\Sigma A^{\dagger}}(\boldsymbol\mu')}} \notag\\
&=\frac{e^{-\boldsymbol\mu^{\dagger} (A^{-1})^{\dagger}\Sigma^{-1}A^{-1}\boldsymbol\mu}}{\sum_{\mb{z} \in \Lambda + \mb{c}} e^{-(A\mb{z})^{\dagger} (A^{-1})^{\dagger}\Sigma^{-1}A^{-1}(A\mb{z})}} \notag\\
&=\frac{e^{-\boldsymbol\mu^{\dagger} (A^{-1})^{\dagger}\Sigma^{-1}A^{-1}\boldsymbol\mu}}{\sum_{\mb{z} \in \Lambda + \mb{c}} e^{-\mb{z}^{\dagger} \Sigma^{-1}\mb{z}}}. \tag*{\IEEEQED}
\end{align}

\subsection{Proof of Lemma \ref{lemma_subgaussian}} \label{proof_subgaussian}
We have
\begin{align*}
&\mathbb{E}\left[ e^{2\Re(\mb{t}^{\dagger}A\mb{x})}\right]= \sum_{\mb{x} \in \Lambda + \mb{c}} D_{\Lambda + \mb{c}, \sigma}(\mb{x}) e^{2\Re(\mb{t}^{\dagger} A \mb{x})}\\
&= \sum_{\mb{x} \in \Lambda + \mb{c}} \frac{f_{\sigma}(\mb{x})}{f_{\sigma}(\Lambda+\mb{c})} e^{\Re(\mb{t}^{\dagger} A \mb{x})}.
\end{align*}
Therefore we can write
\begin{align*}
f_{\sigma}(\Lambda + \mb{c}) \mathbb{E}\left[ e^{2\Re(\mb{t}^{\dagger}A\mb{x})}\right]=\sum_{\mb{x} \in \Lambda + \mb{c}} \frac{1}{(\pi \sigma^2)^k} e^{-\frac{\norm{\mb{x}}^2}{\sigma^2}+2\Re(\mb{t}^{\dagger}A\mb{x})}.
\end{align*}
Using the identity
$$ \norm{\frac{\mb{x}}{\sigma} - \sigma A^{\dagger} \mb{t}}^2=\frac{\norm{\mb{x}}^2}{\sigma^2} - 2 \Re(\mb{t}^{\dagger}A \mb{x}) + \sigma^2 \norm{A^{\dagger} \mb{t}}^2,$$
we can rewrite the last expression as 
\begin{align*}
&\sum_{\mb{x} \in \Lambda + \mb{c}} \frac{1}{(\pi \sigma^2)^k} e^{-\norm{\frac{\mb{x}}{\sigma}-\sigma A^{\dagger} \mb{t}}^2+\sigma^2 \norm{A^{\dagger} \mb{t}}^2}\\
&=e^{\sigma^2 \norm{A^{\dagger}\mb{t}}^2} f_{\sigma}(\Lambda +\mb{c} -\sigma^2 A^{\dagger} \mb{t}).
\end{align*}
Thus we have 
$$\mathbb{E}\left[e^{2 \Re(\mb{t}^{\dagger}A\mb{x})}\right]=e^{\sigma^2 \norm{A^{\dagger} \mb{t}}^2} \frac{f_{\sigma}(\Lambda +\mb{c} - \sigma^2 A^{\dagger} \mb{t})}{f_{\sigma}(\Lambda + \mb{c})}$$
Adapting \cite[Lemma 4]{LLBS} to the complex case, we find that $\forall \mb{c} \in \C^k$ 
$$\frac{f_{\sigma,\mb{c}}(\Lambda)}{f_{\sigma}(\Lambda)} \in \left[\frac{1-\epsilon_{\Lambda}(\sigma)}{1+\epsilon_{\Lambda}(\sigma)},1\right].$$
Replacing $\mb{t}$ by $\mb{t}/2$, we obtain
\begin{equation}
\mathbb{E}\left[e^{\Re(\mb{t}^{\dagger}A\mb{x})}\right]=\frac{1+\epsilon_{\Lambda}(\sigma)}{1-\epsilon_{\Lambda}(\sigma)}e^{\frac{\sigma^2}{4} \norm{A^{\dagger} \mb{t}}^2}. \tag*{\IEEEQED}
\end{equation}

\subsection{Proof of Lemma~\ref{complex} }\label{app:numberfield}
Before giving the proof we need some notation.

Given an ideal $I$ of $F$,  the \emph{complementary ideal} of $I$ is defined as 
$I^{\vee}=\{ x \in F: {\Tr}_{F/\Q}(xI) \subseteq \mathbb{Z}\}.$
It is always an ideal of $F$. 


With this notation we have that
\begin{equation} \label{dual_ideal}
\psi(I)^*=2\overline{\psi(I_F^{\vee})},
\end{equation}
where overline means complex conjugation element wise\footnote{This result is well known but we do prove a more general version of it in Appendix \ref{proof_codifferent}.}.


\begin{IEEEproof}
Let us first assume that $\mathcal{I}$ is an integral ideal. In this case a classical result from algebraic number theory states that
\begin{equation*}
V{(\psi(\mathcal{I}))}=[\OO_F:\mathcal{I}] 2^{-k}\sqrt{|d_F|}.
 \end{equation*}
 Noticing that  $\sqrt{|\mathrm{N}_{F/\Q}(x)|}=|\mathrm{p}(\psi(x))|$ and using  the definition of the product distance we have that
$$
\delta(\psi(\mathcal{I}))=\frac{2^{\frac{k}{2}}}{|d_F|^{\frac{1}{4}}}\mathrm{min}(\mathcal{I}),
$$
where $\mathrm{min}(\mathcal{I}):=\underset{x \in \mathcal{I} \setminus \{0\}}{\mathrm{min}}\sqrt{\frac{|\mathrm{N}_{F/\Q}(x)|}{\mathrm{N}(\mathcal{I})}}$ and $\mathrm{N}(\mathcal{I})=[\OO_F:\mathcal{I}]$ is the norm of the ideal $\mathcal{I}$. From basic algebraic number theory we have that for  any element of $a \in \mathcal{I}$,    $|\mathrm{N}_{F/\Q}(a)|\,\mid \,\mathrm{N}(\mathcal{I})$ and the first claim follows.

Let us now assume that $\mathcal{I}$ is a genuine fractional ideal. In this case we can choose  an integer $n$ such that
$nI$ is an integral ideal. The extension to fractional ideals now follows as  for any lattice $\Lambda$ we have  $\delta(n\Lambda)=\delta(\Lambda)$. 

In \eqref{dual_ideal} we  saw that $\psi(\mathcal{I})^*$ is just a complex conjugated version of 
fractional ideal lattice $2\psi(\mathcal{I}^{\vee})$. Therefore  the 
last claim follows from the first one.
\end{IEEEproof}

\subsection{Proof of Lemma \ref{lemma_codifferent}} \label{proof_codifferent}
Let $x,y \in \mathcal{D}$. Then we have
{
\allowdisplaybreaks
\begin{align*}
&\tr_{\mathcal{D}/\Q}(xy)=\tr_{F/\Q}(\tr_{\mathcal{D}/F}(xy))= \tr_{F/\Q}(\Tr(\phi(xy)))=\\
&=\sum_{i=1}^{2k} \alpha_i(\Tr(\phi(xy)))=\sum_{i=1}^{2k} \Tr(\alpha_i(\phi(xy)))=\\
&=\Tr\Big(\sum_{i=1}^{2k} \alpha_i(\phi(xy))\Big)=2\Re\Tr\Big(\sum_{i=1}^{k} \alpha_i(\phi(xy))\Big)\\
&= 2\Re\Tr\Big(\sum_{i=1}^{k} \alpha_i(\phi(x))\alpha_i(\phi(y))\Big)=2\Re\Tr((\psi(x)^h)^{\dagger}\psi(y)).
\end{align*}
}
By definition,
$$\psi(\Gamma)^*=\{X \in M_{nk \times n}(\C): \;\; \forall y \in \Gamma, \;\;\Re(\Tr(X^{\dagger} \psi(y))) \in \ZZ\}$$
and so $2\psi(\Gamma^{\vee})^{h} \subseteq \psi(\Gamma)^*$. We would like to show that $2\psi(\Gamma^{\vee})^{h}=\psi(\Gamma)^*$.\\
The trace form $\tr_{\mathcal{D}/\Q}: \mathcal{D} \times \mathcal{D} \to \Q$ is a non-degenerate bilinear form on the $\Q$-vector space $\mathcal{D}$. Then, any full $\ZZ$-module in $\mathcal{D}$ has a dual basis in $\mathcal{D}$ \cite{Cohn}. In particular, if $\{w_1,\ldots,w_{2n^2k}\}$ is a basis of $\Gamma$ as $\ZZ$-module, then there exists a dual basis $\{w_1',\ldots,w_{2n^2k}'\}$ in $\mathcal{D}$ such that $\forall i,j \in \{1,\ldots, 2n^2k\}$, we have $\tr_{\mathcal{D}/\Q}(w_i'w_j)=\delta_{ij}$. Therefore,
$$\Re\Tr(2(\psi(w_i')^h)^{\dagger}\psi(w_j))=\tr_{\mathcal{D}/\Q}(w_i' w_j)=\delta_{ij}$$
and by definition of the codifferent, this implies that $\psi(w_i')^h \in \psi(\Gamma^{\vee})$. Since $2 \psi(\Gamma^{\vee})^h \subseteq \psi(\Gamma)^*$ and it contains a dual basis for $\psi(\Gamma)^*$, we can conclude that
$2 \psi(\Gamma^{\vee})^h=\psi(\Gamma)^*$. \hfill \IEEEQED

\subsection{Sketch of the proof of Proposition \ref{prop_weak_secrecy}} \label{weak_secrecy}
The proof follows the same steps as the proof of Theorem \ref{main}. Note that the bound (\ref{leakage_sum}) still holds and 
\begin{flalign*}
&\mathbb{E}_{H_e}\!\left[\frac{1}{k}\I(p_{\M};p_{\mb{z}|H_e})\right]\!\leq R\, \mathbb{P}\Big\{\prod\limits_{i=1}^k \Big(1+\frac{P\abs{h_{e,i}}^2}{\sigma_e^2} \Big)^{\frac{1}{k}}\!>e^{C_e+\delta}\Big\} \notag \\
&\!\!+\frac{1}{k}\mathbb{E}_{H_e}\!\!\left[\I(p_{\M};p_{\mb{z}|H_e}) \;\Big|\; \prod\limits_{i=1}^k \Big(1+\frac{P\abs{h_{e,i}}^2}{\sigma_e^2} \Big)^{\frac{1}{k}} \!\!\leq e^{C_e+\delta}\right] \!\!& 
\end{flalign*}
The first term vanishes because of (\ref{LLN_Eve_weak}) and the second term vanishes using Proposition \ref{prop_leakage_SISO} as before.\\
The proof of reliability is unchanged.  
\hfill \IEEEQED

\subsection{Sketch of the proof of Proposition \ref{rate_Gaussian}} \label{proof_Gaussian}
The proof is very similar to the proof of Theorem \ref{main_general}. We only outline the main steps. 
Note that $\forall \gamma>0$, for sufficiently large $k$, we have $\frac{h(\Lambda_b)}{k} > \mathrm{h}_b^2-\gamma$ and $\frac{h(\Lambda_e)}{k} > \mathrm{h}_e^2-\gamma$.
\subsubsection*{Secrecy} 
With the same notation as in Section \ref{SISO_secrecy}, we have $H_e=I$ and $\Sigma=\frac{P\sigma_e^2}{P+\sigma_e^2}$.
With the same scaling as in equation (\ref{alpha_e}), we can replace the bound (\ref{eta_bound}) with the following:
\begin{align*}
&\eta_{\varepsilon_k}(\Lambda_e) \leq \frac{2\sqrt{k}c}{\lambda_1(\Lambda_e^*)}=\frac{2 \sqrt{k} c}{h(\Lambda_e^*) V(\Lambda_e^*)^{\frac{1}{2k}}}=\frac{2\sqrt{k}cV(\Lambda_e)^{\frac{1}{2k}}}{h(\Lambda_e^*)}\\
&=\frac{2 \sqrt{k} c \sqrt{\pi e P}}{h(\Lambda_e^*) e^{\frac{R'}{2}}} \leq \frac{2 c \sqrt{\pi e P}}{(\mathrm{h}_e-\gamma) e^{\frac{R'}{2}}}.
\end{align*} 
We find that $\epsilon_{\Lambda_e}(\sqrt{\Sigma}) \to 0$ as long as
$$\sqrt{\Sigma}=\frac{\sqrt{P}\sigma_e}{\sqrt{P+\sigma_e^2}} > \frac{2c \sqrt{\pi e P}}{(\mathrm{h}_e-\gamma) \sqrt{2 \pi} e^{\frac{R'}{2}}}.$$
This condition is equivalent to 
\begin{equation} \label{secrecy_gaussian}
R' > \ln \frac{e}{\pi (\mathrm{h}_e-\gamma)} +\ln\left(1+\frac{P}{\sigma_e^2}\right).
\end{equation}
\subsubsection*{Reliability} With the same notation as in Section \ref{SISO_reliability}, we have $R=\sqrt{\frac{1+\rho_b}{\rho_b}}I$. With the scaling (\ref{alpha_e}), the error probability tends to zero if (\ref{condition_tau}) holds, that is 
\begin{align*}
&\frac{1}{4 \pi \sigma_b^2}> \frac{2c^2 k \rho_b}{(1+\rho_b) \lambda_1(\Lambda_b)^2}= \frac{2 c^2 k \rho_b}{h(\Lambda_b)^2 V(\Lambda_b)^{\frac{1}{k}}} \\
&= \frac{2c^2 \rho_b e^{R+R'}}{(1+\rho_b) (\mathrm{h}_b-\gamma) \pi e P}.
\end{align*}
Recalling that $\rho_b=P/\sigma_b^2$, after some elementary calculations we find 
\begin{equation} \label{reliability_gaussian}
R+R' < \ln\left(1+\frac{P}{\sigma_b^2}\right) - \ln \left(\frac{4}{\pi e}\right) +\ln (\mathrm{h}_b-\gamma). 
\end{equation}
Combining equations (\ref{secrecy_gaussian}) and (\ref{reliability_gaussian}), and taking $\gamma \to 0$, we get the desired result. \hfill \IEEEQED

\begin{small}

\bibliographystyle{IEEEtran}  

\bibliography{IEEEabrv,wiretap}

\begin{thebibliography}{10}
\providecommand{\url}[1]{#1}
\csname url@samestyle\endcsname
\providecommand{\newblock}{\relax}
\providecommand{\bibinfo}[2]{#2}
\providecommand{\BIBentrySTDinterwordspacing}{\spaceskip=0pt\relax}
\providecommand{\BIBentryALTinterwordstretchfactor}{4}
\providecommand{\BIBentryALTinterwordspacing}{\spaceskip=\fontdimen2\font plus
\BIBentryALTinterwordstretchfactor\fontdimen3\font minus
  \fontdimen4\font\relax}
\providecommand{\BIBforeignlanguage}[2]{{%
\expandafter\ifx\csname l@#1\endcsname\relax
\typeout{** WARNING: IEEEtran.bst: No hyphenation pattern has been}%
\typeout{** loaded for the language `#1'. Using the pattern for}%
\typeout{** the default language instead.}%
\else
\language=\csname l@#1\endcsname
\fi
#2}}
\providecommand{\BIBdecl}{\relax}
\BIBdecl

\bibitem{LLV_ISIT2016}
L.~Luzzi, C.~Ling, and R.~Vehkalahti, ``Almost universal codes for fading
  wiretap channels,'' in \emph{IEEE International Symposium on Information
  Theory (ISIT)}, Barcelona, Spain, July 2016.

\bibitem{Wyner}
A.~D. Wyner, ``The wire-tap channel,'' \emph{Bell System Technical Journal},
  vol.~54, pp. 1355--1387, Oct. 1975.

\bibitem{BlochBarros2011}
M.~Bloch and J.~Barros, \emph{Physical-layer security: from information theory
  to security engineering}.\hskip 1em plus 0.5em minus 0.4em\relax Cambridge
  University Press, 2011.

\bibitem{Csiszar}
I.~Csisz\'ar, ``{A}lmost {I}ndependence and {S}ecrecy {C}apacity,''
  \emph{{P}roblems of {I}nformation {T}ransmission}, vol.~32, no.~1, pp.
  40--47, 1996.

\bibitem{Bellare_Tessaro_Vardy}
M.~Bellare, S.~Tessaro, and A.~Vardy, ``Semantic security for the wiretap
  channel,'' in \emph{Advances in Cryptology}, ser. Lecture Notes in Computer
  Science, vol. 7417.\hskip 1em plus 0.5em minus 0.4em\relax Springer-Verlag,
  2012, pp. 294--311.

\bibitem{LLBS}
C.~Ling, L.~Luzzi, J.-C. Belfiore, and D.~Stehl\'e, ``Semantically secure
  lattice codes for the {G}aussian wiretap channel,'' \emph{IEEE Trans. Inform.
  Theory}, vol.~60, no.~10, pp. 6399--6416, Oct. 2014.

\bibitem{LeungYanCheong_Hellman}
S.~Leung-Yan-Cheong and M.~Hellman, ``The {G}aussian wire-tap channel,''
  \emph{IEEE Trans. Inf. Theory}, vol.~24, no.~4, pp. 451--456, July 1978.

\bibitem{Liang_Poor_Shamai_2008}
Y.~Liang, H.~Poor, and S.~Shamai, ``Secure {C}ommunication {O}ver {F}ading
  {C}hannels,'' \emph{IEEE Trans. Inf. Theory}, vol.~54, no.~6, pp. 2470--2492,
  June 2008.

\bibitem{Gopala_Lai_ElGamal_2008}
P.~Gopala, L.~Lai, and H.~El~Gamal, ``On the {S}ecrecy {C}apacity of {F}ading
  {C}hannels,'' \emph{IEEE Trans. Inf. Theory}, vol.~54, no.~10, pp.
  4687--4698, Oct. 2008.

\bibitem{Bloch_Laneman_PartialCSI}
M.~Bloch and J.~N. Laneman, ``Exploiting partial channel state information for
  secrecy over wiretap channels,'' \emph{IEEE J. Sel. Areas Commun.}, vol.~31,
  no.~9, pp. 1840--1849, September 2013.

\bibitem{Lin_Lin_MISOSE}
S.-C. Lin and P.-H. Lin, ``On secrecy capacity of fast fading multiple input
  wiretap channels with statistical {CSIT},'' \emph{IEEE Trans. Inf. Forensic
  Secur.}, vol.~8, no.~2, pp. 414--419, February 2013.

\bibitem{Maurer_Wolf}
U.~Maurer and S.~Wolf, ``Information-theoretic key agreement: From weak to
  strong secrecy for free,'' \emph{Lecture Notes in Computer Science}, pp.
  351--368, 2000.

\bibitem{He_Yener_2014}
X.~He and A.~Yener, ``{MIMO} wiretap channels with unknown and varying
  eavesdropper channel states,'' \emph{IEEE Trans. Inf. Theory}, vol.~60,
  no.~11, pp. 6844--6869, Nov. 2014.

\bibitem{Khisti_Wornell_2010}
A.~Khisti and G.~Wornell, ``Secure {T}ransmission {W}ith {M}ultiple {A}ntennas
  {P}art {II}: {T}he {MIMOME} {W}iretap {C}hannel,'' \emph{IEEE Trans. Inf.
  Theory}, vol.~56, no.~11, pp. 5515--5532, Nov. 2010.

\bibitem{Oggier_Hassibi_2011}
F.~Oggier and B.~Hassibi, ``The secrecy capacity of the {MIMO} wiretap
  channel,'' \emph{IEEE Trans Inf Theory}, vol.~57, no.~7, pp. 4961--4972,
  2011.

\bibitem{BustinLiuPoorShamai2009}
R.~Bustin, R.~Liu, H.~V. Poor, and S.~Shamai~(Shitz), ``An {MMSE} approach to
  the secrecy capacity of the {MIMO} {G}aussian wiretap channel,''
  \emph{EURASIP J. Wireless Commun. Netw.}, vol. 2009, no.~1, p. 370970, Jul
  2009.

\bibitem{LiuShamai2009}
T.~Liu and S.~S. (Shitz), ``A note on the secrecy capacity of the multi-antenna
  wiretap channel,'' \emph{IEEE Trans. Inf. Theory}, vol.~55, no.~6, pp.
  2547--2553, Jun 2009.

\bibitem{Khina_Kochman_Khisti}
A.~Khina, Y.~Kochman, and A.~Khisti, ``The {MIMO} wiretap channel decomposed,''
  \emph{IEEE Trans. Inf. Theory}, 2017, to appear.

\bibitem{Lin_Lin}
S.-C. Lin and C.-L. Lin, ``On secrecy capacity of fast fading {MIMOME} wiretap
  channels with statistical {CSIT},'' \emph{IEEE Trans. Wireless Commun.},
  vol.~13, no.~6, pp. 3293--3306, June 2014.

\bibitem{Lapidoth_Narayan}
A.~Lapidoth and P.~Narayan, ``Reliable communication under channel
  uncertainty,'' \emph{IEEE Trans. Inf. Theory}, vol.~44, no.~6, pp.
  2148--2177, Oct 1998.

\bibitem{LiangKramerPoorShamai2009}
Y.~Liang, G.~Kramer, H.~V. Poor, and S.~S. (Shitz), ``Compound wiretap
  channels,'' \emph{EURASIP J. Wireless Commun. Netw., Special issue on
  wireless physical layer security}, pp. 1--13, 2009.

\bibitem{Ekrem_Ulukus_2010}
E.~Ekrem and S.~Ulukus, ``On {G}aussian {MIMO} compound wiretap channels,'' in
  \emph{44th Annual Conference on Information Sciences and Systems (CISS)},
  March 2010, pp. 1--6.

\bibitem{BjelakovicBocheSommerfeld2013}
I.~Bjelakovi\'c, H.~Boche, and J.~Sommerfeld, ``Secrecy results for compound
  wiretap channels,'' \emph{Probl. Inf. Transmission}, vol.~49, no.~1, pp.
  73--98, Mar 2013.

\bibitem{Boche_Schaefer_Poor}
H.~Boche, R.~F. Schaefer, and H.~V. Poor, ``On the continuity of the secrecy
  capacity of compound and arbitrarily varying wiretap channels,'' \emph{IEEE
  Trans. Inf. Forensics and Security}, vol.~10, no.~12, pp. 2531--2546, Dec
  2015.

\bibitem{Han_2003}
T.~S. Han, \emph{Information-Spectrum Methods in Information Theory}.\hskip 1em
  plus 0.5em minus 0.4em\relax New York, NY, USA: Springer, 2003.

\bibitem{Loyka_Charalambous_2016}
S.~Loyka and C.~D. Charalambous, ``A general formula for compound channel
  capacity,'' \emph{IEEE Trans. Inf. Theory}, vol.~62, no.~7, pp. 3971--3991,
  July 2016.

\bibitem{Mahdavifar_Vardy}
H.~Mahdavifar and A.~Vardy, ``Achieving the secrecy capacity of wiretap
  channels using polar codes,'' \emph{IEEE Trans. Inf. Theory}, vol.~57,
  no.~10, Oct 2011.

\bibitem{Subramanian}
A.~Subramanian, A.~Thangaraj, M.~Bloch, and S.~W. McLaughlin, ``Strong secrecy
  on the binary erasure wiretap channel using large-girth {LDPC} codes,''
  \emph{IEEE Trans. Inf. Forensic Secur.}, vol.~6, no.~3, pp. 585--594, 2011.

\bibitem{Gulcu_Barg}
T.~C. Gulcu and A.~Barg, ``Achieving secrecy capacity of the wiretap channel
  and broadcast channel with a confidential component,'' \emph{IEEE Trans. Inf.
  Theory}, vol.~63, no.~2, pp. 1311--1324, Feb. 2017.

\bibitem{Wei_Ulukus}
Y.-P. Wei and S.~Ulukus, ``Polar coding for the general wiretap channel,'' in
  \emph{IEEE Information Theory Workshop (ITW)}, April 2015.

\bibitem{Belfiore_Oggier}
J.~C. Belfiore and F.~Oggier, ``Secrecy gain: A wiretap lattice code design,''
  in \emph{International Symposium On Information Theory Its Applications
  (ISITA)}, Oct 2010, pp. 174--178.

\bibitem{OSB}
F.~Oggier, P.~Sol\'e, and J.-C. Belfiore, ``Lattice codes for the wiretap
  {G}aussian channel: Construction and analysis,'' \emph{IEEE Trans. Inform.
  Theory}, vol.~62, no.~10, pp. 5690--5708, Oct. 2016.

\bibitem{BO_ICC}
J.-C. Belfiore and F.~Oggier, ``Lattice code design for the {R}ayleigh fading
  wiretap channel,'' in \emph{IEEE International Conference on Communications
  (ICC)}, 2011.

\bibitem{BO_TComm}
J.-C. {Belfiore} and F.~{Oggier}, ``An error probability approach to {MIMO}
  wiretap channels,'' \emph{IEEE Trans. Commun.}, vol.~61, no.~8, pp.
  3396--3403, August 2013.

\bibitem{KHHV}
D.~Karpuk, A.-M. Ernvall-Hyt\"onen, C.~Hollanti, and E.~Viterbo, ``Probability
  estimates for fading and wiretap channels from ideal class zeta functions,''
  \emph{Adv. Math. Commun.}, vol.~9, no.~4, pp. 391--413, 2015.

\bibitem{Ong_Oggier}
S.~S. Ong and F.~Oggier, ``Wiretap lattice codes from number fields with no
  small norm elements,'' \emph{Designs, Codes and Cryptography}, vol.~73,
  no.~2, pp. 425--440, 2014.

\bibitem{Bloch_Laneman}
M.~R. Bloch and J.~N. Laneman, ``Strong secrecy from channel resolvability,''
  \emph{IEEE Trans. Inform. Theory}, vol.~59, no.~12, Dec 2013.

\bibitem{Hamed}
H.~Mirghasemi and J.-C. Belfiore, ``Lattice code design criterion for {MIMO}
  wiretap channels,'' in \emph{IEEE Information Theory Workshop (ITW)}, 2015.

\bibitem{Liu_Yan_Ling_secrecy}
L.~Liu, Y.~Yan, and C.~Ling, ``Secrecy-good polar lattices with optimal shaping
  for the {G}aussian wiretap channels,'' in \emph{IEEE Information Theory
  Workshop (ITW)}, April 2015.

\bibitem{Tyagi_Vardy_2014}
H.~Tyagi and A.~Vardy, ``Explicit capacity-achieving coding scheme for the
  {G}aussian wiretap channel,'' in \emph{IEEE International Symposium on
  Information Theory (ISIT)}, June 2014, pp. 956--960.

\bibitem{Tavildar_Viswanath}
S.~Tavildar and P.~Viswanath, ``Approximately universal codes over slow-fading
  channels,'' \emph{IEEE Trans. Inf. Theory}, vol.~52, no.~7, pp. 3233--3258,
  July 2006.

\bibitem{Ordentlich_Erez}
O.~Ordentlich and U.~Erez, ``Precoded integer-forcing universally achieves the
  {MIMO} capacity to within a constant gap,'' \emph{IEEE Trans. Inf. Theory},
  vol.~61, no.~1, pp. 323--340, Jan 2015.

\bibitem{LV}
L.~Luzzi and R.~Vehkalahti, ``Almost universal codes achieving ergodic {MIMO}
  capacity within a constant gap,'' \emph{IEEE Trans. Inform. Theory}, vol.~63,
  no.~5, pp. 3224--3241, May 2017.

\bibitem{Campello_Ling_Belfiore}
A.~Campello, C.~Ling, and J.-C. Belfiore, ``Universal lattice codes for {MIMO}
  channels,'' available at: https://arxiv.org/pdf/1603.09263.pdf.

\bibitem{Campello_Ling_Belfiore_secrecy}
A.~{Campello}, C.~{Ling}, and J.-C. {Belfiore}, ``Semantically secure lattice
  codes for compound {MIMO} channels,'' submitted.

\bibitem{Peikert_Rosen}
C.~Peikert and A.~Rosen, ``Lattices that admit logarithmic worst-case to
  average-case connection factors,'' in \emph{Proc. of the 39-th ACM Symp. on
  the Theory of Computing (STOC)}, 2007, pp. 478--487.

\bibitem{ISIT2015_SISO}
R.~Vehkalahti and L.~Luzzi, ``Number field lattices achieve {G}aussian and
  {R}ayleigh channel capacity within a constant gap,'' in \emph{IEEE
  International Symposium on Information Theory (ISIT)}, June 2015, pp.
  436--440.

\bibitem{Micciancio_Regev}
D.~Micciancio and O.~Regev, ``Worst-case to average-case reductions based on
  {G}aussian measures,'' in \emph{Proc. Ann. Symp. Found. Computer Science},
  Rome, Italy, Oct. 2004, pp. 372--381.

\bibitem{Peikert}
C.~Peikert, ``An efficient and parallel {G}aussian sampler for lattices,'' in
  \emph{Proc. CRYPTO}, vol. 6223.\hskip 1em plus 0.5em minus 0.4em\relax
  Springer-Verlag, 2010, pp. 80--97.

\bibitem{Banaszczyk}
W.~Banaszczyk, ``New bounds in some transference theorems in the geometry of
  numbers,'' \emph{Math. Ann.}, vol. 296, pp. 625--635, 1993.

\bibitem{Re09}
O.~Regev, ``On lattices, learning with errors, random linear codes, and
  cryptography,'' \emph{J. ACM}, vol.~56, no.~6, pp. 34:1--34:40, 2009.

\bibitem{Micciancio_Peikert}
D.~Micciancio and C.~Peikert, ``Trapdoors for lattices: Simpler, tighter,
  faster, smaller,'' in \emph{Advances in Cryptology - EUROCRYPT}, ser. Lecture
  Notes in Computer Science, vol. 7237, 2012, pp. 700--718.

\bibitem{Cohn}
P.~M. Cohn, \emph{Algebra}.\hskip 1em plus 0.5em minus 0.4em\relax John Wiley
  and Sons, Aberdeen, 1974, vol.~1.

\bibitem{Xing}
C.~Xing, ``Diagonal lattice space-time codes from number fields and asymptotic
  bounds,'' \emph{IEEE Trans. Inform. Theory}, vol.~53, no.~11, pp. 3921--3926,
  Nov 2007.

\bibitem{Martinet}
J.~Martinet, ``Tours de corps de classes et estimations de discriminants,''
  \emph{Inventiones Mathematicae}, no.~44, pp. 65--73, 1978.

\bibitem{Hajir_Maire_2002}
F.~Hajir and C.~Maire, ``Tamely ramified towers and discriminant bounds for
  number fields - {II},'' \emph{J. Symbolic Computation}, vol.~33, pp.
  415--423, 2002.

\bibitem{GB96}
X.~Giraud and J.-C. Belfiore, ``Constellations matched to the {R}ayleigh fading
  channel,'' \emph{IEEE Trans. Inform. Theory}, vol.~42, no.~1, pp. 106--115,
  Jan. 1996.

\bibitem{Telatar}
E.~Telatar, ``Capacity of multi-antenna {G}aussian channels,'' \emph{Europ.
  Trans. Telecomm.}, vol.~10, no.~6, pp. 585--595, Nov.-Dec. 1999.

\bibitem{Bloch_Laneman_2008}
M.~Bloch and J.~N. Laneman, ``On the secrecy capacity of arbitrary wiretap
  channels,'' in \emph{Allerton Conference on Communication, Control and
  Computing}, September 2008, pp. 818--825.

\bibitem{ElGamal_Caire_Damen}
H.~El~Gamal, G.~Caire, and M.~O. Damen, ``Lattice coding and decoding achieve
  the optimal diversity-multiplexing tradeoff of {MIMO} channels,'' \emph{IEEE
  Trans. Inform. Theory}, vol.~50, no.~6, pp. 968--985, Jun. 2004.

\bibitem{Milnor_Husemoller}
J.~Milnor and D.~Husemoller, \emph{Symmetric bilinear forms}, ser. A Series of
  Modern Surveys in Mathematics.\hskip 1em plus 0.5em minus 0.4em\relax
  Springer-Verlag Berlin Heidelberg, 2014.

\bibitem{OBV}
F.~E. Oggier, J.-C. Belfiore, and E.~Viterbo, \emph{Cyclic division algebras: A
  tool for space-time coding}.\hskip 1em plus 0.5em minus 0.4em\relax
  Foundations and Trends in Communications and Information Theory, 2007,
  vol.~4, no.~1.

\bibitem{Reiner}
I.~Reiner, \emph{Maximal Orders}.\hskip 1em plus 0.5em minus 0.4em\relax
  Academic Press, New York, 1975.

\bibitem{YB07}
S.~Yang and J.~C. Belfiore, ``Optimal space-time codes for the {MIMO}
  amplify-and-forward cooperative channel,'' in \emph{International Zurich
  Seminar on Communications}, 2006, pp. 122--125.

\bibitem{Lu}
H.-F. Lu, ``Constructions of multi-block space-time coding schemes that achieve
  the diversity-multiplexing tradeoff,'' \emph{IEEE Trans. Inform. Theory},
  vol.~54, no.~8, pp. 3790--3796, Aug 2008.

\bibitem{LSV}
B.~Linowitz, M.~Satriano, and R.~Vehkalahti, ``A non-commutative analogue of
  the {O}dlyzko bounds and bounds on performance for space-time lattice
  codes,'' \emph{IEEE Trans. Inform. Theory}, vol.~61, no.~4, pp. 1971--1984,
  Apr 2015.

\bibitem{VHLR}
R.~Vehkalahti, C.~Hollanti, J.~Lahtonen, and K.~Ranto, ``On the densest {MIMO}
  lattices from cyclic division algebras,'' \emph{IEEE Trans. Inform. Theory},
  vol.~55, no.~8, pp. 3751--3780, Aug 2009.

\bibitem{Schaefer_Loyka_2015}
R.~F. Schaefer and S.~Loyka, ``The secrecy capacity of compound {G}aussian
  {MIMO} wiretap channels,'' \emph{IEEE Trans. Inf. Theory}, vol.~61, no.~10,
  pp. 5535--5552, Oct. 2015.

\bibitem{Bjelakovic_Boche_Sommerfeld}
I.~Bjelakovi\'c, H.~Boche, and J.~Sommerfeld, ``Capacity results for
  arbitrarily varying wiretap channels,'' in \emph{Information theory,
  combinatorics, and search theory}.\hskip 1em plus 0.5em minus 0.4em\relax
  Springer, 2013, pp. 123--144.

\bibitem{KKH}
A.~Karrila, D.~Karpuk, and C.~Hollanti, ``On analytical and geometric lattice
  design criteria for wiretap coset codes,'' available at:
  http://arxiv.org/abs/1609.07723.

\end{thebibliography}

\end{small}

\end{document}